\documentclass[twocolumn,showpacs,superscriptaddress,floatfix,prb]{revtex4-2}
\usepackage{graphicx,amsfonts,amssymb,amsmath,enumerate}
\usepackage{braket}
\usepackage{graphicx}
\usepackage{dcolumn}
\usepackage{bm}
\usepackage{color}
\usepackage[colorlinks,allcolors=blue]{hyperref}
\usepackage{physics}
\usepackage{subfigure}
\begin{document}

\preprint{APS/123-QED}

\title{Superfluidity of total angular momentum} 

\author{Yeyang Zhang}
\affiliation{International Center for Quantum Materials, School of Physics, Peking 
 University, Beijing 100871, China}
\author{Ryuichi Shindou}
\email{rshindou@pku.edu.cn}
\affiliation{International Center for Quantum Materials, School of Physics, Peking University, Beijing 100871, China}


\date{\today}

\begin{abstract}
Spontaneous symmetry breaking of a U(1) symmetry 
in interacting systems leads to superfluidity of a 
corresponding conserved charge. We generalize the superfluidity to systems with 
U(1) symmetries acting on both matter fields and 2D spatial coordinates. 
Such systems can be effectively realized in easy-plane ferromagnetic systems with 
spin-orbit coupling where the conserved charge is a total angular momentum. We clarify that 
under a steady injection of spin angular momentum, 
the superfluid of the total angular momentum shows 
spacetime oscillations of the spin density and 
geometry-dependent spin hydrodynamics. 
We also demonstrate that the steady spin injection destabilizes 
the superfluid of total angular momentum, causing a dissipation effect in 
its spin hydrodynamic properties. Although a stability analysis shows that 
the superfluid under the spin injection is nonideal, the unique spin-transport features
persist with weak dissipation of the spin angular momentum. 
Our study broadens the comprehension of superfluidity and sheds new light 
on the interplay between symmetries and phases of matter.
\end{abstract}

\maketitle


\section{\label{sec1}Introduction}
The discovery of superfluidity~\cite{Kapitza1938, Allen1938, London1964} 
is a milestone in the history of physics. 
Exotic macroscopic quantum phenomena in superfluids 
are explained by the condensation of bosonic atoms~\cite{Anderson1995, Davis1995} or neutral Cooper pairs~\cite{Zwierlein2005}. Spontaneous symmetry breaking (SSB) of a U(1) global gauge symmetry leads to Goldstone modes with gapless and linear dispersions~\cite{Goldstone1961, Nambu1961_1, Nambu1961_2}, which enables dissipationless mass currents. By alternative U(1) symmetries, the superfluidity can be generalized to spin~\cite{Vuorio1974, Sonin1978_1, Grein2009, Sonin2010, Takei2014_1, Takei2014_2, Yuan2018, Mao2022} and excitonic~\cite{Lozovik1975, Sonin1977, Sonin1978_2, Eisenstein2004, Li2017, Zhang2022} currents. 
 
General relations between Goldstone modes and SSB of continuous symmetries 
are derived in the literature
~\cite{Nambu2004, Watanabe2011, Watanabe2012, Watanabe2014}, while they 
mostly considered continuous \textit{internal symmetries} that transform only 
field operators locally. \textit{Spacetime symmetries} act on both 
field operators and spacetime coordinates~\cite{Peskin2018}, 
and the symmetries bring about fundamental physical consequences 
such as the relativistic  spin-orbit coupling (SOC). The continuous spacetime symmetries can be spontaneously broken in spinful superfluids in cold-atom systems~\cite{Stanescu2008, Wang2010, Jian2011, Ozawa2013, Zhou2013, Zhai2015}.  
Nonetheless, it remains largely unexplored how the SSB of the continuous 
spacetime symmetries affects the hydrodynamic transport of ``charges" associated 
with the broken spacetime symmetries. 

In this paper, we generalize the concept of superfluidity to the SSB of 
continuous spacetime symmetries. As a physical system, we consider the 
superfluidity of total angular momentum, where a joint U(1) rotational 
symmetry of an in-plane spin vector and two-dimensional (2D) 
spatial coordinates is spontaneously broken. The superfluid of total angular momentum is 
nothing but a spin superfluid~\cite{Vuorio1974, Sonin1978_1, Grein2009, Sonin2010, Takei2014_1,  Takei2014_2, Yuan2018, Mao2022} in the presence of the SOC. It can 
be effectively realized in a ferromagnet and a spin-triplet exciton 
condensate~\cite{Halperin1968, Hakioglu2007, Pikulin2014} with
easy-plane spin anisotropy. 

We derive an effective field theory of a Goldstone mode 
in the total-angular-momentum superfluid and solve its classical 
equation of motion in the presence of a steady injection of spin. 
We find that the total-angular-momentum superfluid 
shows spacetime oscillations of spin density and current  
under the spin injection, which contrasts with conventional 
spin superfluid without SOC~\cite{Vuorio1974, Sonin1978_1, Grein2009, Sonin2010, Takei2014_1, Takei2014_2, Yuan2018, Mao2022}.  We also uncover unique 
geometry dependence and non-reciprocity in its hydrodynamic spin transport, 
which are absent in systems only with discrete internal rotational 
symmetry~\cite{Sonin1977, Sonin1981, Sonin2010}. Especially, 
when the system is in a circular geometry with finite curvature, 
the spin hydrodynamics depends on the direction of the spin flow as 
well as the curvature of the system. The proposed spatial and temporal spin textures can be experimentally detected by magnetic force microscopy~\cite{Milde2013, Seddon2021} and X-ray pump-probe microscopy~\cite{Li2016}, respectively.

We also show that unlike in the conventional spin superfluid, 
the steady spin injection destabilizes the total-angular-momentum superfluid. Landau argued that uniform superfluids moving slower than a critical velocity realize states at local minima of energy, so the superfluidity is protected from any dissipative 
perturbation~\cite{Sonin2010, Landau1941, Girvin2019, Zhai2021}. 
Based on the same spirit as Landau's argument, we demonstrate that the total-angular-momentum superfluid is \textit{not} an energy-local-minimum state in the presence of the spin 
injection and decay processes to lower energy states bring about 
a dissipation effect in the spin hydrodynamic properties of the superfluid. Nonetheless, 
we show that the qualitative behavior of the hydrodynamic spin transport remains unchanged and distinct from a non-superfluid~\cite{Sonin2010}.

The structure of the remaining part of the paper is as follows. 
In Sec.~\ref{sec2}, we introduce the field theory with the U(1) spacetime symmetry and offer possible realizations of the theory in two microscopic models. In Sec.~\ref{sec3}, we show spin and orbital parts of Noether's current corresponding to the U(1) spacetime symmetry. In Sec.~\ref{sec4}, we propose a spin-injection model with different geometries. 
The spacetime distribution of a Goldstone mode of the theory are proposed from a 
classical solution of the mesoscopic model under a steady injection of spin current. In Sec.~\ref{sec5}, we demonstrate the possibility of energy dissipation by a stability analysis against a local deformation. Thereby, we show that the energy of the classical solution 
can be further lowered by local deformations, indicating a decay process to  
lower energy states. The effect of the dissipation on the classical motion of the Goldstone 
mode is also discussed. 
Sec.~\ref{sec6} is devoted to a summary.

A number of appendices are offered to help understand the main text of the paper. 
In Appendix \ref{appendixA}, we provide detailed derivations of the field theory 
from the microscopic models proposed in Sec.~\ref{sec2}.
Detailed derivations and solutions for Secs.~\ref{sec3} and ~\ref{sec4} are displayed in Appendices \ref{appendixB} and \ref{appendixC}, respectively.  
Appendices \ref{appendixD} and \ref{appendixE} provide instrumental 
details of Sec.~\ref{sec5}, where we thoroughly discuss the possibility of the 
dissipation proposed in Sec.~\ref{sec5} and its effects on the equation of 
motion (EOM). In Appendix \ref{appendixF}, we discuss solutions of 
the spin-injection model at some special parameter points. 
In Appendix \ref{appendixG}, we use the same stability analysis as 
in Sec.~\ref{sec5} and Appendix~\ref{appendixD}, and derive the Landau criterion of a conventional 
superfluid. We use this classic and simple example to demonstrate 
the validity of our stability analysis in Sec.~\ref{sec5}.
In Appendix \ref{appendixH}, we present how to construct the local 
deformation of the classical solution of the EOM.


\section{\label{sec2}Model}
Consider a complex bosonic field $\phi\equiv\phi_x+i\phi_y$ 
in three dimensions (3D), 
where the 2D real and time-reversally-odd vector field $(\phi_x,\phi_y)$ and two of three spatial coordinates $(x,y)$ transform under a joint U(1) rotation around $z$-direction,
\begin{align}
\label{eqn1}
\phi\rightarrow\phi e^{i\epsilon},\quad
x+i y \rightarrow (x+i y) e^{i\epsilon}. 
\end{align}
The vector field here stands for a spin vector in physical systems. In the presence of the time-reversal symmetry, 
$\phi\rightarrow-\phi^\dagger$, $t\rightarrow -t$, $i\rightarrow -i$, the SSB of the joint U(1) symmetry is characterized by a real-time field theory of $\phi$,  
\begin{align}
\label{eqn4}
\mathcal{L}_{\phi}=&\frac{\eta_1^2}{2}(\partial_t\phi^\dagger)(\partial_t\phi)-\frac{\eta_1^2c_\perp^2}{2}(\partial_j\phi^\dagger)(\partial_j\phi)\nonumber\nonumber\\
&-\frac{\eta_1^2c_z^2}{2}(\partial_z\phi^\dagger)(\partial_z\phi)-\frac{\alpha\eta_1^2 c_\perp^2}{4}[(\partial_-\phi)^2+(\partial_+\phi^\dagger)^2]\nonumber\\
&-\frac{U}{2}(\phi^\dagger\phi-\rho_0)^2,
\end{align}
where $\partial_\pm\equiv\partial_x\pm i\partial_y$, $j=x,y$. A global phase of $\phi$ 
is properly chosen so that $\alpha$ is real and 
positive.  $\alpha\in\mathbb{R}$ and $\alpha>0$. 
We assume $0<\alpha<1$ for the stability of the theory. Ground states for $\rho_0>0$ break the U(1) symmetry by uniform field configurations $\phi=\sqrt{\rho_0}e^{i\theta_0}$. 

The joint nature of the rotational symmetry 
results from spin-orbit locking in solid-state materials with SOC. 
An example of the joint rotational 
symmetry breaking is in the $XY$ ferromagnet phase on a 3D trigonal or hexagonal lattice. A localized spin model for spin-orbit coupled magnets generally comprises symmetric 
and antisymmetric exchange interactions, 
\begin{align}
\label{eqn_h_spin}
H_{\rm spin} = \frac{1}{2}\sum_{{\bm i},{\bm j}} \sum_{\mu,\nu=x,y,z} 
\Big({\cal J}_{{\bm i\bm j},\mu\nu} 
+ {\cal D}_{{\bm i \bm j},\mu\nu}\Big) S_{{\bm i},\mu} S_{{\bm j},\nu},
\end{align}
with lattice sites ${\bm i}\equiv ({\bm i}_{\perp},i_z)$, 
${\bm j}\equiv ({\bm j}_{\perp},j_z)$, spin operators $S_{{\bm i},\mu}$ ($\mu=x,y,z$),  
${\cal J}_{{\bm i\bm j},\mu\nu}={\cal J}_{{\bm i\bm j},\nu\mu}$, 
and ${\cal D}_{{\bm i\bm j},\mu\nu}=-{\cal D}_{{\bm i\bm j},\nu\mu}$. 
${\bm i}_{\perp}$ and $i_z$ are $xy$ and $z$ coordinates of the lattice site ${\bm i}$ on the lattices, respectively.
For ${\bm i}={\bm j}$, ${\cal D}_{{\bm i}{\bm i},\mu\nu}=0$ while ${\cal J}_{{\bm i}{\bm i},\mu\nu}$ gives an single-ion spin anisotropy energy. 
Suppose that $H_{\rm spin}$ has an easy-plane spin anisotropy in the $XY$ spin plane, and undergoes a 
quantum phase transition of ferromagnetic ordering of the $XY$ spins, $\vec{S}_{{\bm i},\perp} \equiv (S_{{\bm i},x},S_{{\bm i},y})$. When $H_{\rm spin}$ belongs to a point group of $C_{3i}$, $D_{3d}$, $C_{3v}$, $C_{3h}$, $C_{6h}$, $C_{6v}$, $D_{3h}$, or $D_{6h}$,
spin hydrodynamics of the $XY$ spin near the ferromagnetic transition point is described by Eq.~(\ref{eqn4}), where $\phi({\bm r}_i) \equiv S_{{\bm i},x}+iS_{{\bm i},y}$ and 
a 2 by 2 symmetric matrix comprised of ${\cal J}_{{\bm i \bm j},\mu\nu}$ 
($\mu,\nu=x,y$) determines the strength of the $\alpha$ term. Specifically, for each bond $({\bm i},{\bm j})$, the 2 by 2 matrix ${\cal J}_{{\bm i}{\bm j},\mu\nu}$ has real eigenvalues $\lambda_{{\bm i}{\bm j},m}$ and eigenvectors ${\bm t}_{{\bm i}{\bm j},m}$ ($m=1,2$). Defining $\Delta \lambda_{{\bm i}{\bm j}} \equiv \lambda_{{\bm i}{\bm j},1}-\lambda_{{\bm i}{\bm j},2}$, $a_{{\bm i}{\bm j},\perp} \equiv |{\bm i}_{\perp}-{\bm j}_{\perp}|$, $\Omega$ as the total volume of the material, and $\varepsilon_{{\bm i}{\bm j},\perp}$ as an angle between ${\bm t}_{{\bm i}{\bm j},1}$ in the $XY$ spin plane and ${\bm i}_{\perp}-{\bm j}_{\perp}$ in the $xy$ coordinate plane, $\alpha$ in Eq.~(\ref{eqn4}) is given by a sum of $\Delta\lambda_{{\bm i}{\bm j}}$ with a phase $e^{2i\varepsilon_{{\bm i}{\bm j},\perp}}$ over all the bonds [see Appendix \ref{appendixA1}], 
\begin{align}
\alpha  \eta^2_1 c_\perp^2 = \frac{1}{8\Omega}\sum_{{\bm i},{\bm j}} a^2_{{\bm i}{\bm j},\perp} \!\ \Delta\lambda_{{\bm i}{\bm j}} \!\ e^{2i\varepsilon_{{\bm i}{\bm j},\perp}}. 
\end{align}
The antisymmetric exchange interactions ${\cal D}_{{\bm i \bm j},\mu\nu}$ do not contribute to the spin hydrodynamic equation near the critical point. 
Note that the joint rotational symmetry in solid-state materials 
with periodic lattices must be discrete due to the lattices. In fact, 
${\cal L}_{\phi}$ for the $XY$ ferromagnet on the hexagonal lattices 
generally acquires additional hexagonal easy spin axes 
within the $XY$ spin plane 
in the form of $\tilde{c}_6\phi^6 + {\rm H.c.}$. 
The symmetries also allow $\tilde{\alpha}_6 \phi^3 (\partial^2_{+} \phi) + {\rm H.c.}$ in the action ${\cal L}_{\phi}$. 
Near the ferromagnetic transition point, however, the additional term becomes effectively negligible compared to the $\alpha$ term in a hydrodynamic regime 
with an intermediate crossover length scale [see Appendix \ref{appendixA1}]. 

Another example of Eq.~(\ref{eqn4}) can be found in the triplet exciton condensate phase in semiconductors with Rashba spin-orbit interaction. We consider a 2D model for simplicity. It may be regarded as an effective model of 3D systems. Suppose that electron energy bands near the conduction-band bottom and valence-band top in the semiconductors can be approximated by a model with continuous rotational symmetry, 
\begin{align}
\label{eqn_h_ex}
H_{\mathrm{ex}}=&\int d^2\bm{r}\bm{a}^\dagger[(-\frac{\partial_i^2}{2m_0}+\epsilon_{g0})\bm{\sigma}_0+\xi_{R0}(-i\partial_y\bm{\sigma}_x+i\partial_x\bm{\sigma}_y)]\bm{a}\nonumber\\
&+\int d^2\bm{r}\bm{b}^\dagger[(\frac{\partial_i^2}{2m^{\prime}_0}-\epsilon_{g0})\bm{\sigma}_0\nonumber+\xi^{\prime}_{R0}(i\partial_y\bm{\sigma}_x-i\partial_x\bm{\sigma}_y)]\bm{b}\nonumber\\
&+\int d^2\bm{r}(\Delta_t\bm{a}^\dagger\bm{\sigma}_0\bm{b}+\Delta_t^*\bm{b}^\dagger\bm{\sigma}_0\bm{a})\nonumber\\
&+\frac{g_{s0}}{2}\sum_{\sigma,\sigma'=\uparrow,\downarrow}\int d^2\bm{r}(a^\dagger_\sigma a^\dagger_{\sigma'}a_{\sigma'}a_\sigma+b^\dagger_\sigma b^\dagger_{\sigma'}b_{\sigma'}b_\sigma\nonumber\\
&\hspace{2cm}+2\xi_1 a^\dagger_\sigma b^\dagger_{\sigma'}b_{\sigma'}a_\sigma),
\end{align}
with $i=x,y,z$, ${\bm a} \equiv (a_{\uparrow},a_{\downarrow})$ and ${\bm b}\equiv (b_{\uparrow},b_{\downarrow})$ for spin-$\frac{1}{2}$ electrons in conduction 
and valence bands, respectively. In the presence of the Rashba interactions ($\xi_{R,0}$, $\xi^{\prime}_{R,0}$) and spinless inter-band coupling 
($\Delta_t$, $\Delta^{*}_{t}$), an attractive interaction $g_{s0}$ between conduction electrons and valence holes 
induces a condensation of the $XY$ components of 
the real part of the $s$-wave exciton pairing, 
$O_{\mu} \equiv \langle {\bm b}^{\dagger}\sigma_{j} {\bm a}\rangle$ ($j=x,y$) [see Appendix \ref{appendixA2}]. Thereby, 
the spin hydrodynamics of the $XY$-components can 
be well described by Eq.~(\ref{eqn4}) with 
$\phi \propto {\rm Re}O_x + i{\rm Re}O_y$, where $\alpha$ is determined by the Rashba interactions (Appendix \ref{appendixA2}).

\section{\label{sec3}Classical motion and conserved current}

Motivated by these physical realizations, we study classical motion around the ground states. Taking $\phi=\sqrt{\rho_0+\delta\rho}e^{i\theta}$,
integrating a gapped amplitude mode $\delta\rho$, and neglecting fluctuations along $z$, we obtain a 2D effective 
field theory for a Goldstone mode $\theta$ in the SSB phase,
\begin{align}
\label{eqn5}
&\mathcal{L}=\frac{1}{2}(\partial_t\theta)^2-\frac{1}{2}(\partial_x\theta)^2[1-\alpha\mathrm{cos}(2\theta)]\nonumber\\
&\!\ -\frac{1}{2}(\partial_y\theta)^2[1+\alpha\mathrm{cos}(2\theta)]+\alpha (\partial_x\theta)(\partial_y\theta)\mathrm{sin}(2\theta).
\end{align}
We set $\eta_1=c_\perp=\rho_0=1$ without loss of generality. For a given ground state $\phi=\sqrt{\rho_0}e^{i\theta_0}$, the dispersion of a phase fluctuation $\delta\theta=\theta-\theta_0$ is gapless with a linear dispersion, where velocities are anisotropic and depend on $\theta_0$. Note that the joint U(1) symmetry generally allows higher-order terms in derivatives or fields in the effective theory Eq.~(\ref{eqn4}), while they do not affect the hydrodynamic transport of low-energy excitations near the ground states. 

According to Noether's theorem~\cite{Peskin2018, Srednicki2007}, 
the U(1) continuous spacetime symmetry endows the classical motion with a conserved current of total angular momentum, which  
can be divided 
into a spin part ($j^s_\mu$) and an orbital part ($j^l_\mu$), 
\begin{align}
\label{eqn8}
&j^s_\mu=\frac{\partial\mathcal{L}}{\partial (\partial_\mu\theta)}\Delta\theta,\nonumber\\
&j^l_\mu=[\delta_{\mu\nu}\mathcal{L}-\frac{\partial\mathcal{L}}{\partial (\partial_\mu\theta)}(\partial_\nu\theta)]\Delta x_\nu,
\end{align}
with $\mu,\nu\in\{t,x,y\}$, $\Delta x_\nu\in\{\Delta t,\Delta x,\Delta y\}$, $\Delta \theta=1$, and 
$(\Delta t,\Delta x,\Delta y)=(0,-y,x)$. The two parts are not conserved by themselves, $\partial_\mu j^s_\mu=-\partial_\mu j^l_\mu=G$, where a spin torque $G$ can be defined by the divergence of the spin current. The spin torque ($G$), spin currents ($j^s_x$, $j^s_y$), and a spin angular momentum along $z$-direction ($j^s_t$) are given by the following equations [see Appendix \ref{appendixB}],
\begin{align}
\label{eqn9}
G=-\alpha[(\partial_x\theta)^2-(\partial_y\theta)^2]\mathrm{sin}(2\theta)&+2\alpha(\partial_x\theta)(\partial_y\theta)\mathrm{cos}(2\theta),\nonumber\\
j^s_x=-(\partial_x\theta)[1-\alpha\mathrm{cos}(2\theta)]&+\alpha(\partial_y\theta)\mathrm{sin}(2\theta),\nonumber\\
j^s_y=-(\partial_y\theta)[1+\alpha\mathrm{cos}(2\theta)]&+\alpha(\partial_x\theta)\mathrm{sin}(2\theta),\nonumber\\
s\equiv j^s_t=\partial_t\theta&.
\end{align}
Though the orbital part $j^l_{\mu}$ is non-local [see Eqs.~(\ref{eqn2.13})-(\ref{eqn2.15})], the spin torque $G$ as well as the spin part $j^s_{\mu}$ are local. The locality of the spin torque results from a continuous spacetime translational symmetry of ${\cal L}$ (Appendix~\ref{appendixB}).


\section{\label{sec4}Spin injection and transport}
\begin{figure*}[t]
\centering
\subfigure[ ]{
\label{fig_a}
\begin{minipage}{0.3\linewidth}
\centering
\includegraphics[width=1\linewidth]{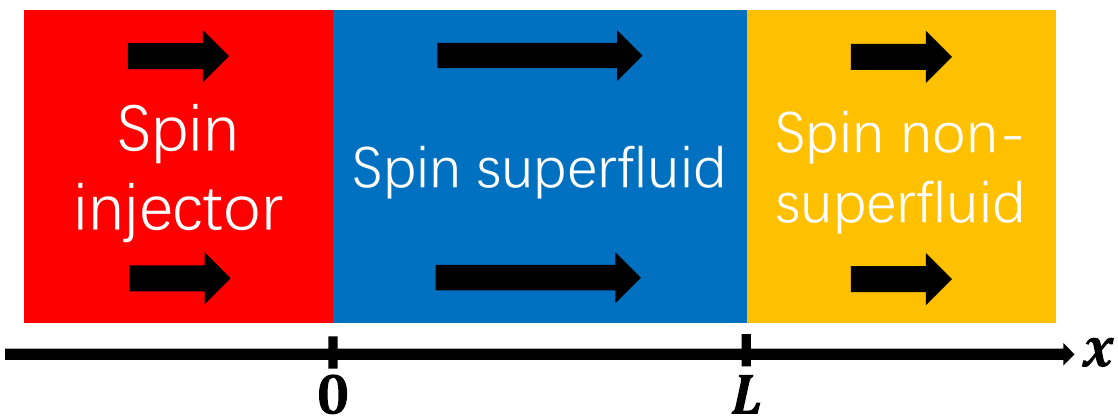}
\end{minipage}
}
\subfigure[ ]{
\label{fig_b}
\begin{minipage}{0.3\linewidth}
\centering
\includegraphics[width=1\linewidth]{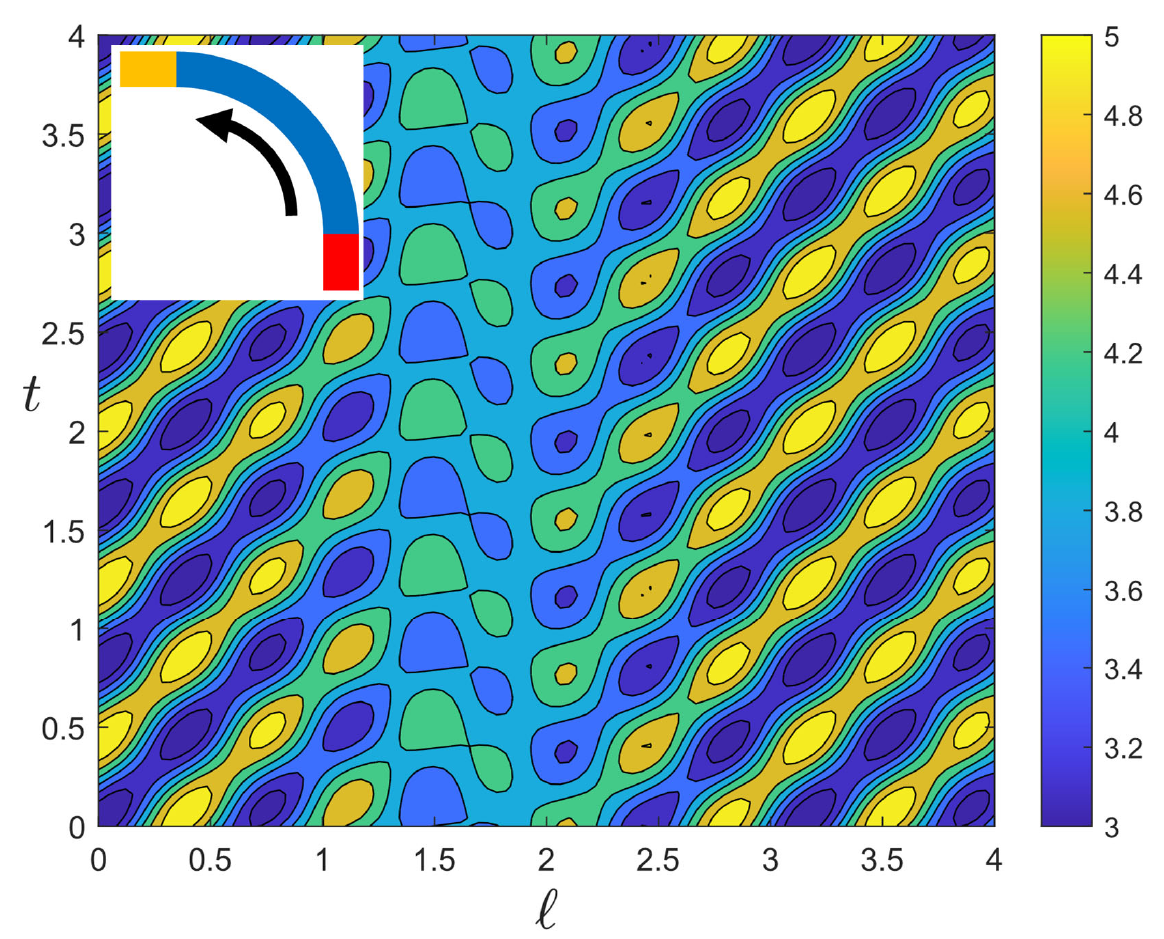}
\end{minipage}
}
\subfigure[ ]{
\label{fig_c}
\begin{minipage}{0.3\linewidth}
\centering
\includegraphics[width=1\linewidth]{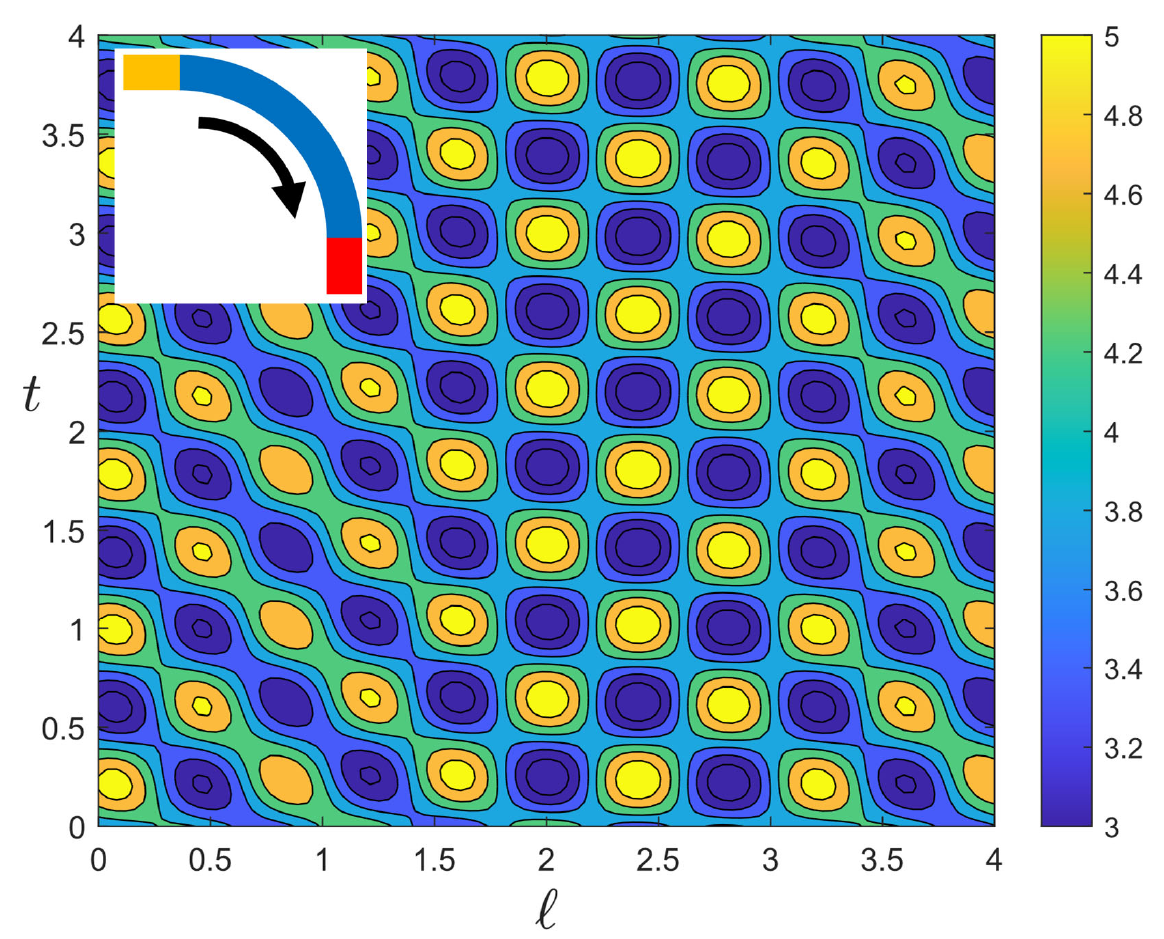}
\end{minipage}
}

\caption{\label{fig} A spin-injection model. A steady spin current $j_0$ is injected from a spin injector (red) to the total-angular-momentum 
superfluid (blue). The spin current passes through the superfluid (blue) and flows into a spin non-superfluid (yellow). The direction of the dc component of the current is indicated by black arrows. (a) A straight geometry. (b) A contour plot of $s(\ell,t)$ in a circular geometry with a positive current $j_0=4$. (c) A contour plot of $|s(\ell,t)|$ in a circular geometry with a negative current $j_0=-4$. $\chi=\chi'=D_s=1$, $T_1'=4$, $\beta_t=2$, $r=1$, $L=4$, $\alpha=0.1$ are used in the contour plots.}
\end{figure*}

To illustrate observables of a total-angular-momentum superfluid,
consider a uniform spin current $j_0$ ($j_0>0$) injected into one end ($x=0$) of the 
superfluid ($0<x<L$). The spin current passes through 
the superfluid and flows into a spin non-superfluid at the other end $x=L$ 
[see Fig.~\ref{fig_a}]~\cite{Sonin1978_1, Sonin2010}. 
The non-superfluid ``lead" has diffusive spin transport. 
Hydrodynamic spin transport in the superfluid is determined 
by a one-dimensional (1D) EOM of the Goldstone 
mode $\theta(x,t)$ in Eq.~(\ref{eqn5}) with $\partial_y \theta=0$,
\begin{align}
\label{eqn10}
&\partial^2_t\theta-(\partial_x^2\theta)[1-\alpha\mathrm{cos}(2\theta)] 
-\alpha(\partial_x\theta)^2\mathrm{sin}(2\theta)=0.  
\end{align}

The EOM Eq.~(\ref{eqn10}) will be solved together with proper 
boundary conditions. To determine the boundary conditions, note that 
spin transport in the non-superfluid ($x>L$) is described by 
diffusion equations~\cite{Sonin1978_1,Sonin2010},
\begin{align}
\label{eqn11}
&\frac{\partial s}{\partial t}+\frac{\partial j^s_x}{\partial x}=-\frac{s}{T'_1}, \ 
j^s_x=-D_s\frac{\partial s}{\partial x},
\end{align}
with relaxation time $T'_1$ and a diffusion coefficient $D_s$. The diffusive spin current is caused by the gradient of the spin density. 
Due to the relaxation time, the density and current decay 
exponentially in space for $L>0$, 
\begin{align}
\label{eqn12}
s(x,t)&\equiv\sum_{c\in\mathbb{R}}s_c(x,t)=
\sum_{c\in\mathbb{R}}a_ce^{ic t}e^{-\sqrt{D_s^{-1}\omega_c}x},\nonumber\\
j^s_x(x,t) &\equiv\sum_{c\in\mathbb{R}}j^s_{x,c}(x,t) 
=\sum_{c\in\mathbb{R}}\sqrt{D_s\omega_c}a_ce^{ic t}e^{-\sqrt{D_s^{-1}\omega_c}x}. 
\end{align}
From Eq.~(\ref{eqn12}) we get $j^s_{x,c}=\sqrt{D_s\omega_c}s_c$, where $j^s_{x,c}$ and $s_c$ are Fourier components of $j^s_x$ and $s$ with a time periodicity $2\pi c^{-1}$. 
Here $\omega_c=ic+\frac{1}{T^{\prime}_1}$, $a_c$ are complex coefficients, and the square roots of $D^{-1}_s \omega_c$ 
take positive real parts.
The spin current is assumed to be continuous 
at the junction between the superfluid and non-superfluid, 
and it is proportional to the gradient of an effective local magnetic field felt 
by the spin density~\cite{Sonin1978_1},
\begin{align}
\label{eqn13}
&j^s_x(x=L-,t)=j^s_x(x=L+,t)\nonumber\\
=&-\beta_t[\frac{1}{\chi'}s(x=L+,t)-\frac{1}{\chi}s(x=L-,t)].
\end{align}
Here $\chi$, $\chi'$ are magnetic susceptibilities 
at $x=L-$ and $x=L+$ respectively, $\beta_t$ is a response coefficient of the junction, and they are all positive. 
Eq.~(\ref{eqn13}) imposes a boundary condition (BC) on the spin density and current at $x=L-$ for each frequency $c$,
\begin{align}
\label{eqn14}
&s_c(x=L-,t)=k_cj^s_{x,c}(x=L-,t),  
\end{align}
with $k_c \equiv 
\frac{\chi}{\chi'}\big{[}D_s(\frac{1}{T'_1}+ic)\big{]}^{-\frac{1}{2}}+\frac{\chi}{\beta_t}$, 
$k_{-c}=k_c^*$, and $\mathrm{Re}(k_c)>0$. The steady injection of 
spin imposes another boundary condition at $x=0+$, $j^s_x(x=0+,t)=j_0$~\cite{Sonin1978_1}. 
In the following, 
the EOM Eq.~(\ref{eqn10}) 
is solved for $\theta(x,t)$ such that $s(x,t)$ and $j^s_x(x,t)$ satisfy the BCs. 

An analytical solution of $\theta(x,t)$ can be obtained perturbatively  
in the SOC. The solution at the first order 
consists of three parts,
\begin{align}
\label{eqn15}
\theta(x,t)=\theta_0(x,t)+\theta_1(x,t)+\theta_2(x,t)+\mathcal{O}(\alpha^2).
\end{align}
$\theta_0$ is the zeroth-order solution satisfying 
the EOM and BCs~\cite{Sonin1978_1, Sonin2010},
\begin{align}
\label{eqn16}
\theta_0(x,t)=k_0j_0t-j_0x, 
\end{align}
with $k_0 = 
\frac{\chi}{\chi'}\sqrt{\frac{T'_1}{D_s}}+\frac{\chi}{\beta_t}$. 
An oscillation is absent at the zeroth order due to the BCs 
with $\mathrm{Re}(k_c)>0$. $\theta_1$ and 
$\theta_2$ are at the first order in $\alpha$. $\theta_1$ is 
a special solution of an inhomogeneous linear differential 
equation,
\begin{align}
\label{eqn17}
\partial_t^2\theta_1-\partial_x^2\theta_1=&-\alpha(\partial_x^2\theta_0)\mathrm{cos}(2\theta_0)
+\alpha(\partial_x\theta_0)^2\mathrm{sin}(2\theta_0).
\end{align}
$\theta_2$ is a solution of a homogeneous linear 
differential equation such that $\theta$ satisfies 
the BCs at the first order in $\alpha$,
\begin{align}
\label{eqn18}
\partial_t^2\theta_2-\partial_x^2\theta_2=0.
\end{align}
The solution at the first order 
oscillates with two spatial wavenumbers, $2j_0$ and 
$2k_0 j_0$, and one temporal frequency $2k_0 j_0$ [see Appendix \ref{appendixC1}],
\begin{align}
\label{eqn19}
&\theta(x,t)=j_0(k_0t-x)-\frac{\alpha}{4(k_0^2-1)}\mathrm{sin}[2j_0(k_0t-x)]\nonumber\\
&\!\ -\frac{\alpha(2k_0^2-1)}{4(k_0^2-1)}\mathrm{cos}(2k_0j_0t)\mathrm{sin}(2k_0j_0x)\nonumber\\
&\!\ +\alpha\mathrm{Im}(\eta)\mathrm{cos}(2k_0j_0t)\mathrm{cos}(2k_0j_0x)\nonumber\\
&\!\ +\alpha\mathrm{Re}(\eta)\mathrm{sin}(2k_0j_0t)\mathrm{cos}(2k_0j_0x)+\mathcal{O}(\alpha^2).
\end{align}
$\eta$ is a constant depending on $k_0$, $k_{c=2k_0j_0}$, and $2j_0L$ 
[see Eqs.~(\ref{eqn3.32},\ref{eqn3.23})]. The two spacetime frequencies $(2j_0, 2k_0j_0)$ and $(2k_0j_0,2k_0j_0)$ come from $\theta_1$ and $\theta_2$ respectively, and are determined by the BCs. Note that the perturbative solution is divergent and fails near a ``resonant" point $k_0=1$~\cite{ Taylor2004} [see Appendix~\ref{appendixF}]. 
The divergence can be resolved by adding a finite dissipation term, $-T_1^{-1}\partial_t\theta$, to the EOM Eq.~(\ref{eqn10}).

Higher-order solutions can be systematically obtained by the perturbative iteration, 
where the spin density and current have the same periodicity in time as the first-order solution, $\pi(k_0j_0)^{-1}$. The time periodicity can be detected by a time-resolved measurement of the spin density in the non-superfluid ``lead", which depends on the injected spin current ($j_0$) and properties of the junction ($k_0$).  
The higher-order solution has no spatial periodicity in general, while its    
Fourier-transform in space has two major peaks at $2j_0$ and $2k_0j_0$ as in the first-order solution. 
The two major wavenumbers can be observed by a local measurement of the spin density in the superfluid. 

%


The spin hydrodynamics under the spin current has a unique geometric effect 
in a geometry with a finite curvature 
(Figs.~\ref{fig_b},\ref{fig_c}). 
To demonstrate this, suppose that the width of the junction in the circular 
geometry is small enough that the radius of the junction is taken as 
a constant $r$ and the field depends only on time and a 1D angular 
coordinate $\vartheta$. With 
$(x,y)=r(\mathrm{cos}\vartheta,\mathrm{sin}\vartheta)$, Eq.~(\ref{eqn5}) 
leads to a 1D Lagrangian [see Appendix \ref{appendixC2}],
\begin{align}
\label{eqn21}
\mathcal{L}=\frac{1}{2}(\partial_t\theta)^2-\frac{1}{2}(\partial_\ell\theta)^2[1+\alpha\mathrm{cos}(2\theta-\frac{2}{r}\ell)],  
\end{align}
where $\ell \equiv r\vartheta$. The corresponding EOM under the injected spin 
current $j_0$ together with the junction parameter $k_0$ has a zeroth-order 
solution, $\theta_0(\ell,t)=k_0j_0t-j_0\ell$, and a first-order solution, 
$\theta_0(\ell,t)+\theta_1(\ell,t)+\theta_2(\ell,t)$. Here $\theta_1$ is a special 
solution of an inhomogeneous differential equation, 
\begin{align}
\label{eqn22}
&\partial_t^2\theta_1-\partial_\ell^2\theta_1\nonumber\\
=&-\alpha j_0(j_0+\frac{2}{r})\mathrm{sin}[2k_0j_0t-2(j_0+\frac{1}{r})\ell].
\end{align}
$\theta_1$ and $\theta_2$ 
introduce two wavenumbers,    
$2j_0 + \frac{2}{r}$ and $2k_0j_0$, in the observables 
respectively, where 
the wavenumber of $\theta_1$ acquires a linear curvature ($\frac{1}{r}$) dependence. 
Due to the 
curvature dependence, 
two 
injected spin currents with 
opposite signs ($j_0\equiv \mathsf{j}_0$ from  Fig.~\ref{fig_b} and $j_0=-\mathsf{j}_0$ from Fig.~\ref{fig_c}) lead to different spatial distributions of the observables (non-reciprocal spin hydrodynamics, see Appendix \ref{appendixC2}). The non-reciprocity in the curved geometry contradicts neither the time-reversal symmetry nor an inversion at the origin ($r=0$), as a uniform circular spin current is even under those symmetries.

When the discreteness of the lattice rotational symmetry becomes relevant, the Lagrangian and EOM acquire $\mathbb{Z}_n$ terms,
where the U(1) spacetime symmetry reduces to the $\mathbb{Z}_n$ spacetime symmetry [see Appendix \ref{appendixA1}]. 
The $\mathbb{Z}_n$ theory leads to a gapped ground state at equilibrium, whose low-energy spin transport is characterized by the dynamics of domain walls. 
A $\mathbb{Z}_n$ term $\tilde{c}_n \sin(n\theta)$ also gives rise to similar spacetime oscillations in the observables under the spin injection~\cite{Sonin1978_1, Sonin1978_2, Sonin2010, Sonin1981}, while there is no geometric dependence. 
On the contrary, as described above, the spacetime oscillations induced by the SOC ($\alpha$) have non-reciprocal and curvature-dependent hydrodynamics in the curved geometry.

\section{\label{sec5}Dissipation effect}
In the presence of the Galilean covariance, 
a uniform superfluid moving slower than the velocity of its Goldstone mode 
achieves a local energy minimum so that it is 
stable against dissipation by local perturbations, e.g.,~elastic scattering by disorder~\cite{Sonin2010}. 
To see the stability of 
a supercurrent state with the broken U(1) spacetime symmetry,  
we compare classical energies of 
the 1D solution $\theta(x,t)$ 
and its local deformation $\theta(x,t)+\delta\theta(x,t)$. 
The deformation $\delta\theta(x,t)$ is induced by local perturbations, so 
the spacetime derivatives of $\delta\theta$ do not contain any uniform component in spacetime. 
$\theta(x,t)+\delta\theta(x,t)$ as well as $\theta(x,t)$ is a classical 
solution of Eq.~(\ref{eqn10}), 
while  
they do not necessarily share the same boundary conditions. 
The classical energy in the 1D model can be evaluated from a Hamiltonian,
\begin{align}
\label{eqn5a}
& H[\theta]  =  \int dx \Big\{\frac{1}{2} (\partial_t\theta)^2 + \frac{1}{2}(\partial_x\theta)^2[1-\alpha\mathrm{cos}(2\theta)] \Big\}. 
\end{align}
As the classical energies are independent of time, for simplicity, we compare time averages of  
the energies (with $k_0\neq 1$) over a large period of time $T$ [see Appendix \ref{appendixD}], 
\begin{align}
\label{eqn23}
&\Delta J\equiv \lim_{T\rightarrow \infty}\frac{1}{T} 
\Big(
\int^T_{0} dt H[\theta+\delta\theta]-\int^T_0 dt  H[\theta]
\Big)\nonumber\\
=& \lim_{T\rightarrow \infty} \frac{1}{T} \int^T_0 dt 
\int dx\{(\partial_t\theta)(\partial_t\delta\theta)+(\partial_x\theta)(\partial_x\delta\theta)\nonumber\\
& \!\ \!\ \!\ \times [1-\alpha\mathrm{cos}(2\theta)] + \alpha 
(\partial_x\theta)^2\mathrm{sin}(2\theta)(\delta\theta)\}+\mathcal{O}((\delta\theta)^2)\nonumber\\
=&  \lim_{T\rightarrow \infty} \frac{2}{T}
\int^T_0 dt \int dx(\partial_x\theta_2) (\partial_x\delta\theta_0)+\mathcal{O}(\alpha^2\delta\theta,(\delta\theta)^2),
\end{align}
with  
$\theta+\delta\theta=\theta_0+\delta\theta_0+\mathcal{O}(\alpha)$. Terms oscillating in space or time vanish after the spacetime integrals. 
$\delta\theta_0$, as well as $\theta_2$,   
is a solution of Eq.~(\ref{eqn18}), and both are given by linear superpositions of $e^{iq(t-x)}$ and $e^{iq(x+t)}$ over $q$. 
Thus, the right-hand side of Eq.~(\ref{eqn23}) indicates that for a given $\theta_2 \ne 0$, 
one can always choose $\delta \theta_0$ with $\Delta J<0$. 
This means that the supercurrent state 
is classically unstable toward other states, and  
energy always dissipates by the local perturbations. The instability results from the absence of the Galilean covariance. The superflow state is distinct from the ground state by the spacetime oscillation 
feature, and the energy of the superflow state 
can be lowered by excitations $\delta\theta_0$ which match the oscillation periodicity.

The dissipation effect on the spin hydrodynamics can be phenomenologically studied 
through addition of the simplest time-reversal breaking 
term, $-T^{-1}_1 \partial_t \theta$, into the classical EOM [see Appendix \ref{appendixE}]. 
With finite $T_1^{-1}$, the zeroth-order solution of spin current acquires linear spatial decay~\cite{Sonin1978_1, Sonin2010}, which contrasts with the exponential decay in the non-superfluid~\cite{Sonin2010}. Thereby, the hydrodynamic feature of spin transport survives against dissipation. When a phase accumulation $\gamma$ due to the spatially dependent current is small, a double expansion in  $\alpha$ and $\gamma$ enables a perturbative solution of $\theta$ systematically [see 
Appendix.~\ref{appendixE}]. The lowest-order solution suggests that the spin density and current remain periodic in time, and they show non-reciprocal and curvature-dependent spin transport in the curved geometry.  



\section{\label{sec6}Summary}
In this paper, we generalize the U(1) internal symmetry in conventional superfluid theories into the U(1) spacetime symmetry. Due to the joint symmetry, the supercurrent state shows geometry-dependent spacetime oscillations, and it is unstable against the dissipation effect. Note that in the spin and excitonic systems, 
the orbital angular momentum density $j^l_t$ has much smaller coupling 
with local magnetic probes than the spin angular 
momentum density $j^s_t$, because both the spin and exciton are charge neutral, and their motions have no direct coupling with a magnetic field. For this reason, we expect that the local measurements of spin angular 
momentum density in the proposed physical systems are experimentally 
feasible. Our study paves the way for further exploration of multiple spacetime symmetries and their coupling with internal symmetries.

\begin{acknowledgments}

We are grateful to Lingxian Kong, Zhenyu Xiao, and Xuesong Hu for their helpful discussions. The work was supported by the National Basic Research 
Programs of China (No.~2019YFA0308401) and the National Natural Science 
Foundation of China (No.~11674011 and No.~12074008).

\end{acknowledgments}

\appendix
\section{\label{appendixA}Microscopic models}

In this appendix, we discuss physical realizations of the U(1) spacetime symmetry in two physical models; (i) an $XY$ spin model for spin-orbit coupled magnets near a critical point, and (ii) a triplet excitonic model for semiconductors. The U(1) spacetime symmetry is a joint continuous rotational symmetry that acts on both matter field and 2D spatial coordinates. The joint nature of the symmetry results from a locking between the rotation of the matter field and that of the spatial coordinate. Such locking is ubiquitous in solid-state materials with relativistic spin-orbit interaction, where spin or an interband component of spin forms the matter field. In solid-state materials with periodic lattices, the spatial rotation must be discrete due to the lattices, so the joint rotation symmetry is also discrete; the U(1) spacetime symmetry cannot be an exact symmetry and it is valid only approximately. Nonetheless, for some solid-state systems, the approximation becomes effective where the difference between discrete and continuous joint rotations becomes irrelevant. 

\subsection{\label{appendixA1} Easy-plane ferromagnetic spin model}
To see the effectiveness of the U(1) theory in magnetic systems, let us consider a $XY$ ferromagnetic spin system in 3D lattices that is symmetric under $C_n$ rotation around a $z$ axis $(n=3,4,6,\cdots)$ and time reversal. We will first impose a spatial inversion symmetry; at the end of this subsection, we will show that the inversion symmetry is not necessary to derive the U(1) theory for some cases. We suppose that the spin system is a spin-orbit coupled magnet with an easy-plane spin anisotropy (an $XY$ plane being the easy plane), and it undergoes a continuous phase transition from a disordered phase to a ferromagnetic ordered phase of $XY$ components of spins, $S_{{\bm i},x}$ and $S_{{\bm i},y}$. In this subsection, we will discuss the effective symmetry of spin hydrodynamics near the phase transition point. 

The second-order phase transition in the $XY$ ferromagnet with the $C_n$ rotation can be described by a partition function $\mathcal{Z}_n$ ($n=3,4,6,\cdots$) with a Ginzburg-Landau (GL) action for a 2D complex variable  $\phi({\bm x})$ ($\hbar=1$), 
\begin{align}
&\mathcal{Z}_n = \int {\cal D}\phi {\cal D}\phi^{\dagger} \exp[-\mathcal{S}_{n,\phi}], \nonumber \\
&\mathcal{S}_{n,\phi} = \int d^3{\bm r} \int^{1/k_{B}T}_{0} d\tau \!\  \!\ s_{n}[\phi({\bm r},\tau)]. 
\end{align}
Here the internal field $\phi({\bm r})$ is a spatial 
average of $S_{{\bm i},x}+iS_{{\bm i},y}$ with respect to a lattice site ${\bm i}$ 
over some hydrodynamic volume element. $i$ is the complex unit. ${\bm r}$ is a spatial 
coordinate of the hydrodynamic volume element. 
The transition can happen either at the zero-temperature $T=0$ critical point (quantum critical point) or at finite-temperature $T \ne 0$ critical point (classical critical point). In this appendix, we consider the spin hydrodynamics near the $T=0$ quantum critical point. The time-reversal symmetry means the absence of an external magnetic (Zeeman) field in the spin model, and the ferromagnetic order breaks the symmetry spontaneously. The time-reversal symmetry in the model requires the dynamical exponent $z$ at the quantum critical point to be one, $z=1$ [see below]. 

For $n=4$, the spin system is defined on a 3D tetragonal lattice with a $C_4$ rotational symmetry around the $z$ axis, such as a layered square lattice. For $n=3$ or $n=6$, the system is defined on a 3D trigonal or hexagonal lattice with a $C_3$ or $C_6$ rotation, e.g., layered honeycomb or triangle lattices. In the following, we first employ a symmetry argument to determine the form of the GL action $s_{n}[\phi]$ for $n=4$, $n=3$, and $n=6$ and show that for the $n=3$ or $n=6$ case, the U(1) joint rotational symmetry is an effective symmetry of the GL action for the $xy$ spins near the critical point, while for the $n=4$ case, the effective symmetry remains discrete (a $\mathbb{Z}_4$ joint rotational symmetry).  
To this end, note that under the $C_n$ rotation around the $z$ axis, the complex field $\phi({\bm r})$ of the $xy$ spins, and the 3D spatial coordinate, ${\bm r}\equiv (x,y,z)$, are rotated together due to the spin-orbit locking, 
\begin{align}
&{\bm r} \rightarrow {\bm r}^{\prime} = (x^{\prime},y^{\prime},z), 
x^{\prime}+iy^{\prime} = e^{\frac{2\pi}{n}}(x+iy), \nonumber \\ 
&\phi({\bm r}) \equiv S_{x}({\bm r}) + iS_{y}({\bm r}) 
\rightarrow \phi^{\prime}({\bm r}^{\prime}) 
= \phi({\bm r}) e^{i\frac{2\pi}{n}}.  
\end{align}
The spatial inversion changes the sign of the coordinate vector, while  
the time-reversal changes the sign of $\phi_x$, $\phi_y$, and $i$. These symmetries 
constrain forms of the actions for the $\phi$ field. 

\subsubsection{GL action for $XY$ ferromagnets with $C_4$ rotation} 
The symmetries of the joint $C_4$ rotation around $z$ and 
time reversal allow the following terms and their complex conjugates 
in the GL action, 
\begin{align}
&|\phi|^2 \equiv \phi^{\dagger}\phi, \quad 
i[\phi^{\dagger} \partial_z \phi - (\partial_z \phi^{\dagger}) \phi], 
\quad  (\partial_{m}\phi^{\dagger})(\partial_{m} 
\phi), \nonumber \\  
&\phi \partial^2_{-} \phi, \quad \phi \partial^2_{+} \phi, \quad 
(|\phi|^2)^2, \quad \phi^4, \quad \cdots,
\end{align}
with $m=x,y,z$, and 
$\partial_{\pm} \equiv \partial_{x}\pm i \partial_y$. Higher-order terms in $\phi$, higher-order spatial gradient terms, and total-derivative 
terms are omitted as `$\cdots$'. The higher-order $\phi$ terms are irrelevant near the critical point where the amplitude of $\phi$ becomes smaller. 
The higher-order spatial gradient terms are irrelevant in the hydrodynamic regime where the volume element over which the spin operator is averaged becomes larger. 
Here the time reversal forbids odd-order terms of $\phi$. The spatial 
inversion further forbids 
$i[\phi^{\dagger} \partial_z \phi - (\partial_z \phi^{\dagger}) \phi]$ 
from the action. Accordingly, the GL functional form allowed by 
the symmetries is given by 
\begin{align} 
\mathcal{Z}_{n=4} &= \int {\cal D}\phi {\cal D}\phi^{\dagger} 
\exp [- \mathcal{S}_{n=4,\phi}], \nonumber \\
\mathcal{S}_{n=4,\phi} &\equiv \int d^3{\bm r} \int^{\hbar/k_{\rm B}T}_{0} d\tau 
\bigg\{ \frac{\eta_1^2}{2}(\partial_\tau\phi^\dagger)(\partial_\tau\phi) \nonumber \\ 
&\hspace{-0.8cm} + \!\ 
\frac{\eta_1^2c_\perp^2}{2}(\partial_j\phi^\dagger)(\partial_j\phi) 
+ \frac{\eta_1^2c_z^2}{2}(\partial_z\phi^\dagger)(\partial_z\phi)
\nonumber \\
& \hspace{-0.8cm} 
+\frac{\eta_1^2 c_\perp^2}{4}[\alpha(\partial_-\phi)^2+\alpha^*(\partial_+\phi^\dagger)^2 
+ \beta (\partial_+\phi)^2+\beta^*(\partial_-\phi^\dagger)^2] \nonumber \\
& \hspace{-0.8cm} + \frac{U}{2}(\phi^\dagger\phi-\rho_0)^2 +\frac{1}{2}[\tilde{c}_4\phi^4+\tilde{c}_4^*(\phi^\dagger)^4] + \cdots \bigg\}, \label{z4}
\end{align}
with $j=x,y$, imaginary time $\tau$, real numbers $\rho_0$, $\eta_1$, $c$, $U$, and complex numbers $\alpha$, $\beta$, $\tilde{c}_4$. Here 
the first-order time-derivative of $\phi$, such as 
$\phi^{\dagger} \partial_{\tau}\phi$, is forbidden by the time reversal symmetry. 
The second-order time-derivative term is induced by an integration of the $z$-component spin. To be specific, one can start from a path integral of a spin Lagrangian with a Wess-Zumino term that ensures correct commutation relations between spin operators. In the presence of easy-plane anisotropy, the out-of-plane component of spin ($S_z$) is gapped, and one can integrate out the gapped degree of freedom, leading to an effective field theory of $\phi=S_x+iS_y$. In the effective field theory, the leading-order time-dependent term allowed by time-reversal symmetry is $\frac{\eta^2_1}{2} (\partial_{\tau} \phi^{\dagger}) (\partial_{\tau} \phi)$. Note that the dynamical exponent $z$ around 
the $T=0$ quantum critical point becomes one, namely $z=1$.  

A partition function $\mathcal{Z}_{n=4}$ describes the phase transition 
from the disordered phase ($\rho_0<0$) 
to the ordered phase of the $xy$ spins ($\rho_0>0$). In the ordered phase, 
a phase of $\phi$ is locked into four minima determined by the 
$\tilde{c}_4$ term. 
At the quantum critical point, 
the terms with higher-order gradient and/or 
$\phi$ terms become irrelevant in the long-wavelength limit, and the effective symmetry of the GL action at the critical point is determined by a gapless free theory part, 
\begin{align}
\overline{\mathcal{S}}_{n=4,\phi} &\equiv \int d^3{\bm r} \int^{\hbar/k_{\rm B}T}_{0} d\tau 
\bigg\{ \frac{\eta_1^2}{2}(\partial_\tau\phi^\dagger)(\partial_\tau\phi) \nonumber \\
&\hspace{-0.8cm}+ 
\frac{\eta_1^2c_\perp^2}{2}(\partial_j\phi^\dagger)(\partial_j\phi)
+\frac{\eta_1^2c_z^2}{2}(\partial_z\phi^\dagger)(\partial_z\phi) \nonumber \\
&\hspace{-0.8cm} + \frac{\eta_1^2 c_\perp^2}{4}[\alpha(\partial_-\phi)^2+\alpha^*(\partial_+\phi^\dagger)^2  + 
\beta (\partial_+\phi)^2+\beta^*(\partial_-\phi^\dagger)^2] \bigg\}. 
\end{align} 
Importantly, though the $\alpha$ term in the second line is symmetric under  
the U(1) spacetime symmetry, ${\partial}_{\pm} \rightarrow \partial^{\prime}_{\pm} = 
e^{\pm i \epsilon}\partial_{\pm}$, $\phi \rightarrow 
\phi^{\prime} = e^{i\epsilon} \phi$ for $\forall \epsilon$, the $\beta$ term is 
symmetric only under the joint $\mathbb{Z}_4$ rotational symmetry, 
\begin{align}
&{\partial}_{\pm} \rightarrow \partial^{\prime}_{\pm} = 
e^{\pm i \frac{\pi}{2}}\partial_{\pm}, \!\ \!\ \phi \rightarrow 
\phi^{\prime} = e^{i\frac{\pi}{2}} \phi, \nonumber \\ 
&\phi^{\dagger} \rightarrow 
(\phi^{\prime})^{\dagger} = e^{-i\frac{\pi}{2}} \phi^{\dagger}. 
\end{align}
Due to the $\beta$ term, the effective symmetry 
at the critical point remains discrete for the $n=4$ case. 

\subsubsection{GL action for $XY$ ferromagnets with $C_3$ or $C_6$ rotation} 
For the hexagonal crystal family, on the contrary, 
the joint $C_3$ or $C_6$ rotational symmetry forbids the $\beta$ term, 
so the corresponding gapless free theory does have the U(1) spacetime symmetry. To this end, we analyze the terms allowed in the action.
The symmetries of the $C_6$ rotation around $z$, spatial inversion, 
and time reversal allow the following terms and their complex conjugates in the action, 
\begin{align}
&|\phi|^2, \!\  
(\partial_{m}\phi^{\dagger})(\partial_{m} 
\phi), \!\ 
\phi \partial^2_{-} \phi, \!\ 
(|\phi|^2)^2, \!\ \phi^3 \partial^2_{+} \phi, \!\ \phi^6, \!\ \cdots, \label{z6-}
\end{align} 
with $m=x,y,z$. 
When the $C_6$ rotation is substituted by the $C_3$ rotation, Eq.~(\ref{z6-}) 
also exhausts all symmetry-allowed terms apart from higher-order $\phi$ terms, higher-order spatial-gradient terms, 
and total derivative terms. In fact, 
the $C_3$ rotational symmetry alone allows $\phi^3$ and $\phi \partial_{+}\phi$, while the $\phi^3$ term is prohibited by the time-reversal symmetry and $\phi\partial_+\phi$ is a total derivative term. Thus, the partition function $\mathcal{Z}_{n=3,6}$ near the critical point is given by
\begin{align}
\mathcal{Z}_{n=3,6} &= \int {\cal D}\phi {\cal D}\phi^{\dagger} 
\exp [- \mathcal{S}_{n=3,6,\phi}] \nonumber \\
\mathcal{S}_{n=3,6,\phi} &\equiv \int d^3{\bm r} \int^{\hbar/k_{\rm B}T}_{0} d\tau 
\bigg\{ \frac{\eta_1^2}{2}(\partial_\tau\phi^\dagger)(\partial_\tau\phi) \nonumber \\
& \hspace{-0.8cm} +  
\frac{\eta_1^2c_\perp^2}{2}(\partial_j\phi^\dagger)(\partial_j\phi)
+ \frac{\eta_1^2c_z^2}{2}(\partial_z\phi^\dagger)(\partial_z\phi) \nonumber \\
&\hspace{-0.8cm} 
+ \frac{\eta_1^2 c_\perp^2}{4}[\alpha(\partial_-\phi)^2+\alpha^*(\partial_+\phi^\dagger)^2 + \tilde{\alpha}_6 \phi^3 (\partial^2_{+} \phi) \nonumber \\
&\hspace{-0.8cm} 
+ \tilde{\alpha}^*_6 (\phi^{\dagger})^3 (\partial^2_{-} \phi^{\dagger})]
 + \frac{U}{2}(\phi^\dagger\phi-\rho_0)^2
 \nonumber \\
 &\hspace{-0.8cm} 
 +\frac{1}{2}[\tilde{c}_6\phi^6+\tilde{c}_6^*(\phi^\dagger)^6] + \cdots \bigg\}, \label{z6}
\end{align}
where $\rho_0>0$ and $\rho_0<0$ correspond to ordered and disordered phases, 
respectively. Importantly, a gapless free theory part $\overline{\mathcal{S}}_{n=3,6,\phi}$ of the action, 
\begin{align}
\overline{\mathcal{S}}_{n=3,6,\phi} &\equiv \int d^3{\bm r} \int^{\hbar/k_{\rm B}T}_{0} d\tau 
\bigg\{ \frac{\eta_1^2}{2}(\partial_\tau\phi^\dagger)(\partial_\tau\phi) \nonumber \\
& \hspace{-0.6cm} +\frac{\eta_1^2c_\perp^2}{2}(\partial_j\phi^\dagger)(\partial_j\phi)
+\frac{\eta_1^2c_z^2}{2}(\partial_z\phi^\dagger)(\partial_z\phi) \nonumber \\
&\hspace{-0.6cm}
+ \frac{\eta_1^2 c_\perp^2}{4}[\alpha(\partial_-\phi)^2+\alpha^*(\partial_+\phi^\dagger)^2] 
\bigg\}, 
\end{align}
is symmetric under the U(1) spacetime symmetry, 
\begin{align}
&{\partial}_{\pm} \rightarrow \partial^{\prime}_{\pm} = 
e^{\pm i \epsilon}\partial_{\pm}, \quad \phi \rightarrow 
\phi^{\prime} = e^{i\epsilon} \phi, \nonumber \\ 
&\phi^{\dagger} \rightarrow 
(\phi^{\prime})^{\dagger} = e^{-i\epsilon} \phi^{\dagger}, \quad 
{\rm for} \quad \forall \!\ \epsilon. 
\end{align}
This contrasts with the free theory for the $n=4$ case which is symmetric only 
under the joint discrete rotational symmetry. 

In the ordered phase ($\rho_0>0$) for the $C_6$ case, a phase of $\phi$ is locked into six minima by 
the $\tilde{c}_6$ term. The $\tilde{c}_6$ and $\tilde{\alpha}_6$ terms reduce  
the symmetry of the whole action into a joint discrete ($\mathbb{Z}_6$) rotational symmetry, 
\begin{align}
&{\partial}_{\pm} \rightarrow \partial^{\prime}_{\pm} = 
e^{\pm i \frac{\pi}{3}}\partial_{\pm}, \quad \phi \rightarrow 
\phi^{\prime} = e^{i\frac{\pi}{3}} \phi, \nonumber \\
&\phi^{\dagger} \rightarrow 
(\phi^{\prime})^{\dagger} = e^{-i\frac{\pi}{3}} \phi^{\dagger}. 
\end{align}
In the ordered phase for the $C_3$ case, the phase of $\phi$ is locked into three minima by the 
$\tilde{c}_6$ term and other higher-ordered terms omitted as `$\cdots$' in Eq.~(\ref{z6}). 
Nonetheless, unlike the $\beta$-term in $\mathcal{Z}_{n=4}$, 
$\tilde{\alpha}_6$ and $\tilde{c}_6$ terms as well as the higher-order terms are 
{\it irrelevant} in the long-wavelength limit at the quantum critical point, since  
their scaling dimensions at the critical point are all negative. 

The scaling dimensions of $\tilde{\alpha}_6$ and $\tilde{c}_6$ 
terms at the critical point, $y_{\alpha_6}$ and $y_{c_6}$, 
can be evaluated from a dimensional analysis of the gapless 
free theory part at $T=0$;  
$y_{\alpha_6} =2-D=-2$ and $y_{c_6} =6-2D=-2$ with $D=3+1$. 
Scaling dimensions of the higher-order $\phi$ terms and 
higher-order spatial gradient terms are also negative 
and smaller than $y_{\alpha_6}$ and $y_{c_6}$. 
When the hydrodynamic volume element becomes larger, the terms with negative scaling dimensions get smaller at the critical point. Thanks to their irrelevance at the critical point, the GL action respects effectively the U(1) spacetime symmetry in the long-wavelength limit. In other words, the spin hydrodynamics at the critical point becomes U(1) spacetime symmetric more effectively for larger hydrodynamic volume element.  
The hydrodynamic regime with the effective U(1)
 symmetry has a lower crossover boundary in its length scale; in order that the hydrodynamics has the effective U(1) spacetime symmetry, the length scale $\Lambda$ of the volume element should be greater than a certain crossover length $\Lambda_{c,1}$, 
\begin{align}
    \Lambda \gg \Lambda_{c,1}.  \label{hydro-1}
\end{align}
The crossover length scale is dependent on $\tilde{c}_6$, $\tilde{\alpha}_6$, and other higher-order terms that manifest the joint discrete rotational symmetries. 
In the $C_6$ case, for example, $\Lambda_{c,1}$ 
is primarily dependent on $\tilde{c}_6$, $\tilde{\alpha}_6$ and their 
scaling dimensions with the following scalings, 
\begin{align}
\Lambda_{c,1} \propto 
|\tilde{c}_6|^{\frac{1}{|y_{c_6}|}} \quad {\rm or} \quad 
|\tilde{\alpha}_6|^{\frac{1}{|y_{\alpha_6}|}}.
\end{align}

When the system is in the ordered phase but close to the critical point ($\rho_0 \gtrsim 0$), the hydrodynamics regime with the effective U(1) symmetry has also an upper bound in its length scale, 
\begin{align}
\Lambda_{c,2} \gg \Lambda \gg \Lambda_{c,1}. \label{hydro-2}
\end{align}
The upper bound is because the ground state for $\rho_0>0$  breaks the 
joint rotational symmetry spontaneously, 
and in this sense, $\tilde{c}_6$, 
$\tilde{\alpha}_6$, and the other higher-order terms  manifesting the discrete symmetry 
are {\it dangerously} irrelevant. In typical renormalization group (RG) flow trajectory, they get smaller around a saddle-point fixed point for the critical point upon the increase of the length scale, while in the very long wavelength limit, they become larger again around another fixed point that describes an ordered phase with broken joint U(1) symmetry (a Nambu-Goldstone fixed point). The upper bound $\Lambda_{c,2}$ defines a length scale for this upturn behavior of the dangerously irrelevant scaling variables. Generally, $\Lambda_{c,2}$ has a complicated scaling form 
 of $\rho$, 
as it also depends on scaling of the coupling constants around the Nambu-Goldstone fixed point. Nonetheless, $\Lambda_{c,2}$ is always greater than the lower bound, $\Lambda_{c,2} \gg \Lambda_{c,1}$, for smaller $\rho_0$. 
In the $C_6$ case, for example, $\Lambda_{c,2} \gg \Lambda_{c,1}$ is sastified when $\rho_0$, $\tilde{c}_6$, 
and $\tilde{\alpha}_6$ are in the following regimes,
\begin{align}
\rho_0^{\frac{|y_{c_6}|}{y_{\rho_0}}} \ll \frac{1}{|\tilde{c}_{6}|}, \quad 
\rho_0^{\frac{|y_{\alpha_6}|}{y_{\rho_0}}} \ll \frac{1}{|\tilde{\alpha}_{6}|}. 
\end{align}
Here $y_{\rho_0}$ is the scaling dimension of $\rho_0$ around the critical point; $y_{\rho_0}=2$. 

\subsubsection{Continuum limit of generic XY ferromagnetic spin models in the 3D hexagonal crystal family}
The above argument is solely based on the symmetry and scaling arguments, 
suggesting that {\it any} $XY$ ferromagnetic spin models 
with the $C_3$ or $C_6$ rotational, spatial inversion, and time reversal symmetries has the effective U(1) spacetime symmetry near the quantum critical point, if the models undergo the continuous phase transition of the ferromagnetic ordering. In the following, we will argue this by deriving explicitly a continuum limit of 
{\it generic} $XY$ ferromagnetic spin models with the symmetries. 

Exchange interactions in spin-orbit coupled magnets generally comprise of symmetric part ${\cal J}_{{\bm i\bm j},\mu\nu}={\cal J}_{{\bm i\bm j},\nu\mu}$ and antisymmetric part ${\cal D}_{{\bm i\bm j},\mu\nu}=-{\cal D}_{{\bm i\bm j},\nu\mu}$ (Eq.~(\ref{eqn_h_spin})), 
\begin{align}
H_{\mathrm{spin}} = \frac{1}{2}\sum_{{\bm i},{\bm j}}\sum_{\mu,\nu} 
S_{{\bm i},\mu} \Big( {\cal J}_{{\bm i \bm j},\mu\nu} + {\cal D}_{{\bm i 
\bm j},\mu\nu}\Big) S_{{\bm j},\nu},  \label{Heisenberg} 
\end{align}
with spin vector ${\bm S}_{{\bm i}}\equiv (S_{{\bm i},x},S_{{\bm i},y},S_{{\bm i},z})$. 
We first consider that the spins live on the 3D hexagonal lattice with $C_6$ rotational and spatial inversion symmetries, namely the lattice 
belongs to either $C_{6h}$ or $D_{6h}$ point group. Further discussion of other possibilities of point groups will be provided below.  
Here, the exchange interactions are not only limited to those between the nearest neighboring  
sites on the lattice, but they can also be between further neighboring sites. 

Near the transition point of the ferromagnetic ordering of the $XY$ spins, the $Z$ component of the spins fluctuates
rapidly in space and time, so that one can legitimately integrate out the $Z$ component, yielding effective spin 
models for the $XY$ spins, 
\begin{align}
H^{\rm eff} &= \frac{1}{2}\sum_{{\bm i},{\bm j}}\sum_{\mu,\nu=x,y} 
S_{{\bm i},\mu} \Big( J_{{\bm i \bm j},\mu\nu} + D_{{\bm i \bm j},\mu\nu}\Big)
S_{{\bm j},\nu} \nonumber \\
& = \frac{1}{2}\sum_{\bm i, \bm j} 
\bigg\{ 
\left(\begin{array}{cc}
S_{{\bm i},x} & S_{{\bm i},y} \\
\end{array}\right) \left(\begin{array}{cc} 
J_{{\bm i \bm j},xx} & J_{{\bm i \bm j},xy} \\
J_{{\bm i \bm j},yx} & J_{{\bm i \bm j},yy} \\
\end{array}\right) \left(\begin{array}{c} 
S_{{\bm j},x} \\
S_{{\bm j},y} \\
\end{array}\right) \nonumber \\
& + D_{{\bm i \bm j},xy} 
\big[S_{{\bm i},x}S_{{\bm j},y}-S_{{\bm i},y}S_{{\bm j},x} \big] \bigg\},   \label{XY}
\end{align}
with 2 by 2 symmetric and antisymmetric interactions,  
$J_{{\bm i \bm j},\mu\nu}=J_{{\bm i \bm j},\nu\mu}$ and 
$D_{{\bm i \bm j},\mu\nu}=-D_{{\bm i \bm j},\nu\mu}$ for $\mu,\nu=x,y$. 
The effective exchange interactions in Eq.~(\ref{XY}) as well as the exchange interactions in Eq.~(\ref{Heisenberg}) respect the joint $C_6$ rotational 
symmetry and inversion symmetry. 
In the following, we show that due to the $C_6$ rotational symmetry, the continuum limit of 
the symmetric exchange interactions
in the effective spin models always take the same form as in Eq.~(\ref{z6}),
\begin{align}
\frac{1}{2}\sum_{{\bm i},{\bm j}} \sum_{\mu,\nu=x,y} 
S_{{\bm i},\mu} J_{{\bm i \bm j},\mu\nu}  
S_{{\bm j},\nu} \simeq & \int 
d {\bm r}^3 \bigg\{ \frac{r}{2} |\phi|^2 \nonumber \\
& \hspace{-4.2cm} + 
\frac{\eta_1^2c^2_z}{2} \partial_{z} \phi^{\dagger} \partial_z \phi + \frac{\eta_1^2c^2_{\perp}}{2} \sum_{i=x,y} 
\partial_{i} \phi^{\dagger} \partial_i \phi \nonumber \\
&\hspace{-4.2cm} 
+ \frac{\eta_1^2c_\perp^2}{4} 
\big[\alpha (\partial_{-} \phi)^2 + \alpha^{*} (\partial_{+} \phi^{\dagger})^2 \big]
+ \cdots \bigg\}. \label{continuumlimit}
\end{align} 

To see this, note first that any bond of two spin sites, $({\bm i}, {\bm j})$, in a sum of 
Eq.~(\ref{XY}) has 5 other bonds in the sum that are derived from the first bond $({\bm i}, {\bm j})$ 
by the $C_6$ rotation, i.e. 
$(C^n_6({\bm i}), C^n_6({\bm j}))$ $(n=1,2,\cdots,5)$. Due to the joint $C_6$ 
rotation symmetry, 
$J_{C^n_6({\bm i}) C^{n}_6({\bm j}),\cdots}$ and $J_{{\bm i} {\bm j},\cdots}$ are related by 
the $C_6$ spin rotation around $z$; 
\begin{widetext}
\begin{align}
\left(\begin{array}{cc} 
J_{C_{6}({\bm i}) C_{6}({\bm j}),xx} & J_{C_{6}({\bm i}) C_{6}({\bm j}),xy} \\
J_{C_{6}({\bm i}) C_{6}({\bm j}),yx} & J_{C_{6}({\bm i}) C_{6}({\bm j}),yy} \\
\end{array}\right) = \left(\begin{array}{cc} 
\cos\frac{\pi}{3} & \sin\frac{\pi}{3} \\
-\sin \frac{\pi}{3} & \cos \frac{\pi}{3} \\
\end{array}\right) \left(\begin{array}{cc} 
J_{{\bm i \bm j},xx} & J_{{\bm i \bm j},xy} \\
J_{{\bm i \bm j},yx} & J_{{\bm i \bm j},yy} \\
\end{array}\right) \left(\begin{array}{cc} 
\cos\frac{\pi}{3} & -\sin \frac{\pi}{3} \\
\sin \frac{\pi}{3} & \cos \frac{\pi}{3} \\
\end{array}\right). \label{exchange-sym}
\end{align}
Then, by using a gradient expansion, $S_{{\bm j},\mu}=S_{{\bm i},\mu} 
+ ({\bm j}-{\bm i})_{\lambda} \partial_{\lambda} S_{{\bm i},\mu} 
+ \frac{1}{2}({\bm j}-{\bm i})_{\lambda}({\bm j}-{\bm i})_{\epsilon} \partial_{\lambda} \partial_{\epsilon} S_{{\bm i},\mu} + \cdots$, one 
can explicitly show that a sum of the symmetric exchange interactions 
over the six bonds reduce to the same form of 
the continuum limit as Eq.~(\ref{continuumlimit}) up to the second order in the gradient expansion;  
\begin{align}
&\frac{1}{2}\sum_{n=0,1,\cdots,5} 
\left(\begin{array}{cc}
S_{C^n_6({\bm i}),x} & S_{C^n_6({\bm i}),y} \\
\end{array}\right) \left(\begin{array}{cc} 
J_{C^n_{6}({\bm i}) C^n_{6}({\bm j}),xx} & J_{C^n_{6}({\bm i}) C^n_{6}({\bm j}),xy} \\
J_{C^n_{6}({\bm i}) C^n_{6}({\bm j}),yx} & J_{C^n_{6}({\bm i}) C^n_{6}({\bm j}),yy} \\
\end{array}\right) \left(\begin{array}{c} 
S_{C^n_6({\bm j}),x} \\
S_{C^n_6({\bm j}),y} \\
\end{array}\right) \nonumber \\
& \hspace{1cm} 
\simeq \frac{3}{2}(\lambda_{{\bm i}{\bm j},1}+\lambda_{{\bm i}{\bm j},2}) \phi^{\dagger} \phi 
+  \frac{3 a^2_{{\bm i}{\bm j},\perp}}{16} \bigg\{ 2(\lambda_{{\bm i}{\bm j},1} + \lambda_{{\bm i}{\bm j},2}) \!\ \phi^{\dagger} (\partial^2_x 
+ \partial^2_y) \phi \nonumber \\
&\hspace{2cm}
+ \sum_{m=1,2} \lambda_{{\bm i}{\bm j},m} e^{2i(\varphi_{{\bm i}{\bm j},\perp}-\psi_{{\bm i}{\bm j},\perp,m})} \!\ 
\phi (\partial_x - i\partial_y)^2 \phi + 
{\rm c.c.} \bigg\} +  \frac{3 a^2_{{\bm i}{\bm j},z}}{4}  
(\lambda_{{\bm i}{\bm j},1}+ \lambda_{{\bm i}{\bm j},2}) 
\!\ \phi^{\dagger} \partial^2_z \phi+... \label{6-pair}
\end{align} 
\end{widetext}
Here $\phi$ on the right-hand side is 
from $\phi({\bm r}_{\bm i}) 
\equiv S_{{\bm i},x}+iS_{{\bm i},y}$ in Eq.~(\ref{XY}), and the higher-order derivative and 
total derivative terms are omitted. We also regard that 
${\bm i}$ and $C^n_6({\bm i})$ ($n=1,\cdots,5$) 
are the same for the argument of $\phi$, because their differences (if exist) can be controlled by the microscopic length. $a_{{\bm i}{\bm j},\perp}$ and $a_{{\bm i}{\bm j},z}$ 
are the spatial length of the bond (${\bm i},{\bm j}$) within 
the $xy$ plane and along $z$ axis, respectively; 
$a_{{\bm i}{\bm j},\perp} \equiv |{\bm i}_{\perp}-{\bm j}_{\perp}|$,  
$a_{{\bm i}{\bm j},z} \equiv |i_z - j_z|$, with 
${\bm i}=({\bm i}_{\perp},i_z)$
and ${\bm j}=({\bm j}_{\perp},j_z)$.
$\varphi_{{\bm i}{\bm j},\perp}$ is the angle between 
${\bm j}_{\perp}-{\bm i}_{\perp}$ and the $x$ axis. 
$\lambda_{{\bm i}{\bm j},m}$ and ${\bm t}_{{\bm i}{\bm j},m}$ 
are real-valued eigenvalues and eigenvectors of 
the 2 by 2 symmetric matrix $J_{\bm i \bm j}$ ($m=1,2$). 
$\psi_{{\bm i}{\bm j},\perp,m}$ is the angle between 
${\bm t}_{{\bm i}{\bm j},m}$ and the $x$-axis in the $xy$ plane. 
As ${\bm t}_{{\bm i}{\bm j},1}$ and ${\bm t}_{{\bm i}{\bm j},2}$ 
are orthogonal to each other, 
$\psi_{{\bm i}{\bm j},2} = \psi_{{\bm i}{\bm j},1} + \pi/2$ and 
$\sum_{m=1,2} \lambda_{{\bm i}{\bm j},m} 
e^{2i(\varphi_{{\bm i}{\bm j},\perp}-\psi_{{\bm i}{\bm j},\perp,m})}
= (\lambda_{{\bm i}{\bm j},1} - \lambda_{{\bm i}{\bm j},2}) 
e^{2i(\varphi_{{\bm i}{\bm j},\perp}-\psi_{\bm i\bm j,\perp,1})}$. 
A sum of Eq.~(\ref{6-pair}) over different types of bonds leads 
to Eq.~(\ref{continuumlimit}), where $\alpha$ is simply given by the sum of 
$a^2_{{\bm i}{\bm j},\perp} \sum_{m=1,2} \lambda_{{\bm i}{\bm j},m} 
e^{2i(\varphi_{{\bm i}{\bm j},\perp}-\psi_{{\bm i}{\bm j},\perp,m})}$. Note 
that in the absence of the spin-orbit interaction, 
$J_{\bm i \bm j}$ are always proportional to the unit matrix, 
where $\lambda_{{\bm i}{\bm j},1}=\lambda_{{\bm i}{\bm j},2}$, 
$(\lambda_{{\bm i}{\bm j},1}-\lambda_{{\bm i}{\bm j},2}) 
e^{2i(\varphi_{{\bm i}{\bm j},\perp}-\psi_{{\bm i}{\bm j},\perp,1})}=0$ 
for any bond $({\bm i},{\bm j})$, and $\alpha$ vanishes.

The continuum limit of the antisymmetric exchange interaction yields 
the first-order spatial gradient terms, 
\begin{align}
&D_{{\bm i \bm j},xy}(S_{{\bm i},x} S_{{\bm j},y} - S_{{\bm i},y} S_{{\bm j},x}) 
\nonumber \\
&= iD_{{\bm i \bm j},xy} ({\bm i}-{\bm j})_{\mu} 
\big(\phi^{\dagger} \partial_{\mu} \phi - (\partial_{\mu} \phi^{\dagger}) \phi\big) 
+ {\cal O}(\partial^3). \label{dm}
\end{align}
In the presence of the spatial inversion, they are cancelled by its 
inversion symmetric counterpart; 
\begin{align} 
&D_{I({\bm i})I({\bm j}),xy}(S_{I({\bm i}),x} S_{I({\bm j}),y} - 
S_{I({\bm i}),y} S_{I({\bm j}),x}) \nonumber \\
&= - iD_{{\bm i \bm j},xy} ({\bm i}-{\bm j})_{\mu} 
\big(\phi^{\dagger} \partial_{\mu} \phi - (\partial_{\mu} \phi^{\dagger}) \phi\big) 
+ {\cal O}(\partial^3). 
\end{align}
with $D_{I({\bm i}) I({\bm j}),xy}=D_{{\bm i} {\bm j},xy}$. Thus,   
the antisymmetric interaction gives only higher-order 
gradient terms in the continuum limit for the GL action; they are all 
irrelevant in the sense that their scaling dimensions around the critical 
point are negative.   

Near the quantum critical point of the ferromagnetic order of the $xy$ spin, the systems effectively have the U(1) spacetime symmetry. 
Note that apart from the $C_{6h}$ and $D_{6h}$ point groups, the hexagonal 
crystal family (including the trigonal crystal system and the hexagonal crystal system) has 10 other point groups: $C_3$, $C_{3i}$, $D_3$, $C_{3v}$, $D_{3d}$ from the trigonal crystal system and $C_6$, $C_{3h}$, $D_6$, $C_{6v}$, $D_{3h}$ from the hexagonal crystal system. 
The GL action for $C_3$, $D_3$, $C_6$, and $D_6$
has an additional term,  
$i\gamma[\phi^{\dagger}\partial_z \phi - 
(\partial_z \phi^{\dagger})\phi]$, that comes from the antisymmetric exchange interaction in 
Eq.~(\ref{dm}). Such a term is prohibited for $C_{3i}$, $D_{3d}$, $C_{6h}$, and $D_{6h}$ because there is the inversion symmetry. The term is also prohibited for $C_{3v}$, $C_{3h}$, $C_{6v}$, and $D_{3h}$. Although there is no inversion symmetry, for $C_{3v}$, $C_{6v}$, and $D_{3h}$, there is a vertical mirror symmetry that reflects only one component of the in-plane spin vector; for $C_{3h}$ and $D_{3h}$, there is a horizontal mirror symmetry which makes the term opposite. In conclusion, a continuum limit of {\it generic} $XY$ spin models in the hexagonal crystal family with a spatial inversion or rotoinversion symmetry as well as the time-reversal symmetry is described by Eq.~(\ref{z6}). For the $C_3$, $D_3$, $C_6$, and $D_6$ cases, the first-order $z$-derivative term can be eliminated by re-definitions of 
$\phi$ and $\rho_0$;  $\phi_{\rm new} = e^{-iz A_z} \phi_{\rm old}$ and 
$-U (\rho_0)_{\rm new} = - U(\rho_0)_{\rm old} 
- (\eta^2_1 c_z^2 A_z^2)/2$ with $\gamma=\eta^2_1 c_z^2 A_z/2$. If $A_z$ is commensurate 
to the phase locking by the $\tilde{c}_6$ term, $3A_za_z = \mathbb{Z}\pi$ with an integer $\mathbb{Z}$ ($a_{\bm{i}\bm{j},z}=a_z$), the partition function has 
no magnetic frustration and it describes the continuous phase transition 
from a disordered phase $(\rho_0<0)$ to the $XY$ 
ferromagnetic order phase with a spin-helix along $z$ $(\rho_0>0)$. As the first-order $z$-derivative term also respects the U(1) spacetime symmetry, 
the systems near  the transition point have also the effective U(1) 
spacetime symmetry for these cases. 

\subsubsection{Hydrodynamics in an intermediate length scale near the quantum critical point}
An analytic continuation of Eq.~(\ref{z6}) at $T=0$ ($\tau \rightarrow i t$) leads to 
the following real-time complex field theory $\tilde{\mathcal{L}}_\phi$, 
\begin{align}
\label{eqn1.71}
\tilde{\mathcal{L}}_\phi=&\frac{\eta_1^2}{2}(\partial_t\phi^\dagger)(\partial_t\phi)-\frac{\eta_1^2c_\perp^2}{2}(\partial_i\phi^\dagger)(\partial_i\phi) \nonumber \\
&\hspace{-0.8cm} 
-\frac{\eta_1^2 c^2}{4}[\alpha(\partial_-\phi)^2+\alpha^*(\partial_+\phi^\dagger)^2] -\frac{\eta_1^2 c^2}{4}[\tilde{\alpha}_6(\partial_+\phi)^2\phi^2 \nonumber \\
&\hspace{-0.8cm} 
+\tilde{\alpha}_6^*(\partial_-\phi^\dagger)^2(\phi^\dagger)^2]-\frac{U}{2}(\phi^\dagger\phi-\rho_0)^2-\frac{1}{2}[\tilde{c}_6\phi^6+\tilde{c}_6^*(\phi^\dagger)^6],
\end{align}
where we take classical solutions of $\phi$ independent of $z$, so the term of $(\partial_z\phi^\dagger)(\partial_z\phi)$ is negligible. Here, for simplicity, let us take $\alpha$, $\tilde{\alpha}_6$ and $\tilde{c}_6$ to be real, and assume that a coupling between the phase mode $\theta$ and the amplitude mode $\rho=\phi^\dagger\phi$ can be neglected. Then, we obtain an effective theory of the phase mode $\theta$,
\begin{align}
\label{eqn1.72}
\tilde{\mathcal{L}}=&\frac{\eta_1^2\rho_0}{2}(\partial_t\theta)^2-\frac{\eta_1^2c_\perp^2\rho_0}{2}(\partial_x\theta)^2[1-\alpha\mathrm{cos}(2\theta)-\tilde{\alpha}_6\rho_0^2\mathrm{cos}(4\theta)]\nonumber\\
&-\frac{\eta_1^2c_\perp^2\rho_0}{2}(\partial_y\theta)^2[1+\alpha\mathrm{cos}(2\theta)+\tilde{\alpha}_6\rho_0\mathrm{cos}(4\theta)]\nonumber\\
&+\eta_1^2c_\perp^2\rho_0(\partial_x\theta)(\partial_y\theta)[\alpha\mathrm{sin}(2\theta)-\tilde{\alpha}_6\rho_0\mathrm{sin}(4\theta)] \nonumber \\
& \ \ -\tilde{c}_6\rho_0^3\mathrm{cos}(6\theta).
\end{align}
As in Eq.~(\ref{z6}), terms with higher-order derivatives or 
higher order in $\rho_0$ are neglected in Eq.~(\ref{eqn1.72}). 
Eq.~(\ref{eqn1.72}) is symmetric under the joint $\mathbb{Z}_6$ rotation, 
\begin{eqnarray}
&\theta \rightarrow \theta + \frac{n \pi}{3}, \nonumber \\
&\left(\begin{array}{c}
x \\
y \\
\end{array}\right) \rightarrow \left(\begin{array}{cc} 
\cos \big(\frac{n \pi}{3}\big) & -\sin \big(\frac{n\pi}{3} \big) \\
\sin \big(\frac{n \pi}{3}\big) & \cos \big(\frac{n\pi}{3} \big) \\
\end{array}\right) \left(\begin{array}{c}
x \\
y \\
\end{array}\right), \ \  
\end{eqnarray}
while in the absence of $\tilde{\alpha}_6$ and $\tilde{c}_6$, 
it is symmetric under the joint U(1) rotation; 
\begin{eqnarray}
\theta \rightarrow \theta + \epsilon, \ \ 
\left(\begin{array}{c}
x \\
y \\
\end{array}\right) \rightarrow \left(\begin{array}{cc} 
\cos \epsilon & -\sin \epsilon \\
\sin \epsilon & \cos \epsilon \\
\end{array}\right) \left(\begin{array}{c}
x \\
y \\
\end{array}\right), \ \  
\end{eqnarray}
for $\forall \epsilon$. 

The U(1) theory becomes a good approximation theory for Eq.~(\ref{eqn1.72}), when $\rho_0$ approaches zero before $\partial_{\mu}\theta$ ($\mu=t,x,y$) approach zero. This is the case for the 3D spin model Eq.~(\ref{hydro-2}) in the intermediate length scale near the quantum critical point.
Thereby, the internal field $\phi$ is introduced as a spatial average of $S_{{\bm i},x}+iS_{{\bm i},y}$ over some hydrodynamic volume element. When the length scale of the volume element increases within the intermediate length scale, 
$\Lambda_{c,1}<\Lambda<\Lambda_{c,2}$, a scaling of $\rho_0$ and $\partial_{\mu}\theta$ is controlled by the quantum critical point; $\rho_0$ gets small faster 
than $\partial_{\mu} \theta$, and the approximation becomes better.  
On the other hand, when the length scale of the element becomes larger than the upper bound $\Lambda_{c,2}$, another scaling law from the Nambu-Goldstone fixed point kicks in, and the 
$\tilde{c}_6$ and $\tilde{\alpha}_6$ terms become relevant again~\cite{blankschtein84,oshikawa00}. Besides, for a 2D quantum spin model, although the $\tilde{c}_6$ term is dangerously marginal instead of dangerously irrelevant from simple dimensional counting, as long as a bare value of $\tilde{c}_6$ is small enough, there is still an intermediate length scale where the U(1) theory is applicable.
To summarize, the U(1) theory is effective near 
 the quantum critical point only when $\theta$ fluctuates over a length in the intermediate length scale. When $\theta$ fluctuates more slowly than $\Lambda_{c,2}$, $\partial_{\mu}\theta$ becomes smaller than a small but finite $\rho_0$, 
and the $\tilde{c}_6$ and $\tilde{\alpha}_6$ terms dominate over the others, giving a 
large contribution to the EOM.

\subsection{\label{appendixA2} Spin-triplet exciton model}
As another example of solid-state materials where the U(1) theory of spin dynamics is applicable, we consider 
semiconductors with electron excitations near a conduction-band bottom and hole excitations near a valence-band top around a high-symmetric $k$ point, e.g., the $\Gamma$ point. Near the band top and bottom, suppose the kinetic-energy bands can be approximately described by a rotational-symmetric continuous theory. The theory with relativistic spin-orbit interaction is expected to have joint continuous rotational symmetry.

To be specific, we consider a condensate of spin-triplet excitons in a 2D semiconductor model with Rashba-type spin-orbit interactions. The semiconductor model 
is given by (Eq.~(\ref{eqn_h_ex})) 
\begin{align}
\label{eqn1.1}
&H_{\mathrm{ex}}=\int d^2\bm{r}\bm{a}^\dagger[(-\frac{\partial_i^2}{2m_0}+\epsilon_{g0})\bm{\sigma}_0+\xi_{R0}(-i\partial_y\bm{\sigma}_x+i\partial_x\bm{\sigma}_y)]\bm{a}\nonumber\\
&+\int d^2\bm{r}\bm{b}^\dagger[(\frac{\partial_i^2}{2m^{\prime}_0}-\epsilon_{g0})\bm{\sigma}_0+\xi^{\prime}_{R0}(i\partial_y\bm{\sigma}_x-i\partial_x\bm{\sigma}_y)]\bm{b}\nonumber  \\
& +\int d^2\bm{r}(\Delta_t\bm{a}^\dagger\bm{\sigma}_0\bm{b}+\Delta_t^*\bm{b}^\dagger\bm{\sigma}_0\bm{a})+\frac{g_{s0}}{2}\sum_{\sigma,\sigma'=\uparrow,\downarrow}\int d^2\bm{r} \nonumber \\
& (a^\dagger_\sigma a^\dagger_{\sigma'}a_{\sigma'}a_\sigma +b^\dagger_\sigma b^\dagger_{\sigma'}b_{\sigma'}b_\sigma+2\xi_1 a^\dagger_\sigma b^\dagger_{\sigma'}b_{\sigma'}a_\sigma),
\end{align}
with $i=x,y$. Here $\bm{a}$ and $\bm{b}$ are spin-$\frac{1}{2}$ electron annihilation operators 
near the $\Gamma$ point in the conduction band and valence band, respectively. We suppose inter-band interaction is smaller than intra-band interaction, namely $0<\xi_1<1$. Due to the attraction between electrons and holes $(\xi_1 g_{s0})$, the quasiparticles form bound states inside a band gap ($\epsilon_{g0}$). The bound states have a spin-singlet component and spin-triplet components. In the presence of Rashba interaction ($\xi_{R0},\xi^{\prime}_{R0}$) and inter-band ``spinless" hopping ($\Delta_t$), the in-plane component of the spin-triplet states undergoes Bose-Einstein condensation at $q=0$. In the following, we will show that this condensation is described by Eq.~(\ref{eqn4}) 
(without the $c_z$ term). 

 For simplicity, we take the electron band and the hole band in a symmetric form, $m_0=m^{\prime}_0$, $\xi_{R0}=\xi^{\prime}_{R0}$, while the derivation can be generalized into the case with $m_0 \xi_{R0}=m^{\prime}_0 \xi^{\prime}_{R0}$. The derivation can be also applicable to a 3D model with a finite effective mass along $z$. 
Due to the Rashba interaction ($\xi_{R0}$) and interband tunneling ($\Delta_t$), the system has only a U(1) rotational symmetry and a time-reversal symmetry,
\begin{align}
\label{eqn1.2}
&\bm{a}\rightarrow e^{-i\epsilon\bm{\sigma}_z/2}\bm{a},\quad \bm{b}\rightarrow e^{-i\epsilon\bm{\sigma}_z/2}\bm{b},\nonumber \\ & \left(\begin{array}{c}
x \\ 
y\end{array}\right)\rightarrow\left(\begin{array}{cc}\mathrm{cos}\epsilon & -\mathrm{sin}\epsilon \\ \mathrm{sin}\epsilon & \mathrm{cos}\epsilon\end{array}\right)\left(\begin{array}{c}x \\ y\end{array}\right),
\end{align}
\begin{align}
\label{eqn1.3}
\bm{a}\rightarrow i\bm{\sigma}_y\bm{a},\quad \bm{b}\rightarrow i\bm{\sigma}_y\bm{b},\quad t\rightarrow -t,\quad i\rightarrow -i.
\end{align}
The quadratic part of the Hamiltonian Eq.~(\ref{eqn1.1}) is diagonalized, 
\begin{align}
\label{eqn1.4}
H_{\mathrm{ex}}=&\sum_{\bm{k}}\{a^\dagger_{\bm{k},\sigma_{\bm{k}}}[\frac{(|\bm{k}|-k_R)^2}{2m_0}+E_{g0}]a_{\bm{k},\sigma_{\bm{k}}} 
\nonumber \\
& \hspace{1cm} -b^\dagger_{\bm{k},\sigma_{\bm{k}}}[\frac{(|\bm{k}|-k_R)^2}{2m_0}+E_{g0}]b_{\bm{k},\sigma_{\bm{k}}}\}\nonumber\\
&+\sum_{\bm{k}}[\Delta_t a^\dagger_{\bm{k},\sigma_{\bm{k}}}b_{\bm{k},\sigma_{\bm{k}}}+\Delta_t^*b^\dagger_{\bm{k},\sigma_{\bm{k}}}a_{\bm{k},\sigma_{\bm{k}}}]
+\frac{g_s}{2}\sum_{\sigma,\sigma'=\uparrow,\downarrow} \nonumber \\
& \int d^2\bm{r} (a^\dagger_\sigma a^\dagger_{\sigma'}a_{\sigma'}a_\sigma+b^\dagger_\sigma b^\dagger_{\sigma'}b_{\sigma'}b_\sigma+2a^\dagger_\sigma b^\dagger_{\sigma'}b_{\sigma'}a_\sigma),
\end{align}
where
\begin{align}
\label{eqn1.5}
k_R=m_0 \xi_{R0},\quad E_{g0}=\epsilon_{g0}-\frac{k_R^2}{2m_0},
\end{align}
$\bm{a}_{\bm{k}}$ and $\bm{b}_{\bm{k}}$ are Fourier transforms of $\bm{a}(\bm{r})$ and $\bm{b}(\bm{r})$, $\sigma_{\bm{k}}$ denotes up spin along %
the direction of 
$\widehat{\bm{z}}\times\widehat{\bm{k}}$, $\widehat{\bm{k}}\equiv\frac{\bm{k}}{|\bm{k}|}$. Here we discard the down-spin bands 
of the conduction and valence bands, because they  
are higher in energy and they do not constitute low-energy exciton levels.  
Since excitons are formed by electrons and holes around the $\Gamma$ point, 
we neglect $|\bm{k}|$-dependence of the hybridization coefficients of the conduction and valence bands,
\begin{align}
\label{eqn1.6}
a_\sigma=\alpha_\sigma\mathrm{cos}\Theta-\beta_\sigma e^{i\Phi}\mathrm{sin}\Theta,\quad b_\sigma=\alpha_\sigma e^{-i\Phi}\mathrm{sin}\Theta+\beta_\sigma\mathrm{cos}\Theta,
\end{align}
where
\begin{align}
\label{eqn1.7}
E_{g0}=\sqrt{E_{g0}^2+|\Delta_t|^2}\mathrm{cos}2\Theta,\quad \Delta_t=\sqrt{E_{g0}^2+|\Delta_t|^2}e^{i\Phi}\mathrm{sin}2\Theta.
\end{align}
Taking Eq.~(\ref{eqn1.6}) into the interaction term 
and taking $\Phi=0$ for simplicity, we get
\begin{align}
\label{eqn1.9}
&g_{s0}(a^\dagger_\sigma a^\dagger_{\sigma'}a_{\sigma'}a_\sigma+b^\dagger_\sigma b^\dagger_{\sigma'}b_{\sigma'}b_\sigma+\xi_1 a^\dagger_\sigma b^\dagger_{\sigma'}b_{\sigma'}a_\sigma+\xi_1 b^\dagger_\sigma a^\dagger_{\sigma'}a_{\sigma'}b_\sigma)\nonumber\\
& \ \ \  =g_s(\alpha^\dagger_{\sigma}\beta^\dagger_{\sigma'}\beta_{\sigma'}\alpha_\sigma+\beta^\dagger_{\sigma}\alpha^\dagger_{\sigma'}\alpha_{\sigma'}\beta_\sigma)
\nonumber \\
& \ \ \ \ \  +wg_s(\alpha^\dagger_\sigma\beta^\dagger_{\sigma'}\alpha_{\sigma'}\beta_\sigma+\beta^\dagger_\sigma\alpha^\dagger_{\sigma'}\beta_{\sigma'}\alpha_\sigma)\nonumber\\
&\ \ \ \ \ \  +w g_s(\alpha^\dagger_\sigma\alpha^\dagger_{\sigma'}\beta_{\sigma'}\beta_\sigma+\beta^\dagger_\sigma\beta^\dagger_{\sigma'}\alpha_{\sigma'}\alpha_\sigma)+....,
\end{align}
with
\begin{align}
\label{eqn1.12}
g_s&=\frac{g_{s0}}{2}[\mathrm{sin}^2(2\Theta)+\xi_1+\xi_1\mathrm{cos}^2(2\Theta)],
\nonumber \\
w&=\frac{(1-\xi_1)\mathrm{sin}^2 (2\Theta)}{\mathrm{sin}^2 (2\Theta)+\xi_1[1+\mathrm{cos}^2 (2\Theta)]}, 
\end{align}
and $0<w<1$.
Here we only keep terms in exciton-pairing channels in the basis of $\alpha$ and $\beta$, $\alpha^\dagger_\sigma\beta^\dagger_{\sigma'}\beta_{\sigma'}\alpha_\sigma$, $\alpha^\dagger_\sigma\alpha^\dagger_{\sigma'}\beta_{\sigma'}\beta_\sigma$, and $\beta^\dagger_\sigma\beta^\dagger_{\sigma'}\alpha_{\sigma'}\alpha_\sigma$. 
Neglected terms contain also hybridization between excitons 
and intraband collective modes, $\alpha^\dagger_\sigma\alpha^\dagger_{\sigma'}\alpha_{\sigma'}\beta_\sigma$, which in the absence of the time-reversal symmetry leads to an additional cubic term $(\partial_+\phi)(\phi^\dagger)^2$ in Eq.~(\ref{eqn4}). The hybridization and other neglected terms can be safely omitted as the intraband collective modes are gapped excitations in the semiconductor. 
The Hamiltonian can be rewritten by the new basis, 
\begin{align}
\label{eqn1.10}
H_{\mathrm{ex}}=&\sum_{\bm{k}}\{\alpha^\dagger_{\bm{k},\sigma_{\bm{k}}}[\frac{(|\bm{k}|-k_R)^2}{2m}+E_{g}]\alpha_{\bm{k},\sigma_{\bm{k}}} \nonumber \\
& 
\hspace{1cm} - \beta^\dagger_{\bm{k},\sigma_{\bm{k}}}[\frac{(|\bm{k}|-k_R)^2}{2m}+E_{g}]\beta_{\bm{k},\sigma_{\bm{k}}}\}\nonumber\\
&+g_s\sum_{\sigma,\sigma'=\uparrow,\downarrow}\int d^2\bm{r}(\alpha^\dagger_\sigma \beta^\dagger_{\sigma'}\beta_{\sigma'}\alpha_\sigma+w\alpha^\dagger_\sigma \beta^\dagger_{\sigma'}\alpha_{\sigma'}\beta_\sigma \nonumber \\
& \ \ \ +\frac{w}{2}\alpha^\dagger_\sigma\alpha^\dagger_{\sigma'}\beta_{\sigma'}\beta_\sigma+\frac{w}{2}\beta^\dagger_\sigma\beta^\dagger_{\sigma'}\alpha_{\sigma'}\alpha_\sigma),
\end{align}
where
\begin{align}
\label{eqn1.11}
&\sqrt{[\frac{(|\bm{k}|-k_R)^2}{2m_0}+E_{g0}]^2+|\Delta_t|^2} \nonumber \\
&\approx\sqrt{E_{g0}^2+|\Delta_t|^2}+\frac{E_{g0}}{\sqrt{E_{g0}^2+|\Delta_t|^2}}\frac{(|\bm{k}|-k_R)^2}{2m_0}\nonumber\\
&=E_g+\frac{(|\bm{k}|-k_R)^2}{2m}.
\end{align}
Exciton operators are defined by $O_\mu=\bm{b}^\dagger\bm{\sigma}_\mu\bm{a}$ where $\mu=0,x,y,z$. In terms of a completeness relation
\begin{align}
\label{eqn1.13}
\frac{1}{2}\sum_{\mu}(\sigma_\mu)_{\alpha\beta}(\sigma_\mu)_{\gamma\delta}=\delta_{\alpha\delta}\delta_{\beta\gamma},
\end{align}
the interaction terms are decomposed as follows, 
\begin{align}
\label{eqn1.14}
\sum_{\sigma,\sigma'}\alpha^\dagger_\sigma \beta^\dagger_{\sigma'}\beta_{\sigma'}\alpha_\sigma=-\sum_{\sigma,\sigma'}\alpha^\dagger_\sigma \beta_{\sigma'}\beta^\dagger_{\sigma'}\alpha_\sigma=-\frac{1}{2}\sum_\mu O^\dagger_\mu O_\mu,
\end{align}
\begin{align}
\label{eqn1.15}
\sum_{\sigma,\sigma'}\alpha^\dagger_\sigma \beta^\dagger_{\sigma'}\alpha_{\sigma'}\beta_\sigma=\sum_{\sigma,\sigma'}\alpha^\dagger_\sigma \beta_{\sigma}\beta^\dagger_{\sigma'}\alpha_{\sigma'}=O^\dagger_0 O_0,
\end{align}
\begin{align}
\label{eqn1.16}
\sum_{\sigma,\sigma'}\alpha^\dagger_\sigma\alpha^\dagger_{\sigma'}\beta_{\sigma'}\beta_\sigma
&\sim-\sum_{\sigma,\sigma'}\alpha^\dagger_\sigma\beta_{\sigma'}\alpha^\dagger_{\sigma'}\beta_\sigma+\sum_{\sigma,\sigma'}\alpha^\dagger_\sigma\beta_\sigma\alpha^\dagger_{\sigma'}\beta_{\sigma'} \nonumber \\
&=-\frac{1}{2}\sum_\mu O^\dagger_\mu O^\dagger_\mu+O_0^\dagger O_0^\dagger.
\end{align}
In Eq.~(\ref{eqn1.16}), we decompose the interaction in two different channels, so we do not 
divide the result by two. 
Adding Eqs.~(\ref{eqn1.14}-\ref{eqn1.16}) together, we get
\begin{align}
\label{eqn1.17}
&g_s\sum_{\sigma,\sigma'}(\alpha^\dagger_\sigma \beta^\dagger_{\sigma'}\beta_{\sigma'}\alpha_\sigma+w\alpha^\dagger_\sigma \beta^\dagger_{\sigma'}\alpha_{\sigma'}\beta_\sigma\nonumber \\
& \ +\frac{w}{2}\alpha^\dagger_\sigma\alpha^\dagger_{\sigma'}\beta_{\sigma'}\beta_\sigma+\frac{w}{2}\beta^\dagger_\sigma\beta^\dagger_{\sigma'}\alpha_{\sigma'}\alpha_\sigma)\nonumber\\
=&g_s[-\frac{1}{2}\sum_\mu O^\dagger_\mu O_\mu+wO^\dagger_0 O_0-\frac{w}{4}\sum_\mu(O^\dagger_\mu O^\dagger_\mu+O_\mu O_\mu) \nonumber \\
& \ +\frac{w}{2}(O^\dagger_0 O^\dagger_0+O_0 O_0)] \nonumber\\
=&-\frac{g_s}{4}[\sum_r(2O^\dagger_r O_r+wO^\dagger_rO^\dagger_r+wO_rO_r) 
\nonumber \\
& +(2-4w)O^\dagger_0 O_0-wO^\dagger_0 O^\dagger_0-wO_0O_0]\nonumber\\
=&-\frac{g_s}{2}[\sum_r(1+w\mathrm{cos}2q_r)P_r^2+(1-2w-w\mathrm{cos}2q_0)P_0^2],
\end{align}
where $r=x,y,z$, $O_\mu\equiv P_\mu e^{iq_\mu}$. Note that due to $w$, the U(1) symmetry of the four-component exciton field reduces to a $\mathbb{Z}_2$ 
symmetry.  
Since $0<w<1$, $q_0=\pm \frac{\pi}{2}$ and $q_r=0,\pi$ are preferred 
by the interaction.  
Fluctuations of $q_\mu$ around the minima are gapped excitations, 
so they can be safely neglected. This leads to 
\begin{align}
\label{eqn1.18}
&g_s\sum_{\sigma,\sigma'}(\alpha^\dagger_\sigma \beta^\dagger_{\sigma'}\beta_{\sigma'}\alpha_\sigma+w\alpha^\dagger_\sigma \beta^\dagger_{\sigma'}\alpha_{\sigma'}\beta_\sigma \nonumber \\
& \ \ -\frac{w}{2}\alpha^\dagger_\sigma\alpha^\dagger_{\sigma'}\beta_{\sigma'}\beta_\sigma-\frac{w}{2}\beta^\dagger_\sigma\beta^\dagger_{\sigma'}\alpha_{\sigma'}\alpha_\sigma)\nonumber\\
&=-\frac{g_s}{2}[\sum_r(1+w)P_r^2+(1-w)P_0^2] \nonumber \\
&=-\frac{g_s}{2}[\sum_r(1+w)\Big(\frac{O_r+O_r^\dagger}{2}\Big)^2+
(1-w)\Big(\frac{O_0-O_0^\dagger}{2i}\Big)^2].
\end{align}
By the Hubbard-Stratonovich transformation, we can introduce real exciton fields $\phi_\mu$,
\begin{align}
\label{eqn1.19}
&\mathrm{exp}\{\int d\tau d^2\bm{r}\frac{g_s}{2}[\sum_r(1+w)P_r^2+(1-w)P_0^2]\} \nonumber \\
&=\int\mathcal{D}\phi_\mu \mathrm{exp}\{-\int d\tau d^2\bm{r}[-\sum_r \phi_r (O_r^\dagger+O_r)\nonumber\\
&-i\phi_0(O_0^\dagger-O_0)+\sum_r\frac{2}{g_s(1+w)}\phi_r^2+\frac{2}{g_s(1-w)}\phi_0^2]\},
\end{align}
where $\phi_r$ and $\phi_0$ have the physical meanings of $\frac{g_s(1+w)}{2}\langle P_r\rangle$ and $\frac{g_s(1-w)}{2}\langle P_0\rangle$, respectively. Since $0<w<1$, the interaction term (Eq.~(\ref{eqn1.18})) favors the triplet excitons ($\phi_r$) over the singlet excitons ($\phi_0$). The quadratic part of Eq.~(\ref{eqn1.10}) also lifts the four-fold degeneracy of $\phi_\mu$, while mass terms for $\phi_x$ and $\phi_y$ remain degenerate.

Due to the adjustment of the conduction band and valence band, $m_0 \xi_{R0}=m^{\prime}_0 \xi^{\prime}_{R0}$, one can expect that momentum-energy dispersions of the exciton bands have minima at $q=0$, so condensation of the exciton fields happen at the zero momentum. In the following, we keep track of all the four components, $\phi_{\mu}$ ($\mu=0,x,y,z$), in the derivation of Eq.~(\ref{eqn4}), to see whether and when $\phi_x$ and $\phi_y$ achieve the lowest energy (smallest mass at $q=0$) among others. 
Fermion fields can be integrated out,
\begin{widetext}
\begin{align}
\label{eqn1.20}
\int\mathcal{D}[\bm{a}^\dagger,\bm{b}^\dagger,\bm{a},\bm{b}]\mathrm{exp}\big{[}-\left(\begin{array}{cc}\bm{a}^\dagger &\bm{b}^\dagger\end{array}\right)\bm{G}^{-1}\left(\begin{array}{c}\bm{a}\\ \bm{b}\end{array}\right)\big{]}=\mathrm{det}(\bm{G}^{-1})=e^{\mathrm{Tr}\mathrm{ln}(\bm{G}^{-1})},
\end{align}
where
\begin{align}
\label{eqn1.21}
\bm{G}^{-1}=\bm{G}_{0}^{-1}+\bm{G}_\phi,
\end{align}
\begin{align}
\label{eqn1.22}
\bm{G}_{0}=&\left(\begin{array}{cc}g^a_{0,k}\bm{P}_{\sigma_{\bm{k}}} & 0 \\ 0 & g^b_{0,k}\bm{P}_{\sigma_{\bm{k}}}\end{array}\right) 
\equiv\left(\begin{array}{cc}[-i\omega_n+\frac{(|\bm{k}|-k_R)^2}{2m}+E_{g}]^{-1}\frac{\bm{\sigma}_0+\bm{\sigma}_{\widehat{\bm{z}}\times\widehat{\bm{k}}}}{2} & 0 \\ 0 & [-i\omega_n-\frac{(|\bm{k}|-k_R)^2}{2m}-E_{g}]^{-1}\frac{\bm{\sigma}_0+\bm{\sigma}_{\widehat{\bm{z}}\times\widehat{\bm{k}}}}{2}\end{array}\right),
\end{align}
\begin{align}
\label{eqn1.23}
\bm{G}_{\phi}=\left(\begin{array}{cc}0 & -\sum_r\phi_r\bm{\sigma}_r-i\phi_0\bm{\sigma}_0 \\ -\sum_r\phi_r\bm{\sigma}_r+i\phi_0\bm{\sigma}_0 & 0\end{array}\right).
\end{align}
$\bm{G}_{0}^{-1}$ and $\bm{G}_{\phi}$ are block-diagonal in the momentum-frequency space and the coordinate space, respectively. $\bm{G}_0$ is diagonal in spin along 
$\hat{\bm z}\times \hat{\bm k}$ and its diagonal element is zero for the down spin, 
\begin{align}
\label{eqn1.24}
\bm{P}_{\sigma_{\bm{k}}}=\frac{1}{2}(\bm{\sigma}_0+\bm{\sigma}_{\widehat{\bm{z}}\times\widehat{\bm{k}}})=\frac{1}{2}(\bm{\sigma}_0-\bm{\sigma}_x\mathrm{sin}\theta_{\widehat{\bm{k}}}+\bm{\sigma}_y\mathrm{cos}\theta_{\widehat{\bm{k}}}).
\end{align}
The integration leads to an effective theory of the exciton fields,
\begin{align}
\label{eqn1.25}
\mathcal{S}_\phi[\phi_\mu]=-\mathrm{Tr}\mathrm{ln}(\bm{1}+\bm{G}_0\bm{G}_\phi) +\frac{2}{g_s}\int d\tau d^2\bm{r}(\sum_r\frac{1}{1+w}\phi_r^2+\frac{1}{1-w}\phi_0^2),
\end{align}
\begin{align}
\label{eqn1.26}
-\mathrm{Tr}\mathrm{ln}(\bm{1}+\bm{G}_0\bm{G}_\phi)=\frac{1}{2}\mathrm{Tr}(\bm{G}_0\bm{G}_\phi\bm{G}_0\bm{G}_\phi) +\frac{1}{4}\mathrm{Tr}(\bm{G}_0\bm{G}_\phi\bm{G}_0\bm{G}_\phi\bm{G}_0\bm{G}_\phi\bm{G}_0\bm{G}_\phi)+...
\end{align}
Note that ``Tr" stands for 
traces over both spacetime and spin indices, while ``tr" is 
trace over only spin indices [see below]. To determine the form of the effective theory, we use the following relations,
\begin{align}
\label{eqn1.27}
\bm{P}_{\sigma_{\bm{k}}}\bm{\sigma}_z \bm{P}_{\sigma_{\bm{k}}}=0,
\end{align}
\begin{align}
\label{eqn1.28}
\bm{P}_{\sigma_{\bm{k}}}\bm{\sigma}_x \bm{P}_{\sigma_{\bm{k}}}=\bm{P}_{\sigma_{\bm{k}}}(\bm{\sigma}_{\widehat{\bm{k}}}\mathrm{cos}\theta_{\widehat{\bm{k}}}-\bm{\sigma}_{\widehat{\bm{z}}\times\widehat{\bm{k}}}\mathrm{sin}\theta_{\widehat{\bm{k}}})\bm{P}_{\sigma_{\bm{k}}
} =-\mathrm{sin}\theta_{\widehat{\bm{k}}}\bm{P}_{\sigma_{\bm{k}}}\bm{\sigma}_{\widehat{\bm{z}}\times\widehat{\bm{k}}}=-\mathrm{sin}\theta_{\widehat{\bm{k}}}\bm{\sigma}_{\widehat{\bm{z}}\times\widehat{\bm{k}}}\bm{P}_{\sigma_{\bm{k}}},
\end{align}
\begin{align}
\label{eqn1.29}
\mathrm{tr}(\bm{P}_{\sigma_{\bm{k}}}\bm{\sigma}_0 \bm{P}_{\sigma_{\bm{k}}}\bm{\sigma}_0)=\mathrm{tr}(\bm{P}_{\sigma_{\bm{k}}}^2)=\mathrm{tr}(\bm{P}_{\sigma_{\bm{k}}})=1,
\end{align}
\begin{align}
\label{eqn1.30}
&\mathrm{tr}(\bm{P}_{\sigma_{\bm{k}}}\bm{\sigma}_x\bm{P}_{\sigma_{\bm{k}+\bm{q}}}\bm{\sigma}_x) 
\nonumber \\ 
&=\frac{1}{4}\mathrm{tr}(\bm{\sigma}_x\bm{\sigma}_x)+\frac{1}{4}\mathrm{tr}(\bm{\sigma}_{\widehat{\bm{z}}\times\widehat{\bm{k}}}\bm{\sigma}_x\bm{\sigma}_{\widehat{\bm{z}}\times\widehat{\bm{k}+\bm{q}}}\bm{\sigma}_x) 
=\frac{1}{2}+\frac{1}{4}\mathrm{tr}[(-\mathrm{sin}\theta_{\widehat{\bm{k}}}\bm{\sigma}_x+\mathrm{cos}\theta_{\widehat{\bm{k}}}\bm{\sigma}_y)\bm{\sigma}_x(-\mathrm{sin}\theta_{\widehat{\bm{k}+\bm{q}}}\bm{\sigma}_x+\mathrm{cos}\theta_{\widehat{\bm{k}+\bm{q}}}\bm{\sigma}_y)\bm{\sigma}_x]\nonumber\\
&=\frac{1}{2}(1+\mathrm{sin}\theta_{\widehat{\bm{k}}}\mathrm{sin}\theta_{\widehat{\bm{k}+\bm{q}}}-\mathrm{cos}\theta_{\widehat{\bm{k}}}\mathrm{cos}\theta_{\widehat{\bm{k}+\bm{q}}})=\frac{1}{2}[1-\mathrm{cos}(\theta_{\widehat{\bm{k}}}+\theta_{\widehat{\bm{k}+\bm{q}}})],
\end{align}
\begin{align}
\label{eqn1.31}
\mathrm{tr}(\bm{P}_{\sigma_{\bm{k}}}\bm{\sigma}_y\bm{P}_{\sigma_{\bm{k}+\bm{q}}}\bm{\sigma}_y)=\frac{1}{2}[1+\mathrm{cos}(\theta_{\widehat{\bm{k}}}+\theta_{\widehat{\bm{k}+\bm{q}}})],
\end{align}
\begin{align}
\label{eqn1.32}
\mathrm{tr}(\bm{P}_{\sigma_{\bm{k}}}\bm{\sigma}_x\bm{P}_{\sigma_{\bm{k}+\bm{q}}}\bm{\sigma}_y)&=\frac{1}{4}\mathrm{tr}(\bm{\sigma}_{\widehat{\bm{z}}\times\widehat{\bm{k}}}\bm{\sigma}_x\bm{\sigma}_{\widehat{\bm{z}}\times\widehat{\bm{k}+\bm{q}}}\bm{\sigma}_y)
=\frac{1}{4}\mathrm{tr}[(-\mathrm{sin}\theta_{\widehat{\bm{k}}}\bm{\sigma}_x+\mathrm{cos}\theta_{\widehat{\bm{k}}}\bm{\sigma}_y)\bm{\sigma}_x(-\mathrm{sin}\theta_{\widehat{\bm{k}+\bm{q}}}\bm{\sigma}_x+\mathrm{cos}\theta_{\widehat{\bm{k}+\bm{q}}}\bm{\sigma}_y)\bm{\sigma}_y]\nonumber\\
&=-\frac{1}{2}(\mathrm{sin}\theta_{\widehat{\bm{k}}}\mathrm{cos}\theta_{\widehat{\bm{k}+\bm{q}}}+\mathrm{cos}\theta_{\widehat{\bm{k}}}\mathrm{sin}\theta_{\widehat{\bm{k}+\bm{q}}})=-\frac{1}{2}\mathrm{sin}(\theta_{\widehat{\bm{k}}}+\theta_{\widehat{\bm{k}+\bm{q}}}),
\end{align}
\begin{align}
\label{eqn1.33}
\mathrm{tr}(\bm{P}_{\sigma_{\bm{k}}}\bm{\sigma}_y\bm{P}_{\sigma_{\bm{k}+\bm{q}}}\bm{\sigma}_x)=-\frac{1}{2}\mathrm{sin}(\theta_{\widehat{\bm{k}}}+\theta_{\widehat{\bm{k}+\bm{q}}}),
\end{align}
\begin{align}
\label{eqn1.34}
&\mathrm{tr}(\bm{P}_{\sigma_{\bm{k}}}\bm{\sigma}_x\bm{P}_{\sigma_{\bm{k}}}\bm{\sigma}_x\bm{P}_{\sigma_{\bm{k}}}\bm{\sigma}_x\bm{P}_{\sigma_{\bm{k}}}\bm{\sigma}_x)=\mathrm{tr}[(\bm{P}_{\sigma_{\bm{k}}}\bm{\sigma}_x\bm{P}_{\sigma_{\bm{k}}})^4] 
=\mathrm{tr}[(\bm{P}_{\sigma_{\bm{k}}}\bm{\sigma}_{\widehat{\bm{z}}\times\widehat{\bm{k}}}\bm{\sigma}_{\widehat{\bm{z}}\times\widehat{\bm{k}}}\bm{P}_{\sigma_{\bm{k}}})^2]\mathrm{sin}^4\theta_{\widehat{\bm{k}}}=\mathrm{tr}(\bm{P}_{\sigma_{\bm{k}}})\mathrm{sin}^4\theta_{\widehat{\bm{k}}}=\mathrm{sin}^4\theta_{\widehat{\bm{k}}}.
\end{align}
As the exciton field $\phi_{\mu}({\bm q})$ (Fourier transform of $\phi_{\mu}({\bm r})$) at the zero momentum (${\bm q}=0$) is expected to have the smallest energy, we expand the effective theory in terms of small ${\bm q}$. The zeroth order in ${\bm q}$ gives the mass ($M_{\mu}$) and the quartic term ($U$) from the first and the second terms in Eq.~(\ref{eqn1.26}), respectively. Let us first calculate the masses for the four exciton components $(\mu=0,x,y,z)$,
\begin{align}
\label{eqn1.35}
&\frac{1}{2}\mathrm{Tr}(\bm{G}_0\bm{G}_\phi\bm{G}_0\bm{G}_\phi)\supset \sum_\mu\int d\tau d^2\bm{r}\phi_\mu^2[\frac{1}{\beta L^2}\sum_k g^a_{0,k}g^b_{0,k}\mathrm{tr}(\bm{P}_{\sigma_{\bm{k}}}\bm{\sigma}_\mu \bm{P}_{\sigma_{\bm{k}}}\bm{\sigma}_\mu)]\nonumber\\
=&\int d\tau d^2\bm{r}\phi_0^2(\frac{1}{\beta L^2}\sum_k g^a_{0,k}g^b_{0,k})+\frac{1}{2}\int d\tau d^2\bm{r}(\phi_x^2+\phi_y^2)(\frac{1}{\beta L^2}\sum_k g^a_{0,k}g^b_{0,k}),
\end{align}
\end{widetext}
where we used Eqs.~(\ref{eqn1.27},\ref{eqn1.29}-\ref{eqn1.31}) and 
$\sum_{\bm k}\cos(2\theta_{\widehat{\bm{k}}})=
\sum_{\bm k}\sin(2\theta_{\widehat{\bm{k}}})=0$. Taking Eqs.~(\ref{eqn1.25},\ref{eqn1.35}) together, we get
\begin{align}
\label{eqn1.36}
\frac{1}{2}\sum_\mu M_\mu\phi_\mu^2=&\frac{2}{g_s}(\sum_r\frac{1}{1+w}\phi_r^2+\frac{1}{1-w}\phi_0^2) \nonumber \\
& -\frac{D_0}{2}(\phi_x^2+\phi_y^2+2\phi_0^2),
\end{align}
where
\begin{align}
\label{eqn1.37}
D_0=&-\frac{1}{\beta L^2}\sum_k g^a_{0,k}g^b_{0,k}=-\frac{1}{\beta L^2}\sum_{n,\bm{k}}\frac{1}{i\omega_n-\xi_{\bm{k}}}\frac{1}{i\omega_n+\xi_{\bm{k}}}\nonumber\\
=&-\frac{1}{L^2}\sum_{\bm{k}}\frac{n_F(\xi_{\bm{k}})-n_F(-\xi_{\bm{k}})}{2\xi_{\bm{k}}}\nonumber \\
=&\int\frac{d^2\bm{k}}{(2\pi)^2}\frac{\mathrm{tanh}(\frac{1}{2}\beta\xi_{\bm{k}})}{2\xi_{\bm{k}}}>0,
\end{align}
\begin{align}
\label{eqn1.38}
\xi_{\bm{k}}\equiv\frac{(|\bm{k}|-k_R)^2}{2m}+E_{g},\quad n_F(\xi_{\bm{k}})\equiv\frac{1}{e^{\beta\xi_{\bm{k}}}+1},
\end{align}
$\beta$ is the inverse of temperature. So we have
\begin{align}
\label{eqn1.39}
M_z&=\frac{4}{g_s(1+w)}>0,\!\ M_0=\frac{4}{g_s(1-w)}-2D_0, \nonumber \\ 
M_i&=\frac{4}{g_s(1+w)}-D_0,
\end{align}
where $i=x,y$. $M_i<0<M_0$ is realized by  
\begin{align}
\label{eqn1.40}
\frac{1}{1+w}<\frac{1}{4}g_sD_0<\frac{1}{2(1-w)}.
\end{align}
Given that $w>\frac{1}{3}$, 
the condition can be realized by proper $g_s$, $\beta$,  
and $E_g$. Given that the condition is satisfied, we henceforth 
consider the effective theory only of $\phi=\phi_x+i\phi_y$ and neglect the other exciton components as they are gapped modes. 

Eq.~(\ref{eqn4}) takes the following form in 
the imaginary-time representation,
\begin{align}
\label{eqn1.41}
\mathcal{L}_{\phi,E}=&\frac{\eta_1^2}{2}(\partial_\tau\phi^\dagger)(\partial_\tau\phi)+\frac{\eta_1^2c_\perp^2}{2}(\partial_i\phi^\dagger)(\partial_i\phi)\nonumber \\
&+\frac{\eta_1^2 c_\perp^2}{4}[\alpha(\partial_-\phi)^2+\alpha^*(\partial_+\phi^\dagger)^2]+\frac{U}{2}(\phi^\dagger\phi-\rho_0)^2,
\end{align}
where the mass term was already obtained,
\begin{align}
\label{eqn1.42}
2U\rho_0=M_i=D_0-\frac{4}{g_s(1+w)}.
\end{align}
To determine $U$, we calculate the quartic term in $\phi_x$ from the second term 
of Eq.~(\ref{eqn1.26}), using Eq.~(\ref{eqn1.34}) and $\sum_{\bm k}\sin^4\theta_{\widehat{\bm k}}=\frac{3}{8}\sum_{\bm k}$,
\begin{widetext}
\begin{align}
\label{eqn1.43}
&\frac{1}{4}\mathrm{Tr}[(\bm{G}_0\bm{G}_\phi)^4]\supset\frac{1}{\beta L^2}\sum_{q_1,q_2,q_3}\phi_x(q_1)\phi_x(q_2)\phi_x(q_3)\phi_x(-q_1-q_2-q_3)\{\frac{1}{2\beta L^2}\sum_k (g_{0,k}^ag_{0,k}^b)^2\mathrm{tr}[(\bm{P}_{\sigma_{\bm{k}}}\bm{\sigma}_x)^4]\}\nonumber\\
=&\int d\tau d^2\bm{r}\phi_x^4\{\frac{1}{2\beta L^2}\sum_k (g_{0,k}^ag_{0,k}^b)^2\mathrm{tr}[(\bm{P}_{\sigma_{\bm{k}}}\bm{\sigma}_x)^4]\}=\frac{3}{8}\int d\tau d^2\bm{r}\phi_x^4[\frac{1}{2\beta L^2}\sum_k (g_{0,k}^ag_{0,k}^b)^2].
\end{align}
Thus, $U$ is given by 
\begin{align}
\label{eqn1.44}
U=\frac{3}{8\beta L^2}\sum_k (g_{0,k}^a g_{0,k}^b)^2
=&\frac{3}{8}\int\frac{d^2\bm{k}}{(2\pi)^2}\{\frac{d}{dz}[\frac{1}{e^{\beta z}+1}\frac{1}{(z-\xi_{\bm{k}})^2}]|_{z=-\xi_{\bm{k}}}+\frac{d}{dz}[\frac{1}{e^{\beta z}+1}\frac{1}{(z+\xi_{\bm{k}})^2}]|_{z=\xi_{\bm{k}}}]\nonumber\\
=&\frac{3}{8}\int\frac{d^2\bm{k}}{(2\pi)^2}\frac{1}{(2\xi_{\bm{k}})^2}[\frac{\mathrm{tanh}(\frac{1}{2}\beta\xi_{\bm{k}})}{\xi_{\bm{k}}}-\frac{\beta}{1+\mathrm{cosh}(\beta\xi_{\bm{k}})}]>0.
\end{align}
To determine the coefficients of the lowest-order gradient terms, 
we define $\bm{\sigma}_\pm=\frac{1}{2}(\bm{\sigma}_x\pm i\bm{\sigma}_y)$, 
$\phi=\phi_x+i\phi_y$, $\phi^{\dagger}=\phi_x-i\phi_y$, and 
set $\phi_0=\phi_z=0$ in $\bm{G}_{\phi}$ in Eq.~(\ref{eqn1.23}).
Note also that 
\begin{align}
\label{eqn1.45}
\bm{G}_{\phi}=\left(\begin{array}{cc}0 & -(\phi\bm{\sigma}_-+\phi^\dagger\bm{\sigma}_+) \\ -(\phi\bm{\sigma}_-+\phi^\dagger\bm{\sigma}_+) & 0\end{array}\right),
\end{align}
\begin{align}
\label{eqn1.46}
\mathrm{tr}(\bm{P}_{\sigma_{\bm{k}}}\bm{\sigma}_+\bm{P}_{\sigma_{\bm{k}+\bm{q}}}\bm{\sigma}_-)=\frac{1}{8}[1-\mathrm{cos}(\theta_{\widehat{\bm{k}}}+\theta_{\widehat{\bm{k}+\bm{q}}})+1+\mathrm{cos}(\theta_{\widehat{\bm{k}}}+\theta_{\widehat{\bm{k}+\bm{q}}})]=\frac{1}{4},
\end{align}
\begin{align}
\label{eqn1.47}
&\mathrm{tr}(\bm{P}_{\sigma_{\bm{k}}}\bm{\sigma}_+\bm{P}_{\sigma_{\bm{k}+\bm{q}}}\bm{\sigma}_+)=\frac{1}{8}[1-\mathrm{cos}(\theta_{\widehat{\bm{k}}}+\theta_{\widehat{\bm{k}+\bm{q}}})-1-\mathrm{cos}(\theta_{\widehat{\bm{k}}}+\theta_{\widehat{\bm{k}+\bm{q}}})\nonumber\\
&-i\mathrm{sin}(\theta_{\widehat{\bm{k}}}+\theta_{\widehat{\bm{k}+\bm{q}}})-i\mathrm{sin}(\theta_{\widehat{\bm{k}}}+\theta_{\widehat{\bm{k}+\bm{q}}})]=-\frac{1}{4}e^{i(\theta_{\widehat{\bm{k}}}+\theta_{\widehat{\bm{k}+\bm{q}}})},
\end{align}
\begin{align}
\label{eqn1.48}
\mathrm{tr}(\bm{P}_{\sigma_{\bm{k}}}\bm{\sigma}_-\bm{P}_{\sigma_{\bm{k}+\bm{q}}}\bm{\sigma}_-)=-\frac{1}{4}e^{-i(\theta_{\widehat{\bm{k}}}+\theta_{\widehat{\bm{k}+\bm{q}}})}.
\end{align}
In terms of Eq.~(\ref{eqn1.45}), the first term of Eq.~(\ref{eqn1.26}) is given by 
\begin{align}
\label{eqn1.49}
\frac{1}{2}\mathrm{Tr}(\bm{G}_0\bm{G}_\phi\bm{G}_0\bm{G}_\phi)=&\frac{1}{2\beta L^2}\sum_{k,q}(g^a_{0,k}g^b_{0,k+q}+g^b_{0,k}g^a_{0,k+q})[\phi^\dagger_q\phi_q\mathrm{tr}(\bm{P}_{\sigma_{\bm{k}}}\bm{\sigma}_+ \bm{P}_{\sigma_{\bm{k}+\bm{q}}}\bm{\sigma}_-)\nonumber\\
&+\phi_{-q}\phi_{q}\mathrm{tr}(\bm{P}_{\sigma_{\bm{k}}}\bm{\sigma}_- \bm{P}_{\sigma_{\bm{k}+\bm{q}}}\bm{\sigma}_-)]+\mathrm{H.c.} 
\end{align}
By an expansion of small $q\equiv(\bm{q},i\omega_m)$, we get
\begin{align}
\label{eqn1.50}
\eta_1^2=-\frac{1}{4\beta L^2}\partial_{i\omega_m}^2|_{q=0}\sum_k(g^a_{0,k}g^b_{0,k+q}+g^a_{0,k+q}g^b_{0,k})=-\frac{1}{2\beta L^2}\partial_{i\omega_m}^2|_{q=0}\sum_k g^a_{0,k-\frac{q}{2}}g^b_{0,k+\frac{q}{2}},
\end{align}
\begin{align}
\label{eqn1.51}
\eta_1^2 c^2=\frac{1}{4\beta L^2}\partial_{q_x}^2|_{q=0}\sum_k(g^a_{0,k}g^b_{0,k+q}+g^a_{0,k+q}g^b_{0,k})=\frac{1}{2\beta L^2}\partial_{q_x}^2|_{q=0}\sum_k g^a_{0,k-\frac{q}{2}}g^b_{0,k+\frac{q}{2}},
\end{align}
\begin{align}
\label{eqn1.52}
\eta_1^2 c^2\alpha=&\frac{1}{4\beta L^2}\partial_{q_x}^2|_{q=0}\sum_k(g^a_{0,k}g^b_{0,k+q}+g^a_{0,k+q}g^b_{0,k})e^{-i(\theta_{\widehat{\bm{k}}}+\theta_{\widehat{\bm{k}+\bm{q}}})}\nonumber\\
=&\frac{1}{2\beta L^2}\partial_{q_x}^2|_{q=0}\sum_k g^a_{0,k-\frac{q}{2}}g^b_{0,k+\frac{q}{2}}e^{-i(\theta_{\widehat{\bm{k}-\frac{\bm{q}}{2}}}+\theta_{\widehat{\bm{k}+\frac{\bm{q}}{2}}})},
\end{align}
where we used
\begin{align}
\label{eqn1.53}
\sum_k g^a_{0,k}g^b_{0,k+q}=\sum_k g^a_{0,k-q}g^b_{0,k}=\sum_k g^a_{0,k-\frac{q}{2}}g^b_{0,k+\frac{q}{2}}.
\end{align}
Equivalently, we can also use
\begin{align}
\label{eqn1.54}
\sum_k \frac{1}{2}(g^a_{0,k})''g^b_{0,k}=\sum_k \frac{1}{2}g^a_{0,k}(g^b_{0,k})''=\sum_k[\frac{1}{8}(g^a_{0,k})''g^b_{0,k}+\frac{1}{8}g^a_{0,k}(g^b_{0,k})''-\frac{1}{4}(g^a_{0,k})'(g^b_{0,k})'],
\end{align}
or 
\begin{align}
\label{eqn1.55}
\sum_k(g^a_{0,k})''g^b_{0,k}=\sum_kg^a_{0,k}(g^b_{0,k})''=-\sum_k(g^a_{0,k})'(g^b_{0,k})'. 
\end{align}
Here primes and double primes denote first-order and second-order derivatives with respect to one of the spacetime components of $k$. Note that 
Eqs.~(\ref{eqn1.53},\ref{eqn1.55}) are valid
given that associated integrals vanish or are sufficiently small 
in the ultraviolet regime (large $k$ region). 
Using them, we can determine $\eta_1$ and $\eta_1 c_\perp$ as follows, 
\begin{align}
\label{eqn1.56}
\eta_1^2=&\frac{1}{2\beta L^2}\sum_k (\partial_{i\omega_m}|_{q=0}g^a_{0,k+q})(\partial_{i\omega_m}|_{q=0}g^b_{0,k+q})\nonumber\\
=&\frac{1}{2\beta L^2}\sum_k\frac{1}{(i\omega_n-\xi_{\bm{k}})^2}\frac{1}{(i\omega_n+\xi_{\bm{k}})^2}=\frac{1}{2\beta L^2}\sum_k (g_{0,k}^a g_{0,k}^b)^2=\frac{4U}{3}>0,
\end{align}
\begin{align}
\label{eqn1.57}
\eta_1^2c_\perp^2=&-\frac{1}{2\beta L^2}\sum_k (\partial_{q_x}|_{q=0}g^a_{0,k+q})(\partial_{q_x}|_{q=0}g^b_{0,k+q})\nonumber\\
=&\frac{1}{2\beta L^2}\sum_k\frac{1}{(i\omega_n-\xi_{\bm{k}})^2}\frac{1}{(i\omega_n+\xi_{\bm{k}})^2}\frac{(|\bm{k}|-k_R)^2k_x^2}{m^2|\bm{k}|^2}\nonumber\\
=&\frac{1}{2}\int\frac{d^2\bm{k}}{(2\pi)^2}\frac{1}{(2\xi_{\bm{k}})^2}[\frac{\mathrm{tanh}(\frac{1}{2}\beta\xi_{\bm{k}})}{\xi_{\bm{k}}}-\frac{\beta}{1+\mathrm{cosh}(\beta\xi_{\bm{k}})}]\frac{(|\bm{k}|-k_R)^2}{2m^2}>0.
\end{align}

To determine the coefficient of the spin-coordinate coupling term ($\eta^2_1 c_\perp^2 \alpha$), We can use a similar trick as Eqs.~(\ref{eqn1.53},\ref{eqn1.55}) to simplify Eq.~(\ref{eqn1.52}),
\begin{align}
\label{eqn1.58}
\eta_1^2 c_\perp^2\alpha=-\frac{1}{2\beta L^2}\sum_k \partial_{q_x}|_{q=0}[g^a_{0,k+q}e^{-i\theta_{\widehat{\bm{k}+\bm{q}}}}]\partial_{q_x}|_{q=0}[g^b_{0,k+q}e^{-i\theta_{\widehat{\bm{k}+\bm{q}}}}],
\end{align}
where
\begin{align}
\label{eqn1.59}
\partial_{q_x}|_{q=0}e^{-i\theta_{\widehat{\bm{k}+\bm{q}}}}=-ie^{-i\theta_{\widehat{\bm{k}}}}\partial_{q_x}|_{q=0}\mathrm{arctan}\frac{k_y}{k_x+q_x}=ie^{-i\theta_{\widehat{\bm{k}}}}\frac{k_y}{k_x^2+k_y^2}=ie^{-i\theta_{\widehat{\bm{k}}}}\frac{\mathrm{sin}\theta_{\widehat{\bm{k}}}}{|\bm{k}|}.
\end{align}
Then we have
\begin{align}
\label{eqn1.60}
&\eta_1^2c^2\alpha=-\frac{1}{2\beta L^2}\sum_k[-(g^a_{0,k})^2\frac{(|\bm{k}|-k_R)\mathrm{cos}\theta_{\widehat{\bm{k}}}}{m}+ig^a_{0,k}\frac{\mathrm{sin}\theta_{\widehat{\bm{k}}}}{|\bm{k}|}][(g^b_{0,k})^2\frac{(|\bm{k}|-k_R)\mathrm{cos}\theta_{\widehat{\bm{k}}}}{m}+ig^b_{0,k}\frac{\mathrm{sin}\theta_{\widehat{\bm{k}}}}{|\bm{k}|}]e^{-2i\theta_{\widehat{\bm{k}}}}\nonumber\\
=&-\frac{1}{2\beta L^2}\sum_k[-(g^a_{0,k})^2\frac{|\bm{k}|-k_R}{2m}+g^a_{0,k}\frac{1}{2|\bm{k}|}][(g^b_{0,k})^2\frac{|\bm{k}|-k_R}{2m}+g^b_{0,k}\frac{1}{2|\bm{k}|}]\nonumber\\
=&-\frac{1}{8\beta L^2}\sum_k\{-(g^a_{0,k}g^b_{0,k})^2\frac{(|\bm{k}|-k_R)^2}{m^2}+g^a_{0,k}g^b_{0,k}\frac{1}{|\bm{k}|^2}+[g^a_{0,k}(g^b_{0,k})^2-(g^a_{0,k})^2g^b_{0,k}]\frac{|\bm{k}|-k_R}{m|\bm{k}|}\},
\end{align}
where
\begin{align}
\label{eqn1.61}
\sum_n[g^a_{0,k}(g^b_{0,k})^2-(g^a_{0,k})^2g^b_{0,k}]=&\sum_n[\frac{1}{-i\omega_n+\xi_{\bm{k}}}(\frac{1}{-i\omega_n-\xi_{\bm{k}}})^2-(\frac{1}{-i\omega_n+\xi_{\bm{k}}})^2\frac{1}{-i\omega_n-\xi_{\bm{k}}}]\nonumber\\
=&\frac{(i\omega_n+\xi_{\bm{k}})-(i\omega_n-\xi_{\bm{k}})}{(i\omega_n+\xi_{\bm{k}})^2(i\omega_n-\xi_{\bm{k}})^2}=\sum_n 2\xi_{\bm{k}}(g^a_{0,k}g^b_{0,k})^2,
\end{align}
\begin{align}
\label{eqn1.62}
\frac{|\bm{k}|-k_R}{m|\bm{k}|}[2\xi_{\bm{k}}-\frac{(|\bm{k}|-k_R)|\bm{k}|}{m}]=\frac{|\bm{k}|-k_R}{m|\bm{k}|}[2E_g-\frac{(|\bm{k}|-k_R)k_R}{m}].
\end{align}
In terms of Eqs.~(\ref{eqn1.37},\ref{eqn1.44},\ref{eqn1.61},\ref{eqn1.62}), 
we finally determine $\eta^2_1 c^2 \alpha$ as follows, 
\begin{align}
\label{eqn1.63}
\eta_1^2c_\perp^2\alpha=&\frac{1}{2}\int\frac{d^2\bm{k}}{(2\pi)^2}\{\frac{\mathrm{tanh}(\frac{1}{2}\beta\xi_{\bm{k}})}{2\xi_{\bm{k}}}\frac{1}{4|\bm{k}|^2}\nonumber\\
&+\frac{1}{(2\xi_{\bm{k}})^2}[\frac{\mathrm{tanh}(\frac{1}{2}\beta\xi_{\bm{k}})}{\xi_{\bm{k}}}-\frac{\beta}{1+\mathrm{cosh}(\beta\xi_{\bm{k}})}][\frac{(|\bm{k}|-k_R)k_R}{m}-2E_g]\frac{|\bm{k}|-k_R}{4m|\bm{k}|}\}.
\end{align}
\end{widetext} 

To evaluate $\alpha$ and $c_\perp$, note first that the omission 
of the down-spin bands in Eq.~(\ref{eqn1.4}) is justified when  
\begin{align}
\label{eqn1.64}
1\ll\beta E_g\ll\beta\frac{k_R^2}{2m}.
\end{align}
The condition also implies that when the temperature is low enough, 
electrons and holes are excited only around $k_R$. This naturally 
lets us introduce an ``ultraviolet" cutoff $k_g$ in the integral 
over $|{\bm k}|$ in 
Eqs.~(\ref{eqn1.57},\ref{eqn1.63}),
\begin{align}
\label{eqn1.65}
\eta_1^2c_\perp^2&=\int\frac{d^2\bm{k}}{(2\pi)^2}\frac{1}{16\xi_{\bm{k}}^3}\frac{(|\bm{k}|-k_R)^2}{m^2} \nonumber \\
&=\int_{k_R-k_g}^{k_R+k_g}\frac{dk}{2\pi}\frac{k}{16\xi_k^3}\frac{(k-k_R)^2}{m^2},
\end{align}
\begin{align}
\label{eqn1.66}
\eta_1^2c_\perp^2\alpha=&\int\frac{d^2\bm{k}}{(2\pi)^2}\frac{1}{16\xi_{\bm{k}}^3|\bm{k}|^2}\{\frac{k_R|\bm{k}|(|\bm{k}|-k_R)^2}{2m^2} \nonumber \\
& -\frac{E_g |\bm{k}|(|\bm{k}|-k_R)}{m}+[\frac{(|\bm{k}|-k_R)^2}{2m}+E_g]^2\}\nonumber\\
=&\int_{k_R-k_g}^{k_R+k_g}\frac{dk}{2\pi}\frac{1}{16\xi_{k}^3k}\{\frac{(k-k_R)^2(k^2+k_R^2)}{4m^2} \nonumber \\
& \hspace{2cm} -\frac{E_g k_R(k-k_R)}{m}+E_g^2\}.
\end{align}
The cutoff $k_g$ satifies that 
$k_g=\mathcal{O}(\sqrt{2mE_g}) \ll k_R$. 
Note also that without the cutoff the integral in Eq.~(\ref{eqn1.63}) has 
the logarithmic divergence at $k=0$. 
With the cutoff, We finally obtain 
\begin{align}
\label{eqn1.67}
\eta_1^2 c_\perp^2 E_g =&\int_{-k_g}^{k_g}\frac{dk}{2\pi}\frac{(k_R+k)E_g}{16}(\frac{k^2}{2m}+E_g)^{-3}\frac{k^2}{m^2} \nonumber \\
&=\int_{0}^{k_g}\frac{dk}{\pi}\frac{k_RE_g}{16}(\frac{k^2}{2m}+E_g)^{-3}\frac{k^2}{m^2}=\mathcal{O}(\frac{k_R}{k_g}),
\end{align}
\begin{align}
\label{eqn1.68}
\alpha\eta_1^2 c_\perp^2 E_g=&\int_{-k_g}^{k_g}\frac{dk}{2\pi}\frac{(k_R+k)^{-1}E_g}{16}(\frac{k^2}{2m}+E_g)^{-3}
\nonumber \\ 
& \ \  \times \{\frac{k[(k_R+k)^2+k_R^2]}{4m^2}-\frac{E_g k_R k}{m}+E_g^2\}\nonumber\\
=&\frac{1}{2}\eta_1^2 c^2_{\perp} E_g+\mathcal{O}(1)<\frac{1}{2}\eta_1^2 c^2_{\perp} E_g+|\mathcal{O}(\frac{k_R}{k_g})|.
\end{align}
A comparison between Eq.~(\ref{eqn1.67}) and Eq.~(\ref{eqn1.68}) suggests that 
$|\alpha|=\frac{1}{2}<1$ in the limit of Eq.~(\ref{eqn1.64}). $\alpha=\mathcal{O}(1)$ is due to 
the large spin-orbit-coupling limit, while $\alpha \ll \mathcal{O}(1)$
for smaller spin-orbit coupling. Nonetheless, the competition between different components of excitons will be more complicated in the smaller spin-orbit coupling case. 
$\eta_1^2c_\perp^2 E_g\gg 1$ is consistent with the physical picture of Eq.~(\ref{eqn1.64}). In the large spin-orbit coupling limit, We can also simplify Eqs.~(\ref{eqn1.37},\ref{eqn1.56}) and all the coefficients in Eq.~(\ref{eqn4}),
\begin{align}
\label{eqn1.69}
D_0=&\frac{4}{g_s(1+w)}+2U\rho_0=\int\frac{d^2\bm{k}}{(2\pi)^2}\frac{1}{2\xi_{\bm{k}}}\nonumber \\
=&\int_0^{k_g}\frac{dk}{\pi}\frac{1}{2}(\frac{k^2}{2m}+E_g)^{-1},
\end{align}
\begin{align}
\label{eqn1.70}
\eta_1^2=\frac{4U}{3}=\int\frac{d^2\bm{k}}{(2\pi)^2}\frac{1}{8\xi_{\bm{k}}^3}=\int_0^{k_g}\frac{dk}{\pi}\frac{1}{8}(\frac{k^2}{2m}+E_g)^{-3}.
\end{align}

To summarize, the U(1) theory of Eq.~(\ref{eqn4}) can be derived as an effective theory for the spin-triplet exciton condensate phase in semiconductors with Rashba spin-orbit 
interaction. Thereby, the spinless inter-band mixing 
$(\Delta_t, \Delta^*_t)$ and an attractive interaction $g_s$ between electrons in the conduction band and holes 
in the valence band induces a condensation of the 
$XY$-component of the real part of the excitonic pairing, 
$\phi \propto {\rm Re}O_x + i {\rm Re}O_y$ with
$O_{j} \equiv \langle {\bm b}^{\dagger} \sigma_j 
{\bm a}\rangle$ ($j=x,y$). Physically reasonable 
values of the coefficients in the U(1) theory 
are obtained within certain limits.

\section{\label{appendixB}Derivation and conservation of Noether's current}

In this appendix, we derive the spin ($j^s_\mu$) and orbital ($j^l_\mu$) parts of Noether's current in Eqs.~(\ref{eqn8},\ref{eqn9}), and verify that the total angular momentum is conserved. We start with the classical effective theory Eq.~(\ref{eqn5}),
\begin{align}
\label{eqn2.1}
\mathcal{L}=&\frac{1}{2}(\partial_t\theta)^2-\frac{1}{2}(\partial_x\theta)^2[1-\alpha\mathrm{cos}(2\theta)] \nonumber \\
&\hspace{-1cm} 
-\frac{1}{2}(\partial_y\theta)^2[1+\alpha\mathrm{cos}(2\theta)]+\alpha (\partial_x\theta)(\partial_y\theta)\mathrm{sin}(2\theta).
\end{align}
The theory has a U(1) spacetime symmetry,
\begin{align}
\label{eqn2.2}
&\theta\rightarrow\theta+\epsilon\Delta\theta=\theta+\epsilon,\quad x\rightarrow x+\epsilon\Delta x=x-\epsilon y,\nonumber \\
&y\rightarrow y+\epsilon\Delta y=y+\epsilon x,\quad t\rightarrow t+\epsilon\Delta t=t.
\end{align}
With the continuous symmetry Eq.~(\ref{eqn2.2}), Noether's theorem gives a conserved current,
\begin{align}
\label{eqn2.3}
j_\mu=&\frac{\partial\mathcal{L}}{\partial (\partial_\mu\theta)}\Delta\theta+[\delta_{\mu\nu}\mathcal{L}-\frac{\partial\mathcal{L}}{\partial (\partial_\mu\theta)}(\partial_\nu\theta)]\Delta x_\nu \nonumber \\
=&\frac{\partial\mathcal{L}}{\partial (\partial_\mu\theta)}\Delta\theta+T_{\mu \nu}\Delta x_\nu,
\end{align}
where $\mu,\nu\in\{t,x,y\}$, $\Delta x_\nu\in\{\Delta t,\Delta x,\Delta y\}$.
$T_{\mu\nu}\equiv \delta_{\mu\nu}\mathcal{L}-\frac{\partial\mathcal{L}}{\partial (\partial_\mu\theta)}(\partial_\nu\theta)$ is a stress-energy tensor. The conserved current obeys $\partial_\mu j_\mu=0$ as long as an EOM is satisfied,
\begin{align}
\label{eqn2.4}
\partial_\mu[\frac{\partial\mathcal{L}}{\partial(\partial_\mu\theta)}]-\frac{\partial\mathcal{L}}{\partial\theta}=0. 
\end{align}
The EOM is given by
\begin{align}
\label{eqn2.5}
&\hspace{-0.5cm} 
\partial^2_t\theta-(\partial_x^2\theta)[1-\alpha\mathrm{cos}(2\theta)]-2\alpha(\partial_x\theta)^2\mathrm{sin}(2\theta)
\nonumber \\
&\ \  -(\partial_y^2\theta)[1+\alpha\mathrm{cos}(2\theta)]+2\alpha(\partial_y\theta)^2\mathrm{sin}(2\theta)\nonumber\\
& \ \ +2\alpha(\partial_x\partial_y\theta)\mathrm{sin}(2\theta)+4\alpha(\partial_x\theta)(\partial_y\theta)\mathrm{cos}(2\theta) \nonumber \\
& \ \  +\alpha[(\partial_x\theta)^2-(\partial_y\theta)^2]\mathrm{sin}(2\theta)-2\alpha(\partial_x\theta)(\partial_y\theta)\mathrm{cos}(2\theta)\nonumber\\
= & \partial^2_t\theta-(\partial_x^2\theta)[1-\alpha\mathrm{cos}(2\theta)] \nonumber \\
& \ \  -(\partial_y^2\theta)[1+\alpha\mathrm{cos}(2\theta)]+2\alpha(\partial_x\partial_y\theta)\mathrm{sin}(2\theta)\nonumber\\
& \ \ -\alpha[(\partial_x\theta)^2-(\partial_y\theta)^2]\mathrm{sin}(2\theta)+2\alpha(\partial_x\theta)(\partial_y\theta)\mathrm{cos}(2\theta) \nonumber \\ 
= & 0.
\end{align}
The theory has spatial and temporal translational symmetries, which imposes a conservation rule on the stress-energy tensor, 
\begin{align}
\label{eqn2.6}
\partial_\mu T_{\mu\nu}=\partial_\mu [\delta_{\mu\nu}\mathcal{L}-\frac{\partial\mathcal{L}}{\partial (\partial_\mu\theta)}(\partial_\nu\theta)]=0.
\end{align}
The total angular momentum current of Eq.~(\ref{eqn2.3}) can be divided into spin angular momentum current $j_s$ that does not depend on $T_{\mu\nu}$, and orbital angular momentum current $j_l$ that depends on $T_{\mu\nu}$. 
\begin{align}
\label{eqn2.7}
j^s_\mu=\frac{\partial\mathcal{L}}{\partial (\partial_\mu\theta)}\Delta\theta,\quad j^l_\mu=[\delta_{\mu\nu}\mathcal{L}-\frac{\partial\mathcal{L}}{\partial (\partial_\mu\theta)}(\partial_\nu\theta)]\Delta x_\nu.
\end{align}
Let us focus on the spin part. Spin angular momentum density is 
\begin{align}
\label{eqn2.8}
s=j^s_t=\partial_t\theta,
\end{align}
and corresponding spin currents are
\begin{align}
\label{eqn2.9}
j^s_x=-(\partial_x\theta)[1-\alpha\mathrm{cos}(2\theta)]+\alpha(\partial_y\theta)\mathrm{sin}(2\theta),
\end{align}
\begin{align}
\label{eqn2.10}
j^s_y=-(\partial_y\theta)[1+\alpha\mathrm{cos}(2\theta)]+\alpha(\partial_x\theta)\mathrm{sin}(2\theta).
\end{align}
The spin density and current are zero 
at equilibrium, $j^s_\mu=0$. The spin angular momentum 
is not conserved,
\begin{align}
\label{eqn2.11}
\partial_\mu j^s_\mu=G=-\partial_\mu j^l_\mu.
\end{align}
Local sources of the spin angular momentum are given by 
spin torque $G$. The torque represents the mutual conversion between 
orbital and spin angular momenta. Taking the EOM Eq.~(\ref{eqn2.5}) into Eqs.~(\ref{eqn2.8}-\ref{eqn2.10}), we get the 
spin torque as follows, 
\begin{align}
\label{eqn2.12}
G=&\partial_\mu j^s_\mu=\partial_t^2\theta-(\partial_x^2\theta)[1-\alpha\mathrm{cos}(2\theta)] 
-(\partial_y^2\theta)[1+ \alpha\mathrm{cos}(2\theta)] \nonumber \\
& -2(\partial_x\theta)^2\alpha\mathrm{sin}(2\theta)+2(\partial_y\theta)^2\alpha\mathrm{sin}(2\theta)\nonumber\\
&+2\alpha(\partial_x\partial_y\theta)\mathrm{sin}(2\theta)+4\alpha(\partial_x\theta)(\partial_y\theta)\mathrm{cos}(2\theta)\nonumber\\
=&-\alpha[(\partial_x\theta)^2-(\partial_y\theta)^2]\mathrm{sin}(2\theta)+2\alpha(\partial_x\theta)(\partial_y\theta)\mathrm{cos}(2\theta).
\end{align}
The orbital-angular-momentum density and current are given by
\begin{align}
\label{eqn2.13}
l=j^l_t=y(\partial_t\theta)(\partial_x\theta)-x(\partial_t\theta)(\partial_y\theta),
\end{align}
\begin{align}
\label{eqn2.14}
j^l_x=&-y\big{\{}\frac{1}{2}(\partial_t\theta)^2-\frac{1}{2}(\partial_y\theta)^2[1+\alpha\mathrm{cos}(2\theta)] \nonumber \\
& +\frac{1}{2}(\partial_x\theta)^2[1-\alpha\mathrm{cos}(2\theta)]\big{\}}\nonumber\\
&+x\big{\{}(\partial_x\theta)(\partial_y\theta)[1-\alpha\mathrm{cos}(2\theta)]-\alpha(\partial_y\theta)^2\mathrm{sin}(2\theta)\big{\}},
\end{align}
\begin{align}
\label{eqn2.15}
j^l_y=&x\big{\{}\frac{1}{2}(\partial_t\theta)^2-\frac{1}{2}(\partial_x\theta)^2[1-\alpha\mathrm{cos}(2\theta)] \nonumber \\
& +\frac{1}{2}(\partial_y\theta)^2[1+\alpha\mathrm{cos}(2\theta)]\big{\}}\nonumber\\
&-y\big{\{}(\partial_x\theta)(\partial_y\theta)[1+\alpha\mathrm{cos}(2\theta)]-\alpha(\partial_x\theta)^2\mathrm{sin}(2\theta)\big{\}}.
\end{align}
The orbital angular momentum density and current depends explicitly 
on spatial coordinates, and they depend on a choice of the origin for the 
spatial coordinates. 
Besides, the EOM Eq.~(\ref{eqn2.5}) gives $\partial_t^2\theta$ instead of $\partial_t\theta$, 
while $j^s_t$ as well as $\partial_t j^s_{t}$ contains $\partial_t \theta$.  
Nonetheless, we can verify the continuity equation Eq.~(\ref{eqn2.11}) 
directly, using Eqs.~(\ref{eqn2.5},\ref{eqn2.13}-\ref{eqn2.15}). 
$\partial_{\mu} j^l_{\mu}$ is formally given by a term that has no explicit dependence on $x$ and $y$, and terms that depend explicitly and 
linearly on the spatial coordinates. The latter terms vanish thanks to 
Eq.~(\ref{eqn2.5}); 
\begin{align}
\label{eqn2.16}
&\frac{\partial(\partial_\mu j^l_\mu)}{\partial y}\big{|}_{x,\theta,\partial_\mu\theta}=(\partial_t^2\theta)(\partial_x\theta)+(\partial_t\theta)(\partial_x\partial_t\theta)\nonumber\\
& -(\partial_t\theta)(\partial_x\partial_t\theta)+(\partial_y\theta)(\partial_x\partial_y\theta)[1+\alpha\mathrm{cos}(2\theta)] 
\nonumber \\
& \ -\alpha(\partial_y\theta)^2(\partial_x\theta)\mathrm{sin}(2\theta) 
-(\partial_x\theta)(\partial_x^2\theta)[1-\alpha\mathrm{cos}(2\theta)] \nonumber \\
& \ -\alpha(\partial_x\theta)^2(\partial_x\theta)\mathrm{sin}(2\theta) 
-(\partial_x\theta)(\partial_y^2\theta)[1+\alpha\mathrm{cos}(2\theta)]
\nonumber \\
& \ -(\partial_x\partial_y\theta)(\partial_y\theta)[1+\alpha\mathrm{cos}(2\theta)]+2\alpha(\partial_x\theta)(\partial_y\theta)^2\mathrm{sin}(2\theta)\nonumber\\
&+2\alpha(\partial_x\theta)(\partial_x\partial_y\theta)\mathrm{sin}(2\theta)+2\alpha(\partial_x\theta)^2(\partial_y\theta)\mathrm{cos}(2\theta)\nonumber\\
& \ \ \ =(\partial_t^2\theta)(\partial_x\theta)-(\partial_x\theta)(\partial_x^2\theta+\partial_y^2\theta)\nonumber\\
&+\alpha(\partial_x\theta)\mathrm{sin}(2\theta)[(\partial_y\theta)^2-(\partial_x\theta)^2+2(\partial_x\partial_y\theta)]\nonumber\\
&+\alpha(\partial_x\theta)\mathrm{cos}(2\theta)[(\partial_x\theta)^2-(\partial_y\theta)^2+2(\partial_x\theta)(\partial_y\theta)] \nonumber \\
& \ \  \ =0,
\end{align}
\begin{align}
\label{eqn2.17}
&\frac{\partial(\partial_\mu j^l_\mu)}{\partial x}\big{|}_{y,\theta,\partial_\mu\theta}=-(\partial_t^2\theta)(\partial_y\theta)-(\partial_t\theta)(\partial_y\partial_t\theta)\nonumber\\
&+(\partial_t\theta)(\partial_y\partial_t\theta)-(\partial_x\theta)(\partial_x\partial_y\theta)[1-\alpha\mathrm{cos}(2\theta)] \nonumber \\
& - \alpha(\partial_x\theta)^2(\partial_y\theta)\mathrm{sin}(2\theta) 
+(\partial_y\theta)(\partial_y^2\theta)[1+\alpha\mathrm{cos}(2\theta)] \nonumber \\
& -\alpha(\partial_y\theta)^2(\partial_y\theta)\mathrm{sin}(2\theta) 
+(\partial_y\theta)(\partial_x^2\theta)[1-\alpha\mathrm{cos}(2\theta)] \nonumber \\
& +(\partial_x\partial_y\theta)(\partial_x\theta)[1-\alpha\mathrm{cos}(2\theta)]+2\alpha(\partial_y\theta)(\partial_x\theta)^2\mathrm{sin}(2\theta)\nonumber\\
&-2\alpha(\partial_y\theta)(\partial_x\partial_y\theta)\mathrm{sin}(2\theta)-2\alpha(\partial_y\theta)^2(\partial_x\theta)\mathrm{cos}(2\theta)\nonumber\\
& \ \ \ =-(\partial_t^2\theta)(\partial_y\theta)+(\partial_y\theta)(\partial_x^2\theta+\partial_y^2\theta)\nonumber\\
&+\alpha(\partial_y\theta)\mathrm{sin}(2\theta)[(\partial_x\theta)^2-(\partial_y\theta)^2-2(\partial_x\partial_y\theta)]\nonumber\\
&+\alpha(\partial_y\theta)\mathrm{cos}(2\theta)[(\partial_y\theta)^2-(\partial_x\theta)^2-2(\partial_x\theta)(\partial_y\theta)] \nonumber \\
& \ \ \  =0.
\end{align}
The former term is nothing but $-G$,
\begin{align}
\label{eqn2.18}
&\partial_\mu j^l_\mu-\frac{\partial(\partial_\mu j^l_\mu)}{\partial x}\big{|}_{y,\theta,\partial_\mu\theta}-\frac{\partial(\partial_\mu j^l_\mu)}{\partial y}\big{|}_{x,\theta,\partial_\mu\theta}\nonumber\\
=&(\partial_x\theta)(\partial_y\theta)[1-\alpha\mathrm{cos}(2\theta)]-\alpha(\partial_y\theta)^2\mathrm{sin}(2\theta) \nonumber \\
& -(\partial_x\theta)(\partial_y\theta)[1+\alpha\mathrm{cos}(2\theta)]+\alpha(\partial_x\theta)^2\mathrm{sin}(2\theta)\nonumber\\
=&\alpha[(\partial_x\theta)^2-(\partial_y\theta)^2]\mathrm{sin}(2\theta)-2\alpha(\partial_x\theta)(\partial_y\theta)\mathrm{cos}(2\theta).
\end{align}
Thus, the total angular momentum is indeed conserved,
\begin{align}
\label{eqn2.19}
\partial_\mu j^l_\mu 
& =\alpha[(\partial_x\theta)^2-(\partial_y\theta)^2]\mathrm{sin}(2\theta) \nonumber \\
&  \hspace{1cm} 
-2\alpha(\partial_x\theta)(\partial_y\theta)\mathrm{cos}(2\theta) 
=-G.
\end{align}

\section{\label{appendixC}Solutions for the spin-injection model}
In this appendix, we solve $\theta(x,t)$ in the spin-injection model, Eq.~(\ref{eqn10}), 
together with the boundary condition, Eq.~(\ref{eqn13}), 
and $j^s_{x}(x=0,t)=j_0$. We consider a general junction parameter $k_0$ except
$k_0=1$ (straight geometry), $k_0^2=(1+\frac{1}{j_0 r})^2$ (circular geometry), and $k_0=0$, while leaving discussions about solutions at these special parameter points for Appendix \ref{appendixF}.   
\subsection{\label{appendixC1}Straight geometry without curvature}
For the straight geometry without the curvature (Fig.~\ref{fig_a}), 
let us consider the EOM in the 1D system,  
\begin{align}
\label{eqn3.1}
\partial^2_t\theta=(\partial_x^2\theta)[1-\alpha\mathrm{cos}(2\theta)]+\alpha(\partial_x\theta)^2\mathrm{sin}(2\theta).
\end{align}
The boundary conditions are given by
\begin{align}
\label{eqn3.2}
j^s_x(0,t)=j_0,
\end{align}
\begin{align}
\label{eqn3.3}
s_c(L,t)=k_c j_{x,c}^s(L,t), 
\end{align}
with $k_c \equiv \frac{\chi}{\chi^{\prime}}\big[D_s\big(\frac{1}{T^{\prime}_1}+i c\big)\big]^{-1/2} + \frac{\chi}{\beta_t}$. Here 
$``c"$ stands for the frequency of spin density and current at $x=L$,  
and Eq.~(\ref{eqn3.3}) is imposed for each frequency component of 
the density and current. The density and current are given by 
$\theta$ (Eq.~(\ref{eqn9})), 
\begin{align}
\label{eqn3.4}
&s=\partial_t\theta,\quad j^s_x=-(\partial_x\theta)[1-\alpha\mathrm{cos}(2\theta)],\nonumber \\ &j^s_y=\alpha\partial_x\theta\mathrm{sin}(2\theta),\quad G=-\alpha(\partial_x\theta)^2\mathrm{sin}(2\theta).
\end{align}

We solve the EOM by a perturbative expansion of $\alpha$. With $\theta(x,t)=\theta_0(x,t)+\mathcal{O}(\alpha)$, the zeroth order is
\begin{align}
\label{eqn3.5}
\partial_t^2\theta_0=\partial_x^2\theta_0.
\end{align}
The general solution of Eq.~(\ref{eqn3.5}) is
\begin{align}
\label{eqn3.6}
\theta_0(x,t)&=At+Bx+F_0 \nonumber \\
& \hspace{1cm} 
+\sum_{c\in\mathbb{R},c\neq 0}[F_c e^{ic(t-x)}+F'_c e^{ic(t+x)}],
\end{align}
where $A,B,F_c,F'_c$ are constants. Eq.~(\ref{eqn3.2}) leads to
\begin{align}
\label{eqn3.7}
\theta_0(x,t)=At-j_0x+F_0+\sum_{c\in\mathbb{R},c\neq 0}2 F_c\mathrm{cos}(cx)e^{ict}.
\end{align}
Since $\mathrm{Re}(k_c)>0$, Eq.~(\ref{eqn3.3}) requires $F_c=0$ for 
$c \ne 0$, and the zeroth order takes the following form, 
\begin{align}
\label{eqn3.8}
\theta_0(x,t)&=s(x,t)t-j^s_x(x,t)x+F(0)\nonumber \\
& =k_0j_0t-j_0x+F_0.
\end{align}
Here $F_0$ can be absorbed by a time translation so we take $F_0=0$.
Note that without the spin-orbit coupling ($\alpha=0$), the spin 
density and current are static,
\begin{align}
\label{eqn3.9}
&j^s_x(0<x<L)=j_0+\mathcal{O}(\alpha),\nonumber \\
&s(0<x<L)=k_0 j_0+\mathcal{O}(\alpha).
\end{align}

Upon a substitution of Eq.~(\ref{eqn3.9}) into Eq.~(\ref{eqn3.1}) and 
an expansion of Eq.~(\ref{eqn3.1}) in $\alpha$, the first-order correction to 
the solution is given by an inhomogeneous linear differential equation. Thereby, the first-order solution has two parts, $\theta_1$ and $\theta_2$, 
\begin{align}
\label{eqn3.10}
\theta(x,t)=\theta_0(x,t)+\theta_1(x,t)+\theta_2(x,t)+\mathcal{O}(\alpha^2),
\end{align}
and $\theta_1$ is a special solution of the inhomogenous equation,
\begin{align}
\label{eqn3.11}
\partial_t^2\theta_1-\partial_x^2\theta_1=&-\alpha(\partial_x^2\theta_0)\mathrm{cos}(2\theta_0)+\alpha(\partial_x\theta_0)^2\mathrm{sin}(2\theta_0) 
\nonumber \\
=& \alpha j_0^2\mathrm{sin}(2k_0j_0t-2j_0x).
\end{align}
$\theta_2(x,t)$ is a solution of the homogeneous differential equation,
\begin{align}
\label{eqn3.13}
\partial_t^2\theta_2-\partial_x^2\theta_2=0.
\end{align}
With $\theta_1(x,t)$ and $\theta_2(x,t)$, the spin density and current 
should satisfy the BCs up to the first order in $\alpha$. 

Thanks to the linear $x$ and $t$-dependence of $\theta_0(x,t)$ and $k_0\ne 1$, we can find a special solution,
\begin{align}
\label{eqn3.14}
\theta_1(x,t)=-\frac{\alpha}{4(k_0^2-1)}\mathrm{sin}(2k_0j_0t-2j_0x).
\end{align}
$\theta_2(x,t)$ takes the same form as Eq.~(\ref{eqn3.6}). 
With these solutions, the spin density and current are given by 
the following up to the first order in $\alpha$, 
\begin{align}
\label{eqn3.15}
s=&k_0 j_0-\frac{k_0j_0\alpha}{2(k_0^2-1)}\mathrm{cos}(2k_0j_0t-2j_0x) \nonumber \\
&\ \ +\partial_t\theta_2+\mathcal{O}(\alpha^2),
\end{align}
\begin{align}
\label{eqn3.16}
j^s_x=&j_0-\alpha j_0\mathrm{cos}(2k_0j_0t-2j_0x)\nonumber \\
& \ -\frac{j_0\alpha}{2(k_0^2-1)}\mathrm{cos}(2k_0j_0t-2j_0x) - \partial_x\theta_2+\mathcal{O}(\alpha^2)\nonumber\\
=&j_0-\frac{j_0\alpha(2k_0^2-1)}{2(k_0^2-1)}\mathrm{cos}(2k_0j_0t-2j_0x) - \partial_x\theta_2+\mathcal{O}(\alpha^2).
\end{align}
In order that Eqs.~(\ref{eqn3.15},\ref{eqn3.16}) satisfy the BCs, 
$\theta_2(x,t)$ must have the same frequency   
as $\theta_1(x,t)$,
\begin{align}
\label{eqn3.17}
\theta_2(x,t)=\alpha ge^{2ik_0j_0(t-x)}+\alpha g'e^{2ik_0j_0(t+x)}+\mathrm{c.c.}.
\end{align}
Here $g$ and $g'$ are complex constants. By the same reasoning as in the text below Eq.~(\ref{eqn3.7}), other frequency components in $\theta_2(x,t)$ vanish. This leads to 
\begin{align}
\label{eqn3.18}
s=&\frac{1}{2}k_0j_0+\alpha k_0j_0e^{2ik_0j_0t}[2ige^{-2ik_0j_0x}-\frac{e^{-2ij_0x}}{4(k_0^2-1)}]\nonumber\\
&+2i\alpha k_0j_0 g' e^{2ik_0j_0(t+x)}+\mathcal{O}(\alpha^2)+\mathrm{c.c.},
\end{align}
\begin{align}
\label{eqn3.19}
j_x^s=&\frac{1}{2}j_0+\alpha j_0 e^{2ik_0j_0t}[2ik_0ge^{-2ik_0j_0x}-\frac{(2k_0^2-1)e^{-2ij_0x}}{4(k_0^2-1)}]\nonumber\\
&-2i\alpha k_0j_0g'e^{2ik_0j_0(t+x)}+\mathcal{O}(\alpha^2)+\mathrm{c.c.}.
\end{align}
The boundary conditions Eqs.~(\ref{eqn3.2},\ref{eqn3.3}) require
\begin{align}
\label{eqn3.20}
\alpha j_0[2ik_0g-\frac{2k_0^2-1}{4(k_0^2-1)}]-2i\alpha k_0j_0g'=0,
\end{align}
\begin{align}
\label{eqn3.21}
&\alpha j_0[2ik_0ge^{-2ik_0j_0L}-\frac{e^{-2ij_0L}}{4(k_0^2-1)}] +2i\alpha j_0k_0g'e^{2ik_0j_0L} \nonumber  \\
& =k\alpha j_0[2ik_0ge^{-2ik_0j_0L}-\frac{(2k_0^2-1)e^{-2ij_0L}}{4(k_0^2-1)}] \nonumber \\
&\hspace{2.5cm} -2ik\alpha k_0j_0g'e^{2ik_0j_0L},
\end{align}
with 
\begin{align}
\label{eqn3.22}
k_0=\frac{\chi}{\chi'}(\frac{D_s}{T'_1})^{-\frac{1}{2}}+\frac{\chi}{\beta_t},
\end{align}
\begin{align}
\label{eqn3.23}
k\equiv k_{2k_0j_0}=\frac{\chi}{\chi'}\big{[}D_s(\frac{1}{T'_1}+2ik_0j_0)\big{]}^{-\frac{1}{2}}+\frac{\chi}{\beta_t},
\end{align}
and $k_{-c}=(k_c)^*$ for a real number $c$. 
Eqs.~(\ref{eqn3.20},\ref{eqn3.21}) can be simplified, 
\begin{align}
\label{eqn3.24}
2ik_0(g-g')=\frac{2k_0^2-1}{4(k_0^2-1)},
\end{align}
\begin{align}
\label{eqn3.25}
&2ik_0(ge^{-2ik_0j_0L}+g'e^{2ik_0j_0L})-2ikk_0(ge^{-2ik_0j_0L} \nonumber \\
& \hspace{1cm} -g'e^{2ik_0j_0L})=\frac{1-(2k_0^2-1)k}{4(k_0^2-1)}e^{-2ij_0L}.
\end{align}
The two equations Eqs.~(\ref{eqn3.24},\ref{eqn3.25}) determine 
the two coefficients, $g$ and $g^{\prime}$,
\begin{align}
\label{eqn3.26}
&\left(\begin{array}{cc}1&-1\\ (1-k)e^{-ik_0\beta_L} & (1+k)e^{ik_0\beta_L} \end{array}\right)\left(\begin{array}{c}g \\ g'\end{array}\right)\nonumber\\
 & \ =\frac{1}{8ik_0(k_0^2-1)}\left(\begin{array}{c}2k_0^2-1 \\ (1-(2k_0^2-1)k)e^{-i\beta_L} \end{array}\right),
\end{align}
where $\beta_L\equiv 2j_0L$. The solution of the equations is
\begin{align}
\label{eqn3.27}
g=& \frac{(2k_0^2-1)(1+k)e^{ik_0\beta_L}+[1-(2k_0^2-1)k]e^{-i\beta_L}}{(1+k)e^{ik_0\beta_L}+(1-k)e^{-ik_0\beta_L}}\nonumber \\
& \ \ \times \frac{1}{8ik_0(k_0^2-1)},
\end{align}
\begin{align}
\label{eqn3.28}
g'=&\frac{[1-(2k_0^2-1)k]e^{-i\beta_L}-(2k_0^2-1)(1-k)e^{-ik_0\beta_L}}{(1+k)e^{ik_0\beta_L}+(1-k)e^{-ik_0\beta_L}}\nonumber \\
& \ \ \ \times \frac{1}{8ik_0(k_0^2-1)}.
\end{align}
From Eqs.~(\ref{eqn3.8},\ref{eqn3.10},\ref{eqn3.14},\ref{eqn3.17}), we obtain 
\begin{align}
\label{eqn3.30}
&\theta(x,t)=j_0(k_0t-x)-\frac{\alpha}{4(k_0^2-1)}\mathrm{sin}[2j_0(k_0t-x)]\nonumber\\
&+2\alpha\mathrm{Re}(g+g')\mathrm{cos}(2k_0j_0t)\mathrm{cos}(2k_0j_0x) \nonumber \\
& \ \ +2\alpha\mathrm{Re}(g-g')\mathrm{sin}(2k_0j_0t)\mathrm{sin}(2k_0j_0x)\nonumber\\
& \ \ \ -2\alpha\mathrm{Im}(g+g')\mathrm{sin}(2k_0j_0t)\mathrm{cos}(2k_0j_0x) \nonumber \\
&\ \ \ \ +2\alpha\mathrm{Im}(g-g')\mathrm{cos}(2k_0j_0t)\mathrm{sin}(2k_0j_0x)+\mathcal{O}(\alpha^2).
\end{align}
The solution has one frequency ($2k_0j_0$) in time and two 
wavenumbers ($2j_0$, $2k_0j_0$) in space. According to Eqs.~(\ref{eqn3.24},\ref{eqn3.27},\ref{eqn3.28}), Eq.~(\ref{eqn3.30}) is nothing but Eq.~(\ref{eqn19}) with 
\begin{widetext}
\begin{align}
\label{eqn3.32}
\eta\equiv 2i(g+g') 
=\frac{(2k_0^2-1)[(1+k)e^{ik_0\beta_L}-(1-k)e^{-ik_0\beta_L}]+2[1-(2k_0^2-1)k]e^{-i\beta_L}}{4k_0(k_0^2-1)[(1+k)e^{ik_0\beta_L}+(1-k)e^{-ik_0\beta_L}]}.
\end{align}
\end{widetext}

Higher-order solutions can be obtained by the same 
perturbative iteration method. In the solution, the spin density 
and current have the same periodicity in time 
as the first-order solution, $\pi(k_0 j_0)^{-1}$.  
This is because the inhomogeneous terms at every order keep the same 
discrete time translational symmetry as that for the first order. 
For irrational $k_0$, 
the solution is not periodic in the space coordinate $x$ because of 
the superpositions of the two wavenumbers. Nonetheless, the Fourier-transform in the space has two major peaks at $2j_0$ and $2k_0j_0$. 

\subsection{\label{appendixC2}Circular geometry with curvature}
For the 1D circular geometry with a finite radius $r$ of the curvature (insets of Figs.~\ref{fig_b},\ref{fig_c}), the Lagrangian is generalized as follows 
\begin{align}
\label{eqn3.34}
\mathcal{L}=&\frac{1}{2}(\partial_t\theta)^2-\frac{1}{2r^2}(\partial_\vartheta\theta)^2+\frac{\alpha}{2r^2}(\partial_\vartheta\theta)^2 \nonumber \\ 
& \hspace{-0.5cm}  \times [\mathrm{cos}(2\theta)(\mathrm{sin}^2\vartheta-\mathrm{cos}^2\vartheta)-2\mathrm{sin}(2\theta)\mathrm{sin}\theta\mathrm{cos}\theta]\nonumber\\
=&\frac{1}{2}(\partial_t\theta)^2-\frac{1}{2}(\partial_\ell\theta)^2[1+\alpha\mathrm{cos}(2\theta-\frac{2}{r}\ell)],
\end{align}
with a 1D coordinate $\ell\equiv r\vartheta$, and 
\begin{align}
\label{eqn3.33}
&\partial_r\theta(x,y)\equiv\partial_r(r\mathrm{cos}\vartheta,r\mathrm{sin}\vartheta)=0, \nonumber \\
&\partial_x=-\frac{1}{r}(\mathrm{sin}\vartheta)\partial_\vartheta,\quad \partial_y=\frac{1}{r}(\mathrm{cos}\vartheta)\partial_\vartheta.
\end{align}
The Lagrangian gives the classical EOM in the 1D system,  
\begin{align}
\label{eqn3.35}
&\partial^2_t\theta-(\partial^2_\ell\theta)[1+\alpha\mathrm{cos}(2\theta-\frac{2}{r}\ell)]+2\alpha(\partial_\ell\theta) \nonumber \\
& \hspace{0.5cm} 
\times \mathrm{sin}(2\theta-\frac{2}{r}\ell)[(\partial_\ell\theta)-\frac{1}{r}]-\alpha(\partial_\ell\theta)^2  \mathrm{sin}(2\theta-\frac{2}{r}\ell)\nonumber\\
& \ \ \ =\partial^2_t\theta-(\partial^2_\ell\theta)[1+\alpha\mathrm{cos}(2\theta-\frac{2}{r}\ell)] \nonumber \\
& \hspace{0.8cm} +\alpha(\partial_\ell\theta)[(\partial_\ell\theta)-\frac{2}{r}]\mathrm{sin}(2\theta-\frac{2}{r}\ell)=0,  
\end{align}
and the spin density and current,  
\begin{align}
\label{eqn3.36}
s=\partial_t\theta,\quad j^s_\ell=-(\partial_\ell\theta)[1+\alpha\mathrm{cos}(2\theta-\frac{2}{r}\ell)].
\end{align}
The boundary conditions are imposed 
on the spin density and current, 
\begin{align}
\label{eqn3.41}
j^s_\ell(0,t)=j_0,\quad s_c(L,t)=k_c j_{\ell,c}^s(L,t). 
\end{align}
The boundary condition at $\ell=L$ is imposed on every frequency ($c$) 
component of the density and current, and $k_c= \frac{\chi}{\chi^{\prime}}\big[D_s\big(\frac{1}{T^{\prime}_1}+i c\big)\big]^{-1/2} + \frac{\chi}{\beta_t}$. 

The zeroth-order solution of the EOM that satisfies 
the BCs is given by 
\begin{align}
\theta_0(\ell,t)=k_0j_0t-j_0\ell. 
\end{align}
In the 
perturbative iteration method, the first-order solution comprises 
of $\theta_1(\ell,t)$ and $\theta_2(\ell,t)$. 
$\theta_{1}(\ell,t)$ is a special solution of the 
inhomogeneous linear differential equation, Eq.~(\ref{eqn22}), while $\theta_2(\ell,t)$ is a 
solution of the homogeneous linear differential equation. 
For $k^2_0 \ne (1+\frac{1}{j_0r})^2$, we find the special solution, 
\begin{align}
\label{eqn3.37}
\theta_1(\ell,t)=\frac{\alpha(1+\frac{2}{j_0r})}{4[k_0^2-(1+\frac{1}{j_0r})^2]}\mathrm{sin}[2k_0j_0t-2(j_0+\frac{1}{r})\ell],
\end{align} 
together with 
\begin{align}
\theta_2(\ell,t) = &A t + B \ell + F_0 
\nonumber \\
& \ \ + \sum_{c \in \mathbb{R},c\ne 0} 
\big[F_c e^{ic(t-\ell)} + F^{\prime}_c e^{ic(t+\ell)}\big]. 
\end{align}

A substitution of $\theta=\theta_0+\theta_1+\theta_2+{\cal O}(\alpha^2)$ into Eq.~(\ref{eqn3.36}) leads to 
\begin{align}
\label{eqn3.38}
s=& k_0 j_0+\frac{k_0j_0\alpha(1+\frac{2}{j_0 r})}{2[k_0^2-(1+\frac{1}{j_0 r})^2]}\mathrm{cos}[2k_0j_0t-2(j_0+\frac{1}{r})\ell] \nonumber \\
& +\partial_t\theta_2+\mathcal{O}(\alpha^2),
\end{align}
\begin{align}
\label{eqn3.39}
j^s_\ell=&j_0+\frac{\alpha j_0}{2}\frac{2k_0^2-2(1+\frac{1}{j_0r})^2+(1+\frac{1}{j_0r})(1+\frac{2}{j_0r})}{k_0^2-(1+\frac{1}{j_0r})^2} \nonumber \\
& \times 
\mathrm{cos}[2k_0j_0t-2(j_0+\frac{1}{r})\ell]-\partial_x\theta_2+\mathcal{O}(\alpha^2)\nonumber\\
=&j_0+\frac{j_0\alpha[2k_0^2-(1+\frac{1}{j_0r})]}{2[k_0^2-(1+\frac{1}{j_0r})^2]}\mathrm{cos}[2k_0j_0t-2(j_0+\frac{1}{r})\ell] \nonumber \\
& -\partial_x\theta_2+\mathcal{O}(\alpha^2).
\end{align} 
In order that Eqs.~(\ref{eqn3.38},\ref{eqn3.39}) satisfy the BCs, 
$\theta_2(\ell,t)$ must have the same frequency as $\theta_1(\ell,t)$;   
\begin{align}
\label{eqn3.40}
\theta_2(\ell,t)=\alpha ge^{2ik_0j_0(t-\ell)}+\alpha g'e^{2ik_0j_0(t+\ell)}+\mathrm{c.c.}.
\end{align}
The complex constants, $g$ and $g'$, are determined by 
\begin{widetext}
\begin{align}
\label{eqn3.42}
&\left(\begin{array}{cc}1&-1\\ (1-k)e^{-ik_0\beta_L} & (1+k)e^{ik_0\beta_L} \end{array}\right)\left(\begin{array}{c}g \\ g'\end{array}\right) 
=\frac{1}{8ik_0[k_0^2-(1+\frac{1}{j_0r})^2]}\left(\begin{array}{c}2k_0^2-(1+\frac{1}{j_0r}) \\ \{1+\frac{2}{j_0 r}-[2k_0^2-(1+\frac{1}{j_0r})]k\}e^{-i(1+\frac{1}{j_0 r})\beta_L} \end{array}\right), 
\end{align}
\end{widetext}
where $k$ is given by Eq.~(\ref{eqn3.23}). 

$\theta_1(\ell,t)$ vanishes when 
$j_0=-\frac{2}{r}$.
Even when $\theta_1(\ell,t)=0$, 
$\theta_2(\ell,t)\neq 0$ in general. 
The non-zero $\theta_2(\ell,t)$ comes from 
an $\mathcal{O}(\alpha)$ 
contribution of $j^s_\ell$ in Eq.~(\ref{eqn3.36}). 
For $j_0=-\frac{2}{r}$ and $k_0=\frac{1}{2}$, both  $\theta_1(\ell,t)$ and $\theta_2(\ell,t)$ reduce to zero, 
and $\theta_0(\ell,t)$ becomes an ``exact" solution satisfying the BCs. However, the exactness is not protected by the symmetry of the theory, and there will be a finite $\theta_1(\ell,t)$ when higher-order expansion terms are considered in Eq.~(\ref{eqn5}).

Besides, Eqs.~(\ref{eqn3.26},\ref{eqn3.42}) always have unique solutions 
for $g$ and $g^{\prime}$, because 
\begin{align}
\label{eqn3.43}
(1+k)e^{ik_0\beta_L}+(1-k)e^{-ik_0\beta_L}=0
\end{align}
or 
\begin{align}
\label{eqn3.44}
k=\frac{e^{-ik_0\beta_L}+e^{ik_0\beta_L}}{e^{-ik_0\beta_L}-e^{ik_0\beta_L}}=\frac{i}{\mathrm{tan}(2k_0\beta_L)}
\end{align}
contradicts with 
$\mathrm{Re}(k)>0$. The solutions of Eq.~(\ref{eqn3.26}) 
and Eq.~(\ref{eqn3.42}) are divergent at 
$k_0=1$ (straight) and $k_0^2=(1+\frac{1}{j_0r})^2$ (circular), respectively. 
Physically, the divergence could be avoided by 
finite dissipation time $T_1$. 
A more detailed discussion on the divergence is given in Appendix \ref{appendixF}.

\begin{figure*}[t]
\centering
\subfigure[ ]{
\label{SM_fig_a}
\begin{minipage}{0.45\linewidth}
\centering
\includegraphics[width=1\linewidth]{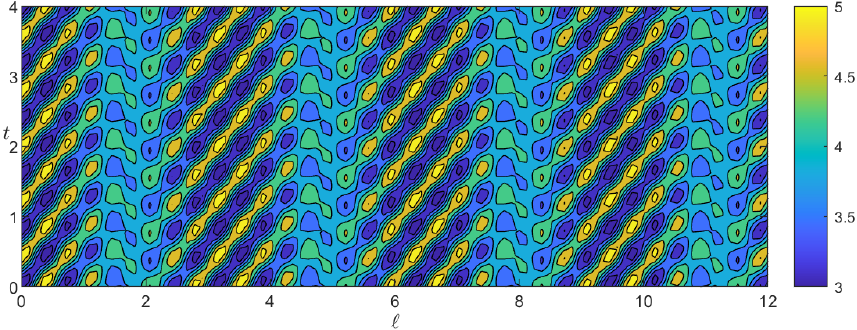}
\end{minipage}
}
\subfigure[ ]{
\label{SM_fig_b}
\begin{minipage}{0.45\linewidth}
\centering
\includegraphics[width=1\linewidth]{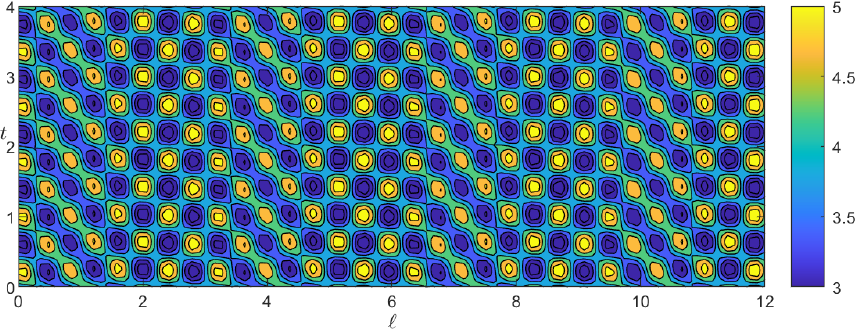}
\end{minipage}
} 
\caption{\label{SM_fig} Contour plots of $\theta(\ell,t)$ in a larger spatial range of $\ell$ with the same parameters in insets of Figs.~\ref{fig_b},\ref{fig_c}. Fig.~(a) and Fig.~(b) above correspond to Fig.~\ref{fig_b} and Fig.~\ref{fig_c}, respectively.}
\end{figure*}

The radius ($r$) dependence of $\theta(\ell,t)$ leads to the non-reciprocity of the hydrodynamic spin transport. We show the non-reciprocity in insets of Figs.~\ref{fig_b},\ref{fig_c}. The non-reciprocity is essentially from $\theta_1(\ell,t)$, as the spatial wavelength of $\theta_1(\ell,t)$ depends on the radius $r$; it also comes from $\theta_2(\ell,t)$ as $\theta_2(\ell,t)$ is different for two opposite currents to satisfy the boundary conditions. From the figures, we can see that $\theta(\ell,t)$ is periodic along $t$ because $\theta_1(\ell,t)$ and $\theta_2(\ell,t)$ share the same temporal frequency; the structure of $\theta(\ell,t)$ along $\ell$ is more complicated because $\theta_1(\ell,t)$ and $\theta_2(\ell,t)$ give two different spatial wavenumbers. To show the spatial structure of $\theta(\ell,t)$ more clearly, we plot them in a larger range of $\ell$ [see Fig.~\ref{SM_fig}], although physically $\ell$ should not be greater than $L$, and $L$ should not be greater than $2\pi r$.

\section{\label{appendixD}Possibility of dissipation}
In this appendix, we study the stability of the superfluid 
state with a finite supercurrent in the 
U(1) spacetime theory. The Lagrangian  
Eq.~(\ref{eqn5}) leads to a classical 
EOM, Eq.~(\ref{eqn2.5}). 
A solution of the EOM was obtained under boundary conditions of a finite current (Sec.~\ref{sec3}). 
The solution characterizes the supercurrent state. 
To study the stability of the supercurrent state, 
we compare the classical energy of the solution $\theta(x,y,t)$ 
with the energy of other solutions of the EOM 
with different BCs, say $\theta(x,y,t)+\delta\theta(x,y,t)$. 
Here, we consider that $\delta \theta$ is a deformation induced by spatially local perturbations. Thus, the spacetime derivatives of $\delta \theta(x,y,t)$ do not contain any uniform component in spacetime. 
The classical energy is evaluated by a Hamiltonian that 
corresponds to the Lagrangian Eq.~(\ref{eqn5}), 
\begin{align}
H[\theta] = &\int d^2{\bm r} 
\Big\{ \frac{1}{2}(\partial_t \theta)^2 + \frac{1}{2} (\partial_x \theta)^2 
\big[1-\alpha \cos(2\theta) \big] 
\nonumber \\
& + \frac{1}{2} (\partial_y \theta)^2 
\big[1+\alpha \cos(2\theta) \big] - \alpha (\partial_x \theta) 
(\partial_y \theta) \sin(2\theta) \Big\}. 
\end{align}
The solution for the supercurrent state with 
broken U(1) spacetime symmetry depends on time, e.g., Eq.~(\ref{eqn19}), while the Hamiltonian of $\theta$ and $\theta+\delta \theta$ are conserved, i.e. time-independent. Thus, for clarity of calculation, we compare the ``time averages" of the classical energies 
over a large period of time $T$, 
\begin{align}
\label{eqn4.1a}
\Delta J = \lim_{T \rightarrow \infty} \frac{1}{T} 
\Big(\int^T_{0} H[\theta+\delta\theta] dt - \int^T_{0} H[\theta] dt \Big).  
\end{align}
When the classical energy of $\theta$ is lower than $\theta+\delta \theta$ for arbitrary small $\delta \theta$, the 
supercurrent state of $\theta(x,t)$ is stable 
against the local perturbation. If it is not for some $\delta \theta$, the supercurrent state is no longer stable, and it must experience energy dissipation.  
Effects of the energy dissipation can be included as finite relaxation time into the 
classical EOM [see Appendix \ref{appendixE}]. 
To demonstrate the validity of our method used in this appendix, 
we also apply the same method to a conventional superfluid moving at a finite velocity and derive its Landau criterion 
[see Appendix \ref{appendixG}]. 

As explained above, $\delta\theta$ is a deformation induced by 
the local perturbations and the spacetime derivatives of $\delta\theta$ are considered to be always zero on average.
The locality of $\delta \theta$ is crucial in the following argument. 
For example, a finite average of the space derivative of 
$\delta\theta$ changes a uniform current, and 
such $\delta \theta$ should be excluded from the local deformation. This is because even for the conventional superfluid, the classical energy with a smaller velocity of the supercurrent will always decrease. In the derivation of the Landau criterion in Appendix \ref{appendixG}, only single excitations with $(k,\omega)$ are considered;  perturbation that lowers the average velocity is excluded implicitly.  

In this appendix, we apply the stability analysis to the 
total-angular-momentum superfluid 
in the spin-injection model. We consider a 
general value of the junction parameter $k_0$, 
except for $k_0=1$ (straight geometry),  
$k^2_0 = (1+\frac{1}{j_0r})^2$ (circular geometry), and $k_0=0$, while leaving discussions about some of 
these points for Appendix \ref{appendixF}. In the following, 
let us study the straight geometry case.  


In the 1D spin injection model, $\theta$ depends only on $x$ and $t$; $\theta(x,t)$ and $\theta(x,t)+\delta\theta(x,t)$. The energy difference between $\theta$ and $\theta+\delta \theta$ will be evaluated 
order by order in powers of the SOC ($\alpha$). To this end, we expand $\theta$ and $\delta\theta$ in powers of $\alpha$, 
\begin{align}
\label{eqn4.34}
\theta=\theta_0(x,t)+\theta'_1(x,t)+\mathcal{O}(\alpha^2),
\end{align}
\begin{align}
\label{eqn4.35}
\delta\theta(x,t)=\delta\theta_0(x,t)+\delta\theta_1(x,t)+\mathcal{O}(\alpha^2),
\end{align}
where
\begin{align}
\label{eqn4.36}
\theta'_1(x,t)=\theta_1(x,t)+\theta_2(x,t),
\end{align}
$\delta\theta_1(x,t)=\mathcal{O}(\alpha)$. 
$\delta\theta_0(x,t)$, $\theta_0(x,t)$, and  $\theta_2(x,t)$ are solutions of Eq.~(\ref{eqn3.5}). Since the spacetime derivatives of $\delta\theta(x,t)$ is not uniform, $\delta\theta_0(x,t)$, as well as $\theta_2(x,t)$,  is given by linear superpositions of $e^{iq(x-t)}$ and $e^{iq(x+t)}$ over $q$, e.g.,  
\begin{align}
\label{eqn4.37}
\delta\theta_0(x,t)=\frac{1}{\sqrt{L}}\sum_q [\delta d_qe^{iq(x-t)}+\delta d_q'e^{iq(x+t)}],
\end{align}
with the system length $L$. In Appendix \ref{appendixH}, we give a perturbation theory in the SOC ($\alpha$) that determines the higher order 
corrections [e.g., $\delta \theta_1$) for a given $\delta \theta_0$ in the form of Eq.~(\ref{eqn4.37}].

Given the $\alpha$-expansions of $\theta$ and $\theta+\delta \theta$, we now evaluate their energy difference order by order in the power of $\alpha$. We expand $\Delta J$ in Eq.~(\ref{eqn4.1a}),
\begin{align}
\label{eqn4.38}
\Delta J=\delta J+\frac{1}{2}\delta^2 J+\mathcal{O}((\delta\theta)^2).
\end{align}
Here $\delta J$ and $\delta^2 J$ are at the first and second order in $\delta \theta$, respectively. The first-order variation is, i.e. Eq.~(\ref{eqn23})
\begin{align}
\label{eqn4.39}
\delta J=&\frac{1}{T}\int^{T}_{0} dt \int dx\{(\partial_t\theta)(\partial_t\delta\theta)+(\partial_x\theta)(\partial_x\delta\theta)
\nonumber \\
& \times 
[1 - \alpha\mathrm{cos}(2\theta)]+\alpha
(\partial_x\theta)^2\mathrm{sin}(2\theta)(\delta\theta)\}\nonumber\\
=&\frac{1}{T}\int^T_0 dt \int dx[(\partial_t\theta_0)(\partial_t\delta\theta_0)+(\partial_t\theta'_1)(\partial_t\delta\theta_0) \nonumber \\
& \ \ +(\partial_t\theta_0)(\partial_t\delta\theta_1) 
+(\partial_x\theta_0)(\partial_x\delta\theta_0) 
+(\partial_x\theta'_1)(\partial_x\delta\theta_0) \nonumber \\
& \ \ \ \ +(\partial_x\theta_0)(\partial_x\delta\theta_1) 
-\alpha(\partial_x\theta_0)(\partial_x\delta\theta_0)\mathrm{cos}(2\theta_0) \nonumber \\
& \ \ \ \ \ \ +\alpha
(\partial_x\theta_0)^2
\mathrm{sin}(2\theta_0)(\delta\theta_0)]+\mathcal{O}(\alpha^2)\nonumber\\
=& \frac{1}{T}\int^T_0 dt  \int dx 
\{(\partial_{t}\theta^{\prime}_1) (\partial_t \delta \theta_0) 
+ (\partial_{x}\theta^{\prime}_1) (\partial_x \delta \theta_0) \nonumber \\
& \ + \alpha[(\partial^2_x \theta_0) \cos(2\theta_0) - (\partial_x \theta_0)^2 
\sin(2\theta_0) ] \delta\theta_0 \} 
+ \mathcal{O}(\alpha^2) \nonumber \\
=& \frac{1}{T}\int^T_0 dt \int dx [(\partial_{t}\theta^{\prime}_1) (\partial_t \delta \theta_0) 
+ (\partial_{x}\theta^{\prime}_1) (\partial_x \delta \theta_0)  \nonumber \\
& \ \ + (-\partial^2_t \theta_1 + \partial^2_x \theta_1) \delta\theta_0 ] + \mathcal{O}(\alpha^2)
\nonumber \\
=&\frac{2}{T}\int^T_0 dt \int dx(\partial_x 
\theta_2)(\partial_x\delta\theta_0)+\mathcal{O}(\alpha^2), 
\end{align}
where we neglect boundary contributions in the 
right-hand side, e.g.,  
\begin{align}
\frac{1}{T}\int^T_{0} dt (\partial_t \theta_0) (\partial_t \delta \theta_1) 
&= \frac{k_0j_0}{T} \int^T_{0}  dt (\partial_t \delta \theta_1) 
= {\cal O}(T^{-1}), \nonumber \\
 \int dx (\partial_x \theta_0) (\partial_x \delta \theta_1) 
&= - j_0 \int dx (\partial_x \delta \theta_1)  \nonumber \\
& = {\cal O}(1) \ll {\cal O}(L).
\end{align}
From the 3rd line to the 4th line, we use Eq.~(\ref{eqn3.11}). 
From the 4th line to the last line, we neglect terms that contain 
$\theta_1$ and $\delta \theta_0$, because for $k_0 \ne 1$, $\theta_1$ and $\delta \theta_0$ have different velocities 
(ratios between the frequency and wavenumber), and their product 
must vanish under the spacetime integral. 
The second-order variation is 
\begin{align}
\label{eqn4.40}
& \frac{1}{2}\delta^2 J \nonumber \\ 
& =\frac{1}{T}\int dtdx\{\frac{1}{2}(\partial_t\delta\theta)^2+\frac{1}{2}(\partial_x\delta\theta)^2[1-\alpha\mathrm{cos}(2\theta)] \nonumber \\
& \ \  -\alpha(\delta\theta)^2[(\partial_x\theta)^2\mathrm{cos}(2\theta)+(\partial^2_x\theta)\mathrm{sin}(2\theta)]\}\nonumber\\
 & =\frac{1}{T} \int dtdx\{\frac{1}{2}(\partial_t\delta\theta)^2+\frac{1}{2}(\partial_x\delta\theta)^2-\frac{\alpha}{2}(\partial_x\delta\theta_0)^2\mathrm{cos}(2\theta_0)\nonumber\\
&  \ \ -\alpha(\delta\theta_0)^2[(\partial_x\theta_0)^2\mathrm{cos}(2\theta_0)+(\partial^2_x\theta_0)\mathrm{sin}(2\theta_0)]\}+\mathcal{O}(\alpha^2)\nonumber\\
& =\frac{1}{T} \int dtdx\{\frac{1}{2}(\partial_t\delta\theta)^2+\frac{1}{2}(\partial_x\delta\theta)^2 -\frac{\alpha}{2}(\partial_x\delta\theta_0)^2\mathrm{cos}(2\theta_0) \nonumber \\
&\ \ -\alpha(\delta\theta_0)^2(\partial_x\theta_0)^2\mathrm{cos}(2\theta_0)\}+\mathcal{O}(\alpha^2).
\end{align}
Note that $\frac{1}{2}\delta^2 J\geq 0$ 
at $\mathcal{O}(\alpha)$. This is because the leading-order term (${\cal O}(1)$-term) is positive semi-definite, and negative contributions come from ${\cal O}(\alpha)$ terms. Besides, under the spacetime integral, the ${\cal O}(\alpha)$ terms can be nonzero only if $\delta \theta_0$ in Eq.~(\ref{eqn4.37}) comprises of (more than) two Fourier components, $q_1$, $q_2$, $\cdots$, and an oscillation 
function from $(\delta\theta_0)^2$ and that from $\cos(2\theta_0) 
= \cos(2j_0 k_0 t - 2j_0 x)$ cancel each other, e.g.,  
\begin{align}
\label{eqn4.41}
q_1-q_2=\pm j_0k_0,\quad q_1+q_2=\pm j_0.
\end{align}
In the presence of such components in $\delta \theta_0$, however, 
the leading-order term is positive definite.

$\delta \theta_0$, as well as $\theta_2$, is a solution of 
Eq.~(\ref{eqn3.5}); both are given by linear superpositions 
of $e^{iq(t+x)}$ and $e^{iq(t-x)}$ over $q$. Thus, for 
given $\theta_2 \ne 0$, one can always choose $\delta \theta_0$ 
such that the spacetime integral of 
$(\partial_x \theta_2) (\partial_x \delta \theta_0)$ remains non-zero and negative, $\delta J<0$.  
This suggests that the superflow state is stable only when $\theta_2(x,t)=0$, while it is not stable 
when $\theta_2(x,t)\neq 0$. 

The same conclusion 
holds true in the spin-injection model with the circular geometry.
In the 1D spin-injection model with finite curvature, 
the Hamiltonian is given by 
\begin{align}
\label{eqn4.45}
\mathcal{H}=\int d\ell \bigg\{ 
\frac{1}{2}(\partial_t\theta)^2+\frac{1}{2}(\partial_\ell\theta)^2[1+\alpha\mathrm{cos}(2\theta-\frac{2}{r}\ell)] 
\bigg\}, 
\end{align}
where $\theta$ and $\theta+\delta \theta$ depend only on 
$\ell$ and $t$. Their energy difference $\Delta J$ can be expanded in the powers of small local deformation $\delta \theta$. The first- and second-order variations of the energy in $\delta \theta$ are 
\begin{align}
\label{eqn4.46}
\delta J=&\frac{1}{T}\int^T_0 dt \int d\ell\bigg\{(\partial_t\theta)(\partial_t\delta\theta)+(\partial_\ell\theta)(\partial_\ell\delta\theta) \nonumber \\
& \hspace{-0.5cm} \times 
[1+\alpha\mathrm{cos}(2\theta-\frac{2}{r}\ell)]-\alpha(\partial_\ell\theta)^2\mathrm{sin}(2\theta-\frac{2}{r}\ell)(\delta\theta)\bigg\},
\end{align}
\begin{align}
\label{eqn4.47}
\frac{1}{2}\delta^2 J=&\frac{1}{T}\int^T_0  dt \int d\ell\bigg\{\frac{1}{2}(\partial_t\delta\theta)^2+\frac{1}{2}(\partial_\ell\delta\theta)^2\nonumber\\
&+\alpha(\partial_\ell\delta\theta)^2\mathrm{cos}(2\theta-\frac{2}{r}\ell)
+ \alpha (\delta\theta)^2 \{[
(\partial_{\ell} \theta)^2 \nonumber \\
& \  -\frac{2}{r}(\partial_\ell\theta)] \mathrm{cos}(2\theta 
- \frac{2}{r}\ell) 
+ (\partial^2_{\ell} \theta) \mathrm{sin}(2\theta - \frac{2}{r}\ell) \}\bigg\}, 
\end{align}
respectively. Eqs.~(\ref{eqn4.46},\ref{eqn4.47}) have a similar structure as Eqs.~(\ref{eqn4.39},\ref{eqn4.40}), respectively. For $k_0^2 \ne (1 + \frac{1}{j_0r})^2$ (resonance point), one can use the $\alpha$-expansion of $\theta$ and $\delta\theta$, and the expressions support the same conclusion in the circular geometry case; $\delta J <0$ for some $\delta \theta_0$ and $\frac{1}{2}\delta^2 J\geq 0$.

In summary, contrary to the 
conventional superfluid with $\theta_1=\theta_2=0$, 
the supercurrent state with the broken U(1) spacetime symmetry is classically unstable toward other states,  
and it must experience the energy dissipation by local perturbation.
Physically speaking, the difference in the stabilities between these two types of superfluids comes from the fact that  
the spin-injection boundary condition does not break the U(1) symmetry of the conventional superfluid, but it breaks the U(1) spacetime symmetry of the total-angular-momentum superfluid; under the U(1) spacetime rotation, the whole junction should also be rotated. Effects of the energy dissipation can be included as 
finite relaxation time $T^{-1}_1$, while the motion of $\theta$ with vanishing or small $T_1^{-1}$ can be realized only in a superclean limit.

A conventional superfluid described by a non-relativistic complex field has a critical velocity given by the Landau criterion. Below the critical velocity, a supercurrent is stable. The Landau criterion should be derived from a theory of the complex field instead of an effective theory of a Goldstone mode. This is because when the velocity approaches the critical value, a low-energy condition is already violated. Our analyses only study the stability in the low-energy limit where a non-relativistic complex field and a relativistic complex field both lead to a Goldstone mode with linear dispersion.


\section{\label{appendixE} Effects of dissipation in the classical EOM}

Note first that the classical equation, Eq.~(\ref{eqn2.5}), 
as well as its 1D descendants, Eqs.~(\ref{eqn3.1},\ref{eqn3.35}), are all invariant under the time-reversal operation; 
$t \rightarrow -t$, and 
$\theta \rightarrow \theta + \pi$. 
In the previous appendix, we demonstrate that the classical 
energy of the spin supercurrent state is higher than 
other states due to the finite $\alpha$. This suggests 
that the supercurrent state decays into other states 
with lower energy. Such an energy-non-conserving decay 
process generally breaks the time-reversal symmetry of 
the classical equation. To study the effect of 
the decay process into the spin hydrodynamics 
predicted in Sec.~\ref{sec5}, we include the simplest time-reversal-breaking term, $\partial_t \theta$, into the classical equation; 
\begin{widetext}
\begin{align}
&\partial^2_t \theta - (\partial^2_{x}) \theta 
[1-\alpha \cos(2\theta)] - (\partial^2_y \theta) [1+\alpha \cos(2\theta)] 
+ 2\alpha (\partial_x \partial_y \theta) \sin(2\theta) \nonumber \\  
& \ \ \ -\alpha [(\partial_x\theta)^2 -(\partial_y\theta)^2] 
\sin(2\theta) + 2\alpha (\partial_x \theta) (\partial_y \theta) 
\cos(2\theta) = -\frac{1}{T_1} \partial_t \theta. 
\label{add1}
\end{align}
From the symmetry point of view, one could also add other
time-reversal breaking terms that respect the 
U(1) spacetime symmetry but breaks the time-reversal symmetry, e.g.,  
\begin{align}
\cdots = -\frac{1}{T_1} \partial_t \theta - 
\frac{1}{T_2} \partial_t \theta \times (\partial^2_x \theta + \partial^2_y 
\theta) - \frac{1}{T_3} \partial_t \theta \times \big\{ 
(\partial^2_x \theta -  \partial^2_y \theta) \cos(2\theta) + 
2\partial_x \partial_y \theta \sin(2\theta) \big\} + \cdots 
\label{add2}
\end{align}
\end{widetext}
Nonetheless, the first term on the right-hand side 
always dominates the others in the hydrodynamic regime, 
since the physical variable $\theta$ changes much more 
slowly than any microscopic length scales in the hydrodynamic 
regime, and in this sense, the other terms in Eq.~(\ref{add2}) are 
higher-order spatial gradient terms than the first term in  Eq.~(\ref{add2}). 
In this appendix, we will solve Eq.~(\ref{add1}) or its 1D descendant 
in the spin-injection model with the straight 
geometry, 
\begin{align}
\partial^2_t \theta - 
(\partial^2_x  \theta)[ 1-\alpha\cos(2\theta)] 
- \alpha (\partial_x \theta)^2 
\sin(2\theta) = -\frac{1}{T_1} \partial_t \theta. 
\end{align}   

$\theta_0(x,t)$ with the dissipation term was previously solved by Ref.~\cite{Sonin1978_1, Sonin2010}. It satisfies
\begin{align}
\label{eqn3.45}
\partial_t^2\theta_0+\frac{1}{T_1}\partial_t\theta_0=\partial_x^2\theta_0.
\end{align}
The general solution (up to a constant $F_0$) of Eq.~(\ref{eqn3.45}) is
\begin{align}
\label{eqn3.46}
\theta_0(x,t)=& At+\frac{A}{2T_1}x^2+Bx  \nonumber \\
&+ \sum_{c\in\mathbb{R},c\neq 0}[F_c e^{ict-i\kappa_c x}+F'_c e^{ict+i\kappa_c x}],
\end{align}
where $A,B,F_c,F'_c$ are constants and
\begin{align}
\label{eqn3.47}
\kappa_c=\sqrt{c^2+i\frac{c}{T_1}}.
\end{align}
The boundary conditions Eqs.~(\ref{eqn3.2}) and (\ref{eqn3.3}) are satisfied by 
\begin{align}
\label{eqn3.48}
B=-j_0,\quad A=-k_0(\frac{A}{T_1}L+B), \quad F_c = F^{\prime}_{c}=0, 
\end{align}
which leads to
\begin{align}
\label{eqn3.49}
A=(1+\frac{k_0L}{T_1})^{-1}k_0j_0.
\end{align}
This gives the zeroth-order solution of the EOM with the BCs, 
\begin{align}
\label{eqn3.50}
\theta_0(x,t)=&\frac{T_1k_0j_0}{T_1+k_0L}t-j_0[1-\frac{k_0x}{2(T_1+k_0L)}]x \nonumber \\
\equiv &\tilde{k}_0 j_0t-j_0 h(x)x,
\end{align}
\begin{align}
\label{eqn3.51}
&j^s_x(0<x<L)=\frac{T_1+k_0(L-x)}{T_1+k_0L}j_0+\mathcal{O}(\alpha),\nonumber \\
& s(0<x<L)=\frac{T_1k_0j_0}{T_1+k_0L}+\mathcal{O}(\alpha).
\end{align}
In the conventional spin superfluid with $T^{-1}_1 \ne 0$, 
the spin density and the spin current are static. 
Different from the dissipationless case ($T^{-1}_1=0$), the spin current decreases linearly in the 1D coordinate $x$, while the spin density is 
uniform in $x$.

Due to nonlinear $x$-dependence of $\theta_0(x,t)$, 
the perturbative analyses in the SOC ($\alpha$) becomes harder. To obtain the solution of 
the EOM analytically, we consider a limit that 
a phase accumulation $\gamma$ is small when the spatial dependence of the current is small,
\begin{align}
\label{eqn3.52}
\gamma\equiv[-\frac{d h(x)}{d x}L]j_0L=\frac{k_0j_0L^2}{2(T_1+k_0L)}\ll 1.
\end{align}
The small $\gamma$ limit can be achieved by a small dissipation term or a short propagation distance. The zeroth-order solution in the small $\gamma$ and $\alpha$ limit is  
\begin{align}
\label{eqn3.53}
\theta_0(x,t)=&\tilde{k}_0j_0(1-\frac{k_0L}{T_1})t-j_0 h(x)x
\nonumber \\
=&\tilde{k}_0j_0t-j_0x+\mathcal{O}(\gamma).
\end{align}
We will solve $\theta(x,t)$ up to the first order in $\alpha$ or in $\gamma$, namely
$\mathcal{O}(\alpha,\gamma)$. Thereby, we keep the zeroth order of $\theta_0(x,t)$ when solving $\theta_1(x,t)$ and $\theta_2(x,t)$. Eq.~(\ref{eqn3.11}) is slightly modified and becomes
\begin{align}
\label{eqn3.54}
\partial_t^2\theta_1+\frac{1}{T_1}\partial_t\theta_1-\partial_x^2\theta_1 =&\alpha j_0^2\mathrm{sin}(2\tilde{k}_0j_0t-2j_0x)\nonumber  \\
=&\frac{\alpha j^2_0}{2i}e^{2i\tilde{k}_0j_0t-2ij_0x}+\mathrm{c.c.}.
\end{align}
Eq.~(\ref{eqn3.54}) has a special solution,
\begin{align}
\label{eqn3.55}
\theta_1(x,t)=&\frac{\alpha j_0^2 e^{2i\tilde{k}_0j_0t-2ij_0x}}{2i(-4\tilde{k}_0^2j_0^2-\frac{2i}{T_1}\tilde{k}_0j_0+4j_0^2)}+\mathrm{c.c.}\nonumber \\
=&-\frac{\alpha j_0T_1 e^{2i\tilde{k}_0j_0t-2ij_0x}}{8ij_0T_1(\tilde{k}_0^2-1)+4\tilde{k}_0}+\mathrm{c.c.}\nonumber\\
\equiv& \alpha g_0e^{2ij_0(\tilde{k}_0t-x)}+\mathrm{c.c.},
\end{align}
where
\begin{align}
\label{eqn3.56}
g_0=-\frac{j_0T_1}{8ij_0T_1(\tilde{k}_0^2-1)+4\tilde{k}_0}.
\end{align}
Eq.~(\ref{eqn3.17}) holds true, while Eqs.~(\ref{eqn3.18},\ref{eqn3.19}) are modified,
\begin{align}
\label{eqn3.57}
s=&\frac{1}{2}\tilde{k}_0j_0+\alpha \tilde{k}_0j_0e^{2i\tilde{k}_0j_0t}
[2ige^{-2i\tilde{k}_0j_0x} \nonumber \\
& -\frac{j_0T_1e^{-2ij_0x}}{4j_0T_1(\tilde{k}_0^2-1)-2i\tilde{k}_0}] +2i\alpha \tilde{k}_0j_0 g' \nonumber \\ & \ \ \ \times e^{2i\tilde{k}_0j_0(t+x)} 
+\mathcal{O}(\alpha^2,\gamma^2,\alpha\gamma)+\mathrm{c.c.},
\end{align}
\begin{align}
\label{eqn3.58}
j_x^s=&\frac{1}{2}j_0+\alpha j_0 e^{2i\tilde{k}_0j_0t}[2i\tilde{k}_0ge^{-2i\tilde{k}_0j_0x} \nonumber \\
& -\frac{j_0T_1e^{-2ij_0x}}{4j_0T_1(\tilde{k}_0^2-1)-2i\tilde{k}_0}-\frac{e^{-2ij_0x}}{2}] 
-2i\alpha \tilde{k}_0j_0g' \nonumber \\
& \ \ \  \times e^{2i\tilde{k}_0j_0(t+x)} 
+{\cal O}(\alpha,\gamma,\gamma j_0L)+\mathrm{c.c.}\nonumber\\
=&\frac{1}{2}j_0+\alpha j_0 e^{2i\tilde{k}_0j_0t}\{2i\tilde{k}_0ge^{-2i\tilde{k}_0j_0x} \nonumber \\
& -\frac{[j_0T_1(2\tilde{k}_0^2-1)-i\tilde{k}_0]e^{-2ij_0x}}{4j_0T_1(\tilde{k}_0^2-1)-2i\tilde{k}_0}\} 
-2i\alpha \tilde{k}_0j_0g' \nonumber \\ 
& \ \ \  \times e^{2i\tilde{k}_0j_0(t+x)} 
+\mathcal{O}(\alpha^2,\gamma^2,\alpha\gamma)+\mathrm{c.c.}.
\end{align}
\begin{widetext}
The boundary conditions Eqs.~(\ref{eqn3.2},\ref{eqn3.3}) leads to the following secular equation (cf. Eq.~(\ref{eqn3.26})),
\begin{align}
\label{eqn3.59}
\left(\begin{array}{cc}1&-1\\ (1-k)e^{-i\tilde{k}_0\beta_L} & (1+k)e^{i\tilde{k}_0\beta_L} \end{array}\right)\left(\begin{array}{c}g \\ g'\end{array}\right)=\frac{1}{8i\tilde{k}_0j_0T_1(\tilde{k}_0^2-1)+4\tilde{k}_0^2} \left(\begin{array}{c}j_0T_1(2\tilde{k}_0^2-1)-i\tilde{k}_0 \\ (j_0T_1-j_0T_1(2\tilde{k}_0^2-1)k+i\tilde{k}_0k)e^{-i\beta_L} \end{array}\right).
\end{align}
The solution of Eq.~(\ref{eqn3.59}) is
\begin{align}
\label{eqn3.60}
g=\frac{[j_0T_1(2\tilde{k}_0^2-1)-i\tilde{k}_0](1+k)e^{i\tilde{k}_0\beta_L}+[j_0T_1-j_0T_1(2\tilde{k}_0^2-1)k+i\tilde{k}_0k]e^{-i\beta L}}{[(1+k)e^{i\tilde{k}_0\beta_L}+(1-k)e^{-i\tilde{k}_0\beta_L}][8i\tilde{k}_0j_0T_1(\tilde{k}_0^2-1)+4\tilde{k}_0^2]},
\end{align}
\begin{align}
\label{eqn3.61}
g'=\frac{[j_0T_1-j_0T_1(2\tilde{k}_0^2-1)k+i\tilde{k}_0k]e^{-i\beta_L}-[j_0T_1(2\tilde{k}_0^2-1)-i\tilde{k}_0](1-k)e^{-i\tilde{k}_0\beta L}}{[(1+k)e^{i\tilde{k}_0\beta_L}+(1-k)e^{-i\tilde{k}_0\beta_L}][8i\tilde{k}_0j_0T_1(\tilde{k}_0^2-1)+4\tilde{k}_0^2]}.
\end{align}
Similar to Eq.~(\ref{eqn3.30}), the solution of $\theta(x,t)$ is
\begin{align}
\label{eqn3.62}
\theta(x,t)=&\tilde{k}_0 j_0t-j_0[1-\frac{k_0x}{2(T_1+k_0L)}]x
+2\alpha\mathrm{Re}(g_0)\mathrm{cos}[2j_0(\tilde{k}_0t-x)]
  -2\alpha\mathrm{Im}(g_0)\mathrm{sin}[2j_0(\tilde{k}_0t-x)]\nonumber\\
&  +2\alpha\mathrm{Re}(g)\mathrm{cos}[2\tilde{k}_0j_0(t-x)]
  +2\alpha\mathrm{Re}(g')\mathrm{cos}[2\tilde{k}_0j_0(t+x)]
 -2\alpha\mathrm{Im}(g)\mathrm{sin}[2\tilde{k}_0j_0(t-x)] \nonumber \\
& \ +2\alpha\mathrm{Im}(g')\mathrm{sin}[2\tilde{k}_0j_0(t+x)] +\mathcal{O}(\alpha^2,\gamma^2,\alpha\gamma),
\end{align}
\end{widetext}
where $g_0,g,g'$ are given by Eqs.~(\ref{eqn3.56},\ref{eqn3.60},\ref{eqn3.61}). $k_0$, $k$, and $\tilde{k}_0$ are given by Eqs.~(\ref{eqn3.22},\ref{eqn3.23},\ref{eqn3.50}). 
Compared to Eqs.~(\ref{eqn3.26},\ref{eqn3.42}), 
Eqs.~(\ref{eqn3.60},\ref{eqn3.61},\ref{eqn3.62}) have no divergence 
due to finite $T_1$. The situation is analogous to periodically driven harmonic oscillators, where the dissipation removes divergence due to resonance~\cite{Taylor2004}.
Note also that the solution Eq.~(\ref{eqn3.62}) has a periodicity in time, $\pi (\tilde{k}_0j_0)^{-1}$, and two characteristic wavelengths, $2j_0$, and $2\tilde{k}_0j_0$. A solution at higher order in $\gamma$ has all spatial Fourier components, while 
it is still periodic in time with the same periodicity.  

\section{\label{appendixF} 
Special parameter points in the spin-injection model}

In this appendix, we study some special parameter points in the spin-injection 
model without the spin relaxation term, where solutions in Appendix \ref{appendixC} do 
not apply directly and need careful investigations.

\subsection{$k_0=1$ (straight geometry) and $k^2_0 = (1+\frac{1}{j_0 r})^2$ 
(circular geometry)} 
Consider $k_0=1$ in the straight geometry and 
$k^2_0=(1+\frac{1}{j_0r})^2$ in the circular geometry. 
Naive substitutions of $k_0=1$ into Eqs.~(\ref{eqn3.14},\ref{eqn3.26}) and 
of $k^2_0=(1+\frac{1}{j_0r})^2$ into 
Eqs.~(\ref{eqn3.37},\ref{eqn3.42}) lead to divergences in $\theta_1$ 
and $\theta_2$, respectively. It seems that the divergences in $\theta_1$ and $\theta_2$ cancel each other.  For example, Eqs.~(\ref{eqn3.32}) and (\ref{eqn19}) at $k_0=1$ become 
\begin{align}
\label{eqn5.1}
\eta|_{k_0=1}&=\frac{[(1+k)e^{i\beta_L}-(1-k)e^{-i\beta_L}]+2(1-k)e^{-i\beta_L}}{4(k_0^2-1)[(1+k)e^{i\beta_L}+(1-k)e^{-i\beta_L}]} \nonumber \\
&=\frac{1}{4(k_0^2-1)},
\end{align}
\begin{align}
\label{eqn5.2}
\theta(x,t)|_{k_0=1}=&j_0(t-x)-\frac{\alpha}{4(k_0^2-1)}\mathrm{sin}[2j_0(t-x)] \nonumber \\
& -\frac{\alpha}{4(k_0^2-1)}\mathrm{cos}(2j_0t)\mathrm{sin}(2j_0x)\nonumber\\
&+\frac{\alpha}{4(k_0^2-1)}\mathrm{sin}(2j_0t)\mathrm{cos}(2j_0x)+\mathcal{O}(\alpha^2)\nonumber\\
=&j_0(t-x)+\mathcal{O}(\alpha^2),  
\end{align}
respectively, where the final result of $\theta(x,t)$ is apparently finite. However, 
Eq.~(\ref{eqn5.2}) is not a solution to the EOM Eq.~(\ref{eqn3.1}). In fact, 
with light-cone coordinates 
\begin{align}
\label{eqn5.3}
\xi=t-x,\quad \zeta=t+x,
\end{align}
Eq.~(\ref{eqn3.11}) at $k_0=1$ becomes
\begin{align}
\label{eqn5.4}
[(\partial_\xi+\partial_\zeta)^2-(\partial_\xi-\partial_\zeta)^2]\theta_1=&4\partial_\xi\partial_\zeta\theta_1 \nonumber \\
=&\alpha j_0^2\mathrm{sin}(2j_0\xi).
\end{align}
Equation~(\ref{eqn5.4}) has a special solution which is not consistent to Eq.~(\ref{eqn5.2}),
\begin{align}
\label{eqn5.5}
\theta_1=-\frac{\alpha\zeta}{16}\mathrm{cos}(2j_0\xi)=-\frac{\alpha(t+x)}{16}\mathrm{cos}[2j_0(t-x)].
\end{align}
Note that $|\theta_1|$ in Eq.~(\ref{eqn5.5}) is not bounded for large 
$\zeta=t+x$. This indicates that the perturbation with respect to 
$\alpha$ becomes invalid at $k_0=1$, leading to the discrepancy.  
The divergences at $k_0=1$ can be regarded as the resonance of the inhomogeneous linearized differential equation~\cite{Taylor2004}, and one can expect that the SOC has 
non-perturbative effects around $k_0=1$. 

To understand the origin of the non-perturbative effect of $\alpha$, let us consider a set of solutions of Eq.~(\ref{eqn3.1}) that depends on $x$ and $t$ only through 
$x-vt$. For the later comparison to a special solution developed in Appendix \ref{appendixC}, $\theta_0+\theta_1+{\cal O}(\alpha^2)$, let $v$ to be $k_0$, 
\begin{align}
\label{eqn5.6}
\theta(x,t)=\theta(x-k_0t),\quad \theta'\equiv\partial_x\theta=-\frac{1}{k_0}\partial_t\theta. 
\end{align}
Here, the prime denotes an $x$-derivative. Eq.~(\ref{eqn3.1}) 
effectively becomes an ordinary differential equation,
\begin{align}
\label{eqn5.7}
(k_0^2-1)\theta''=-\alpha\theta''\mathrm{cos}(2\theta)+\alpha\theta'^2\mathrm{sin}(2\theta).
\end{align}
To solve this equation, use its analogy to 1D classical mechanics, where the phase $\theta(x)$ as a function of $x$ corresponds to a 1D coordinate  
as a function of time.
The classical  mechanics for the 1D coordinate has a Lagrangian whose variation gives Eq.~(\ref{eqn5.7}) as a classical EOM,
\begin{align}
\label{eqn5.8}
L_{\rm 1D}=\frac{1}{2}(k_0^2-1)\theta'^2+\frac{\alpha}{2}\theta'^2\mathrm{cos}(2\theta).
\end{align}
The classical mechanics has a canonical momentum conjugate to the coordinate,
\begin{align}
\label{eqn5.9}
\pi=\frac{\partial L_{\rm 1D}}{\partial\theta'}=(k_0^2-1)\theta'+\alpha\theta'\mathrm{cos}(2\theta), 
\end{align}
as well as a conserved Hamiltonian,
\begin{align}
\label{eqn5.10}
H_{\rm 1D}=\pi\theta'-L_{\rm 1D}= \frac{1}{2}(k_0^2-1)\theta'^2+\frac{\alpha}{2}\theta'^2\mathrm{cos}(2\theta).
\end{align}
Utilizing the $x$-independence (``time"-independence) of $H_{\rm 1D}$, we can solve the  
EOM from Eq.~(\ref{eqn5.10}), 
\begin{align}
\label{eqn5.11}
\frac{d\theta}{dx}=\pm\sqrt{\frac{2H_{\rm 1D}}{k_0^2-1+\alpha\mathrm{cos}(2\theta)}}.
\end{align}
Its formal solution is given by 
\begin{align}
\label{eqn5.12}
\pm (x-x_0)=\int_{\theta(x_0)}^{\theta(x)}\sqrt{\frac{k_0^2-1+\alpha\mathrm{cos}(2\theta)}{2H_{\rm 1D}}}d\theta.
\end{align}
With Eq.~(\ref{eqn5.6}), we get a set of (1+1)-dimensional solutions,
\begin{align}
\label{eqn5.13}
\pm (x-k_0t-C_0)=\int_{\theta(x_0,t_0)}^{\theta(x,t)}\sqrt{\frac{k_0^2-1+\alpha\mathrm{cos}(2\theta)}{2H_{\rm 1D}}}d\theta,
\end{align}
where $C_0=x_0-k_0t_0$. 

Eq.~(\ref{eqn5.13}) is inclusive of those perturbative solutions in Sec.~\ref{sec3} that depend on $x$ and $t$ only through $x-k_0 t$, i.e. $\theta_0+\theta_1$ given by  
Eqs.~(\ref{eqn3.8},\ref{eqn3.14}). 
Namely, when $\alpha\ll |k_0^2-1|$, we can apply an expansion in $\alpha$,
\begin{align}
\label{eqn5.15}
\pm(x-k_0t-C_0)=& \int_{\theta(x_0,t_0)}^{\theta(x,t)}\sqrt{\frac{k_0^2-1}{2H_{\rm 1D}}} [1+ \nonumber \\
& \frac{\alpha}{2(k_0^2-1)}\mathrm{cos}(2\theta)]d\theta+\mathcal{O}(\alpha^2). 
\end{align}
From this, we obtain
\begin{align}
\label{eqn5.16}
\theta(x,t)=&\pm\sqrt{\frac{2H_{\rm 1D}}{k_0^2-1}}(x-k_0t-C_0')\nonumber \\
& -\frac{\alpha}{4(k_0^2-1)}\mathrm{sin}(2\theta)+\mathcal{O}(\alpha^2)\nonumber\\
=&\theta_0-\frac{\alpha}{4(k_0^2-1)}\mathrm{sin}(2\theta_0)+\mathcal{O}(\alpha^2),
\end{align}
where
\begin{align}
\label{eqn5.17}
\theta_0=\pm\sqrt{\frac{2H_{\rm 1D}}{k_0^2-1}}(x-k_0t-C_0'),
\end{align}
\begin{align}
\label{eqn5.18}
C_0'=C_0\mp\sqrt{\frac{k_0^2-1}{2H_{\rm 1D}}}\{\theta(x_0,t_0)+\frac{\alpha}{4(k_0^2-1)}\mathrm{sin}[2\theta(x_0,t_0)]\}.
\end{align}
Note that due to the absence of $\theta_2(x,t)$ in its $\alpha$-expansion, 
Eq.~(\ref{eqn5.16}) does not satisfy the boundary conditions in the spin-injection model in general. 

Nonetheless, Eq.~(\ref{eqn5.13}) is still 
useful to see that the expansion in $\alpha$ is invalid 
when $|k_0^2-1| < \mathcal{O}(\alpha)$. Take $k_0=1$ in Eq.~(\ref{eqn5.13}) as an example. Thereby, the sign of $\mathrm{cos}(2\theta)$ is conserved from Eq.~(\ref{eqn5.10}). 
Then, the integral of Eq.~(\ref{eqn5.13}) gives
\begin{align}
\label{eqn5.19}
\pm\sqrt{\frac{2H_{\rm 1D}}{\alpha}}(x-k_0t-C_0)=E(\theta(x,t),2)-E(\theta(x_0,t_0),2),
\end{align}
where $E(\theta,m)$ is the elliptic integral of the second kind. This special solution can be made independent from $\alpha$ because $\alpha$ can be absorbed into another parameter $H_{\rm 1D}$. The $\alpha$-independence as well as the conserved sign of $\cos(2\theta)$ are not consistent with $\theta_0(x,t)+\theta_1(x,t)$ in Eqs.~(\ref{eqn3.8},\ref{eqn3.14}). This suggests that the 
expansion in $\alpha$ is invalid, particularly at $k_0=1$. 
 
\subsection{$k_0=0$ (straight geometry)} 
Another special parameter point is $k_0=0$. $k_0=0$ is a limit where $\chi$ in Eq.~(\ref{eqn3.22}) is much smaller than $\chi'\sqrt{\frac{D_s}{T_1'}}$ and $\beta_t$, and $k_c$ goes to zero for any $c$;
\begin{align}
k_c \equiv \frac{\chi}{\chi^{\prime}} \sqrt{\frac{1}{D_s(\frac{1}{T^{\prime}_{1}}+ic)}} 
+ \frac{\chi}{\beta_t} \rightarrow 0. 
\end{align}
We first consider $k_0=0$ in the straight geometry 
($r^{-1}=0$). $\theta_0$ at $k_0=0$ has no time dependence in Eq.~(\ref{eqn3.8}), so that the phase $F_0$ in Eq.~(\ref{eqn3.8}) cannot be 
absorbed into the time. Meanwhile, $s(x=L-)=0$ because $k_c=0$ for any $c$,    
and $s(x) \equiv \partial_t \theta =0$ can be always satisfied by 
a time-independent $\theta$. Thus,  
we have only to make Eqs.~(\ref{eqn3.8},\ref{eqn3.10}) 
with an additional $F_0$ to satisfy the other 
boundary condition, $j^s_x(x=0+)=j_0$. Firstly, let us choose 
$\theta_0=-j_0x+F_0$ that satisfies the boundary condition. The substitution into Eq.~(\ref{eqn3.11}) gives $\theta_1 = \frac{\alpha}{4} \sin (-2j_0 x + F_0)$. To satisfy the boundary condition up to the 
1st order in $\alpha$ [see also Eq.~(\ref{eqn3.4})], 
\begin{equation}
\label{eqn5.20}
j^s_x(x=0+)=j_0-\frac{j_0\alpha}{2}\mathrm{cos}(2F_0)-\partial_x\theta_2+\mathcal{O}(\alpha^2)=j_0, 
\end{equation}
we have
\begin{equation}
\label{eqn5.21}
\theta_2(x)=-\frac{j_0\alpha}{2}\mathrm{cos}(2F_0)x.
\end{equation}
Here, $|\theta_2(x)|$ for large $x$ is not bounded, where the perturbation in $\alpha$ breaks down. 
An alternative way to get a consistent perturbative solution is to require $\theta_2(x)=0$ and take $\theta_0(x)=-\tilde{j}_0x+F_0$ where $\tilde{j}_0\neq j_0$. Then we have
\begin{align}
\label{eqn5.22}
j^s_x(x=0+)=\tilde{j}_0-\frac{\tilde{j}_0\alpha}{2}\mathrm{cos}(2F_0)-\partial_x\theta_2+\mathcal{O}(\alpha^2)=j_0.
\end{align}
From this, we get
\begin{align}
\label{eqn5.23}
\tilde{j}_0=j_0[1-\frac{\alpha}{2}\mathrm{cos}(2F_0)]^{-1}=j_0[1+\frac{\alpha}{2}\mathrm{cos}(2F_0)]+\mathcal{O}(\alpha^2).
\end{align}
At ${\cal O}(\alpha)$, this solution is equivalent to 
absorbing $\theta_2(x)$ in Eq.~(\ref{eqn5.21}) into $\theta_0(x)$. 
Note also that from Eq.~(\ref{eqn4.39}), $\theta_2(x)=0$ implies that 
the steady (but not uniform) current without spin accumulation has no energy dissipation. However, a higher-order calculation 
in $\alpha$ suggests that $\delta J$ can be non-zero and negative. This is because $\delta\theta_1$ and $\theta_1$ can share
identical Fourier components [see Eq.~(\ref{eqn4.30})  
in Appendix \ref{appendixH}].

\subsection{$k_0=0$ (circular geometry)}
Let us next consider the circular geometry with $k_0=0$. We take $\theta_2(\ell)=0$ and $\theta_0(\ell)=-\tilde{j}_0\ell+F_0$ with  $\tilde{j}_0\neq j_0$. Then, 
similar to Eq.~(\ref{eqn22}), 
we obtain 
\begin{align}
\label{eqn5.24}
-\partial_\ell^2\theta_1=-\alpha \tilde{j}_0(\tilde{j}_0+\frac{2}{r})\mathrm{sin}(-2\tilde{j}_0\ell-\frac{2}{r}\ell+2F_0).
\end{align}
When $\tilde{j}_0\neq-\frac{1}{r}$, Eq.~(\ref{eqn5.24}) leads to
\begin{align}
\label{eqn5.25}
\theta_1(\ell)=\frac{\alpha \tilde{j}_0(\tilde{j}_0+\frac{2}{r})}{4(\tilde{j}_0+\frac{1}{r})^2}\mathrm{sin}(2\tilde{j}_0\ell+\frac{2}{r}\ell-2F_0).
\end{align}
Note that when $\tilde{j}_0=-\frac{2}{r}$, $\theta_1(\ell)=0$, and $\theta(\ell)=\frac{2\ell}{r}+F_0$ ($F_0\in\mathbb{R}$) becomes an ``exact" solution of Eq.~(\ref{eqn3.35}). The solution is not exact in the presence of higher-order derivatives in the model Eq.~(\ref{eqn5}).

From Eq.~(\ref{eqn5.25}) and $\theta_2=0$, 
the boundary condition at $\ell=0+$ reads 
\begin{align}
\label{eqn5.27}
j^s_\ell(\ell=0+)=j_0=&\tilde{j}_0+\alpha \tilde{j}_0\mathrm{cos}(2F_0) \nonumber \\
& -\frac{\alpha \tilde{j}_0(\tilde{j}_0+\frac{2}{r})}{2(\tilde{j}_0+\frac{1}{r})}\mathrm{cos}(2F_0)+\mathcal{O}(\alpha^2)\nonumber\\
=&\tilde{j}_0+\frac{\alpha \tilde{j}_0^2}{2(\tilde{j}_0+\frac{1}{r})}\mathrm{cos}(2F_0)+\mathcal{O}(\alpha^2).
\end{align}
So we have
\begin{align}
\label{eqn5.28}
\tilde{j}_0=j_0-\frac{\alpha \tilde{j}_0^2}{2(\tilde{j}_0+\frac{1}{r})}\mathrm{cos}(2F_0)+\mathcal{O}(\alpha^2).
\end{align}
As in the previous case, the steady current without spin accumulation has an energy dissipation.   

\subsubsection{$k_0=0$, $\tilde{j}_0=-\frac{1}{r}$ (circular geometry)}
When $\tilde{j}_0=-\frac{1}{r}$, Eq.~(\ref{eqn5.24}) leads to
\begin{align}
\label{eqn5.29}
\theta_1=-\frac{\ell^2}{2}\alpha \tilde{j}_0(\tilde{j}_0+\frac{2}{r})\mathrm{sin}(2F_0)=\frac{\ell^2\alpha}{2r^2}\mathrm{sin}(2F_0).
\end{align}
When $F_0 \ne \frac{N}{2}\pi$, 
the perturbation of $\alpha$ becomes invalid for larger $\ell$, where 
$|\theta_1(\ell)|$ is not bounded. 
When $F_0=\frac{N}{2}\pi$, $\theta_1(\ell)=0$, and the boundary condition requires
\begin{align}
\label{eqn5.30}
j_0=\tilde{j}_0[1+\alpha(-1)^N]=-\frac{1}{r}[1+\alpha(-1)^N].
\end{align}
This solution ($\theta(\ell,t)=\frac{\ell}{r} + \frac{N\pi}{2}$) can exist only at $k_0=0$, while it is not continuously connected to a solution at finite small $k_0$ [see below].

\subsubsection{$k_0 \rightarrow 0$, $j_0=-\frac{1}{r}+\mathcal{O}(\alpha)$  (circular geometry)}
To see that the solution Eq.~(\ref{eqn5.30}) is not continuously connected to a perturbative solution at finite small $k_0$, let us keep a finite $k_0$ and choose $j_0=-\frac{1}{r}+\mathcal{O}(\alpha)$; 
$\theta_0(\ell)=\frac{1}{r}(k_0 t-\ell)+\mathcal{O}(\alpha)$. In the perturbation theory, we should neglect $\mathcal{O}(\alpha^2)$ contributions to $\theta_1$, so $\theta_1$ is not affected by the $\mathcal{O}(\alpha)$ component in $j_0$. Then 
Eq.~(\ref{eqn22}) becomes
\begin{align}
\label{eqn5.31}
\partial_t^2\theta_1-\partial_\ell^2\theta_1=-\frac{\alpha}{r^2}\mathrm{sin}(-\frac{2k_0}{r}t),
\end{align}
which leads to a solution,
\begin{align}
\label{eqn5.32}
\theta_1=-\frac{\alpha}{4k_0^2}\mathrm{sin}(\frac{2k_0}{r}t).
\end{align}
When $k_0\rightarrow 0$, the solution has the divergence.  
Physically speaking, when $k_0\rightarrow 0$,
$F_0$ changes slowly with respect to time, so we cannot 
fix the phase $F_0$ in Eq.~(\ref{eqn5.29}).

\section{\label{appendixG} Derivation of the Landau criterion}

In this appendix, we use the same framework as in Sec.~\ref{sec5} and derive the Landau criterion. A similar argument can be found in Ref.~\cite{Sonin2010}, while the derivation here is more formal than  Ref.~\cite{Sonin2010}. 
We begin with 
a 1D superfluid model,
\begin{align}
\label{eqnA.1}
\tilde{\mathcal{L}}_{\phi}=i\hbar\phi^\dagger \partial_t\phi-\frac{\hbar^2}{2m}(\partial_x\phi^\dagger)(\partial_x\phi)-\frac{U}{2}(\phi^\dagger\phi)^2+\mu\phi^\dagger\phi,
\end{align}
with its classical EOM,
\begin{align}
\label{eqnA.2}
i\hbar\partial_t\phi=(-\frac{\hbar^2}{2m}\partial_x^2+U\phi^\dagger\phi-\mu)\phi.
\end{align}
Thanks to the Galilean covariance of 
the Lagrangian Eq.~(\ref{eqnA.1}), 
our discussion will be easier than Sec.~\ref{sec5}, 
and we do not need to expand the variation of the 
motion as in Sec.~\ref{sec5}. 
Namely, using the Galilean covariance, 
we can directly obtain two motions that are close to each other, and compare their energies from the corresponding Hamiltonian, 
\begin{align}
\tilde{H}_{\phi}[\phi] = \int dx[\frac{\hbar^2}{2m}(\partial_x\phi^\dagger)(\partial_x\phi)+\frac{U}{2}(\phi^\dagger\phi)^2-\mu\phi^\dagger\phi]. 
\end{align}
Consider a steady flow $\phi_0(x,t)$,
\begin{align}
\label{eqnA.3}
\phi_0(x,t)=\sqrt{\rho_0}\exp[\frac{i}{\hbar}(mvx-\frac{mv^2}{2}t)].
\end{align}
Let us assume that the following motion $\phi(x,t)$, 
as well as $\phi_0(x,t)$, satisfies the EOM, Eq.~(\ref{eqnA.2}),
\begin{align}
\label{eqnA.4}
\phi(x,t)=&\phi'(x',t')\exp[\frac{i}{\hbar}(mvx-\frac{mv^2}{2}t)] \nonumber \\
=&\phi'(x-vt,t)\exp[\frac{i}{\hbar}(mvx-\frac{mv^2}{2}t)], 
\end{align}
with $x^{\prime} = x - v t$ and $t^{\prime}=t$. 
Here $|\phi'(x',t')-\sqrt{\rho_0}|\ll \sqrt{\rho_0}$ and $\phi(x,t)$ is close to $\phi_0(x,t)$. 
Using 
a Galilean transformation, 
\begin{align}
\label{eqnA.5}
x'=x-vt,\quad t'=t,
\end{align}
\begin{align}
\label{eqnA.6}
\partial_x=\partial_{x'},\quad \partial_t=\partial_{t'}-v\partial_x,
\end{align}
we can see that $\phi^{\prime}(x^{\prime},t^{\prime})$ must 
satisfy a similar equation as Eq.~(\ref{eqnA.2}), 
\begin{align}
\label{eqnA.7}
(i\hbar\partial_t+\frac{1}{2}mv^2)\phi'(x',t')= & [\frac{1}{2m}(-i\hbar\partial_x+mv)^2 \nonumber \\
& \hspace{-0.5cm} 
+U\phi'^\dagger\phi'-\mu]\phi'(x',t'),
\end{align}
or equivalently, 
\begin{align}
\label{eqnA.8}
i\hbar\partial_{t'}\phi'(x',t')=(-\frac{\hbar^2}{2m}\partial_{x'}^2+U\phi'^\dagger\phi'-\mu)\phi'(x',t').
\end{align}

Now we compare the energies of $\phi(x,t)$ and $\phi_0(x,t)$. The energy of $\phi_0(x,t)$ is,
\begin{align}
\label{eqnA.9}
\tilde{H}_{\phi}[\phi_0]=& \int dx[\frac{\hbar^2}{2m}(\partial_x\phi_0^\dagger)(\partial_x\phi_0)+\frac{U}{2}(\phi_0^\dagger\phi_0)^2 
\nonumber \\
& -\mu\phi_0^\dagger\phi_0]=E_0+\frac{1}{2}mv^2Q_0,
\end{align}
where
\begin{align}
\label{eqnA.10}
E_0=&\int dx(\frac{U}{2}\rho_0^2-\mu\rho_0)=-\frac{\mu}{2}\int dx\rho_0,\nonumber \\
Q_0=&\int dx \rho_0.
\end{align}
The energy of $\phi(x,t)$ is,
\begin{align}
\label{eqnA.11}
\tilde{H}_{\phi}[\phi]=&\int dx[\frac{\hbar^2}{2m}(\partial_x\phi^\dagger)(\partial_x\phi)+\frac{U}{2}(\phi^\dagger\phi)^2-\mu\phi^\dagger\phi]\nonumber\\
=&\int dx\phi'^\dagger(x-vt,t)[\frac{1}{2m}(-i\hbar\partial_x+mv)^2 \nonumber \\
& \hspace{0.5cm} +\frac{U}{2}\phi'^\dagger\phi'-\mu]\phi'(x-vt,t)\nonumber\\
=&\int dx\phi'^\dagger(x,t)[\frac{1}{2m}(-i\hbar\partial_x+mv)^2 \nonumber \\
& \hspace{0.5cm} 
+\frac{U}{2}\phi'^\dagger\phi'-\mu]\phi'(x,t)\nonumber\\
=&\int dx\phi'^\dagger(x,t)[-\frac{\hbar^2}{2m}\partial_x^2-iv\hbar\partial_x+\frac{1}{2}mv^2 \nonumber \\
& \hspace{0.5cm} +\frac{U}{2}\phi'^\dagger\phi'-\mu]\phi'(x,t)\nonumber\\
=&\tilde{H}_\phi[\phi']+vP_\phi[\phi']+\frac{1}{2}mv^2Q_\phi[\phi'],
\end{align}
where
\begin{align}
\label{eqnA.12}
& P_\phi[\phi']=\int dx\phi'^\dagger(x,t)(-i\hbar\partial_x)\phi'(x,t),\nonumber \\
& Q_\phi[\phi']=\int dx\phi'^\dagger(x,t)\phi'(x,t).
\end{align}
Thus, the energy difference between  $\phi(x,t)$ and $\phi_0(x,t)$ is
\begin{align}
\label{eqnA.13}
&\Delta E_\phi[\phi,\phi_0]=\tilde{H}_{\phi}[\phi]-\tilde{H}_{\phi}[\phi_0] \nonumber \\
& \ =(\tilde{H}_\phi[\phi']-E_0)+vP_\phi[\phi']+\frac{1}{2}mv^2(Q_\phi[\phi']-Q_0).
\end{align}
For $\phi'$ that satisfies the EOM Eq.~(\ref{eqnA.8}), energy $H_{\phi}[\phi^{\prime}]$, momentum 
$P_{\phi}[\phi^{\prime}]$, and U(1) charge $Q_{\phi}[\phi^{\prime}]$ must be all conserved. This is because 
from Eq.~(\ref{eqnA.8}), $\phi^{\prime}(x,t)$ is a 
solution of Eq.~(\ref{eqnA.2}). 


The average velocity of $\phi(x,t)$ is not small, while 
$\phi'(x,t)$ can be assumed at the near-equilibrium 
limit~\cite{Zhai2021}. 
In this limit, the EOM of $\phi'=\sqrt{\rho_0+\delta\rho'}e^{i\theta'}$ 
can be described by a wave equation of $\delta \rho^{\prime}$ 
and $\theta^{\prime}$,
\begin{align}
\label{eqnA.15}
\partial_t^2\theta'(x,t)-\frac{\rho_0 U}{m}\partial_x^2\theta'(x,t)=0,
\end{align}
\begin{align}
\label{eqnA.16}
\delta\rho'(x,t)=-\frac{\hbar}{U}\partial_t\theta'(x,t).
\end{align}
Thereby, $\theta'(x,t)$ is given by a superposition of oscillations,
\begin{align}
\label{eqnA.17}
\theta'(x,t)=&\frac{1}{\sqrt{L}}\sum_q [f_qe^{iq(x-v_ct)}+f_q'e^{iq(x+v_ct)}],\nonumber \\
v_c=&\sqrt{\frac{\rho_0U}{m}}.
\end{align}
In the near-equilibirum limit of $\phi^{\prime}$, 
we evaluate the energy difference in the leading order in small $f_q$ and $f^{\prime}_q$,
\begin{align}
\label{eqnA.18}
\tilde{H}_\phi[\phi']-E_0=&\int dx[\frac{\hbar^2\rho_0}{2m}(\partial_x\theta')^2+\frac{U}{2}(\delta\rho')^2]
\nonumber \\
=&\int dx[\frac{\hbar^2\rho_0}{2m}(\partial_x\theta')^2+\frac{\hbar^2}{2U}(\partial_t\theta')^2]\nonumber\\
=&\frac{\hbar^2\rho_0}{m}\sum_q q^2(|f_q|^2+|f'_q|^2),
\end{align}
\begin{align}
\label{eqnA.19}
P_\phi[\phi']=&\int dx\hbar\delta\rho'(\partial_x\theta')=-\frac{\hbar^2}{U}\int dx(\partial_t\theta')(\partial_x\theta')
\nonumber \\
=&\frac{\hbar^2\rho_0}{mv_c}\sum_q q^2(|f_q|^2-|f'_q|^2),
\end{align}
\begin{align}
\label{eqnA.20}
Q_\phi[\phi']-Q_0=\int dx \delta\rho'=0.
\end{align}
Taking Eqs.~(\ref{eqnA.18}-\ref{eqnA.20}) into Eq.~(\ref{eqnA.13}), 
we have
\begin{align}
\label{eqnA.21}
\Delta E_\phi=\frac{\hbar^2\rho_0}{m}\sum_q q^2[(1+\frac{v}{v_c})|f_q|^2+(1-\frac{v}{v_c})|f_q'|^2].
\end{align}
To make $\Delta E_\phi\geq 0$ for any (small) $f_q$ and $f_q'$, we obtain 
the Landau criterion, 
\begin{align}
\label{eqnA.22}
|v|\leq v_c=\sqrt{\frac{\rho_0 U}{m}}.
\end{align}
 
\section{\label{appendixH} Local deformations of classical solutions of the EOM}  
The superfluid state with a finite supercurrent 
is characterized by  
the solution $\theta(x,t)$ of the classical EOM 
in the 1D spin-injection model [e.g.,  with 
the straight geometry]. In Sec.~\ref{sec5}, 
we introduced its local deformation 
$\theta(x,t)+\delta\theta(x,t)$ as another solution 
of the EOM with different boundary conditions. 
We regarded that $\delta\theta(x,t)$, 
as well as $\theta(x,t)$, can be determined perturbatively 
in the SOC ($\alpha$). At the zeroth order in SOC, 
$\theta+\delta\theta$, as well as $\theta$, is a solution 
of $\partial^2_t \theta - \partial^2_x \theta=0$, and  
so is $\delta \theta$. Since $\delta\theta(x,t)$ 
is a local deformation and 
its spacetime derivatives  
should not contain any uniform components in space, 
the zeroth order of $\delta\theta(x,t)$ 
must be given by Eq.~(\ref{eqn4.37});
\begin{align}
\label{B-start}
\delta \theta_0(x,t) = \frac{1}{\sqrt{L}} 
\sum_q \big[\delta d_q e^{iq(x-t)} + 
\delta d^{\prime}_q e^{iq(x+t)} \big]. 
\end{align}
In this appendix, for a given form of 
Eq.~(\ref{eqn4.37}) as the zeroth order,  
we will show how to determine the first order  
of $\delta \theta(x,t)$;  
\begin{align}
\delta \theta (x,t) = \delta \theta_0(x,t) 
+ \delta\theta_1(x,t) + {\cal O}(\alpha^2).  
\end{align}

We first give a general framework to determine $\delta \theta$. 
$\theta+\delta\theta$, as well as $\theta$, is 
a local minimum of the action, $S =\int d^3r\mathcal{L} 
\equiv \int dt d^2r \mathcal{L}$, 
and $\delta \theta$ is infinitesimally small. Thus, we take  
a $\delta \theta$-variation of $S$;
\begin{align}
S[\theta] \equiv S_{xx}[\theta] + S_{yy}[\theta] + 
S_{xy}[\theta] + \frac{1}{2} 
\int d^3 r  (\partial_t \theta)^2,
\end{align}
\begin{align}
S_{xx} \equiv& - \frac{1}{2} 
\int d^3 r  (\partial_x \theta)^2 [1-\alpha \cos(2\theta)], \nonumber \\ 
S_{yy}  \equiv& \frac{1}{2} 
\int d^3 r  (\partial_y \theta)^2 [1+\alpha \cos(2\theta)], \nonumber \\ 
S_{xy} \equiv& \alpha \int d^3 r (\partial_x \theta) 
(\partial_y \theta) \sin(2\theta). 
\end{align}
The first-order variation just gives the EOM Eq.~(\ref{eqn2.5}),
\begin{align}
\label{eqn4.1}
\delta S=\delta S_{xx}+\delta S_{yy}+\delta S_{xy}-\int d^3r(\delta\theta)(\partial_t^2\theta),
\end{align}
\begin{align}
\label{eqn4.2}
\delta S_{xx}\equiv&\int d^3r(\delta\theta)\{(\partial_x^2\theta)[1-\alpha\mathrm{cos}(2\theta)]
\nonumber \\
& \ +\alpha(\partial_x\theta)^2\mathrm{sin}(2\theta)\},
\end{align}
\begin{align}
\label{eqn4.3}
\delta S_{yy}\equiv&\int d^3r(\delta\theta)\{(\partial_y^2\theta)[1+\alpha\mathrm{cos}(2\theta)]
\nonumber \\
&\ -\alpha(\partial_y\theta)^2\mathrm{sin}(2\theta)\},
\end{align}
\begin{align}
\label{eqn4.4}
\delta S_{xy}\equiv&-2\alpha\int d^3r(\delta\theta)[(\partial_x\partial_y\theta)\mathrm{sin}(2\theta)
\nonumber \\
& \ \ +(\partial_x\theta)(\partial_y\theta)\mathrm{cos}(2\theta)].
\end{align}
$\delta S$ vanishes 
since $\theta$ is an extremum or a saddle point of $S$. The second-order variation $\delta^2 S$ determines small deformation $\delta\theta$ in such a way that $\theta+\delta \theta$ is an extremum or a saddle point of $S$. $S_{xx}$ gives
\begin{align}
\label{eqn4.5}
\delta^2 S_{xx}=&\int d^3r(\delta\theta)\{(\partial_x^2\delta\theta)[1-\alpha\mathrm{cos}(2\theta)]\nonumber \\
&+2\alpha(\partial_x^2\theta)\mathrm{sin}(2\theta)(\delta\theta)+2\alpha(\partial_x\theta)(\partial_x\delta\theta)\mathrm{sin}(2\theta) \nonumber \\
& \ \ +2\alpha(\partial_x\theta)^2\mathrm{cos}(2\theta)(\delta\theta)\}\nonumber\\
=&-\int d^3r[(\partial_x\delta\theta)^2
+\alpha\mathrm{cos}(2\theta)(\delta\theta)\partial_x^2(\delta\theta)
\nonumber \\
& -2\alpha\mathrm{sin}(2\theta)(\partial_x^2\theta)(\delta\theta)^2 
-2\alpha(\partial_x\theta)(\delta\theta)(\partial_x\delta\theta)\mathrm{sin}(2\theta) \nonumber \\
&\ \ \ -2\alpha(\partial_x\theta)^2\mathrm{cos}(2\theta)(\delta\theta)^2],
\end{align}
where
\begin{align}
\label{eqn4.6}
&\int d^3r\alpha\mathrm{cos}(2\theta)(\delta\theta)\partial_x^2(\delta\theta)
\nonumber \\
& =-\int d^3r\alpha\partial_x[\mathrm{cos}(2\theta)(\delta\theta)](\partial_x\delta\theta)\nonumber\\
& =\int d^3r[2\alpha\mathrm{sin}(2\theta)(\partial_x\theta)(\delta\theta)(\partial_x\delta\theta)-\alpha\mathrm{cos}(2\theta)(\partial_x\delta\theta)^2].
\end{align}
Taking Eq.~(\ref{eqn4.6}) into Eq.~(\ref{eqn4.5}), we get
\begin{align}
\label{eqn4.7}
\delta^2 S_{xx}=&-\int d^3r\{(\partial_x\delta\theta)^2[1-\alpha\mathrm{cos}(2\theta)] \nonumber \\
& \hspace{-0.5cm} 
-2\alpha(\delta\theta)^2[\mathrm{cos}(2\theta)(\partial_x\theta)^2+\mathrm{sin}(2\theta)(\partial_x^2\theta)]\}.
\end{align}
Similarly, we get, 
\begin{align}
\label{eqn4.8}
\delta^2 S_{yy}=& -\int d^3r\{(\partial_y\delta\theta)^2[1+\alpha\mathrm{cos}(2\theta)] \nonumber \\
& \hspace{-0.5cm} +2\alpha(\delta\theta)^2[\mathrm{cos}(2\theta)(\partial_y\theta)^2+\mathrm{sin}(2\theta)(\partial_y^2\theta)]\}.
\end{align}
$S_{xy}$ gives
\begin{align}
\label{eqn4.9}
\delta^2 S_{xy}=&-2\alpha\int d^3r(\delta\theta)[(\partial_x\partial_y\delta\theta)\mathrm{sin}(2\theta) \nonumber \\
& +2(\delta\theta)\mathrm{cos}(2\theta)(\partial_x\partial_y\theta)
+(\partial_x\delta\theta)(\partial_y\theta)\mathrm{cos}(2\theta)
\nonumber \\
& +(\partial_y\delta\theta)(\partial_x\theta)\mathrm{cos}(2\theta)-2(\delta\theta)\mathrm{sin}(2\theta)(\partial_x\theta)(\partial_y\theta)],
\end{align}
where
\begin{align}
\label{eqn4.10}
&\hspace{-0.5cm}-\alpha\int d^3r(\delta\theta)(\partial_x\partial_y\delta\theta)\mathrm{sin}(2\theta)\nonumber \\
& =\alpha\int d^3r(\partial_y\delta\theta)[(\partial_x\delta\theta)\mathrm{sin}(2\theta)+2(\delta\theta)\mathrm{cos}(2\theta)(\partial_x\theta)]\nonumber\\
& =\alpha\int d^3r(\partial_x\delta\theta)[(\partial_y\delta\theta)\mathrm{sin}(2\theta)+2(\delta\theta)\mathrm{cos}(2\theta)(\partial_y\theta)].
\end{align}
Taking Eq.~(\ref{eqn4.10}) into Eq.~(\ref{eqn4.9}), we get 
\begin{align}
\label{eqn4.11}
\delta^2 S_{xy}=&2\alpha\int d^3r\{(\partial_x\delta\theta)(\partial_y\delta\theta)\mathrm{sin}(2\theta) \nonumber \\
& \hspace{-1.5cm}
+2(\delta\theta)^2[\mathrm{sin}(2\theta)(\partial_x\theta)(\partial_y\theta)-\mathrm{cos}(2\theta)(\partial_x\partial_y\theta)]\}.
\end{align}
Besides, we have
\begin{align}
\label{eqn4.12}
-\delta\int d^3r(\delta\theta)(\partial_t^2\theta)=\int d^3r(\delta\partial_t\theta)^2.
\end{align}
Combining Eqs.~(\ref{eqn4.7},\ref{eqn4.8},\ref{eqn4.11},\ref{eqn4.12}) together, we obtain the second-order variation of 
$S$ with respect to small $\delta \theta$, 
\begin{align}
\label{eqn4.13}
S_{\delta\theta}=\frac{1}{2}\delta^2 S=&\int d^3r\big{\{}\frac{1}{2}(\partial_t\delta\theta)^2-\frac{1}{2}(\partial_x\delta\theta)^2[1-\alpha\mathrm{cos(2\theta)}]\nonumber\\
& \hspace{-2cm} 
-\frac{1}{2}(\partial_y\delta\theta)^2[1+\alpha\mathrm{cos}(2\theta)]+\alpha(\partial_x\delta\theta)(\partial_y\delta\theta)\mathrm{sin}(2\theta)\nonumber\\
& \hspace{-2cm}
+\alpha(\delta\theta)^2\{2\mathrm{sin}(2\theta)(\partial_x\theta)(\partial_y\theta)+\mathrm{cos}(2\theta)[(\partial_x\theta)^2-(\partial_y\theta)^2]\nonumber\\
&\hspace{-2cm} 
+\mathrm{sin}(2\theta)(\partial_x^2\theta-\partial_y^2\theta)-2\mathrm{cos}(2\theta)(\partial_x\partial_y\theta)\}\big{\}}.
\end{align}
Here $\theta$ is a solution of the classical EOM. Given such $\theta$, we have only to find those $\delta \theta$ that makes $\delta S_{\delta \theta}[\delta\theta]=0$.

In the following, we focus on the 1D solution    
($\partial_y\theta= \partial_y\delta\theta=0$) for simplicity,   
and neglect the integral over $y$, while the following derivation 
can be generalized to $\partial_y\delta\theta\neq 0$. 
The action becomes in the 1D model 
\begin{align}
\label{eqn4.14}
K\equiv & S_{\delta\theta}|_{\partial_y\theta 
=\partial_y\delta\theta=0}\nonumber\\
=&\int^{\infty}_{-\infty} dt \int_L dx \!\ 
\{\frac{1}{2}(\partial_t\delta\theta)^2-\frac{1}{2}(\partial_x\delta\theta)^2[1-\alpha\mathrm{cos}(2\theta)] \nonumber \\
& +\alpha(\delta\theta)^2[(\partial_x\theta)^2\mathrm{cos}(2\theta)+(\partial^2_x\theta)\mathrm{sin}(2\theta)]\}.  
\end{align}
We substitute into $K$ a perturbative solution of $\theta$ in $\alpha$, e.g.,  Eq.~(\ref{eqn19}), and expand $K$ in powers of $\alpha$. This 
gives
\begin{align}
\label{eqn4.14a}
K =K_0+K_1+K_2+\mathcal{O}(\alpha^3),
\end{align}
where $K_n=\mathcal{O}(\alpha^n)$. $K$ is a quadratic 
function of $\delta \theta$. In terms of a Fourier transform of $\delta\theta_{q,\omega}$, 
\begin{align}
\label{eqn4.15}
\delta\theta_{q,\omega}=\frac{1}{\sqrt{L}}\int_L dx 
\int^{\infty}_{-\infty} dt  \!\ \!\ 
\delta\theta(x,t) \!\ e^{-iqx+i\omega t}, 
\end{align}
the quadratic function can be characterized by matrix elements among wavenumber $q$ and 
frequency $\omega$;
\begin{align}
\label{B-eqK}
K =\frac{1}{L} \sum_{q,q^{\prime}}  \int \frac{d\omega}{2\pi} 
\int \frac{d\omega^{\prime}}{2\pi} 
\!\ \delta\theta_{q,\omega}^\dagger K_{q,\omega;q',\omega'}[\theta] \!\ \delta\theta_{q',\omega'}. 
\end{align}

Let us take a solution $\theta(x,t)$ for the spin-injection model with the straight geometry as an example. We first take a part of $\theta(x,t)$ with only one spatial wavelength from Eqs.~(\ref{eqn3.8},\ref{eqn3.14}), 
\begin{align}
\label{eqn4.16}
\theta(x,t)=\theta_0(x,t)+\theta_1(x,t)+\mathcal{O}(\alpha^2),
\end{align}
\begin{align}
\label{eqn4.17}
&\theta_0(x,t)=-j_0x+j_0k_0t, \nonumber \\
&\theta_1(x,t)=\frac{\alpha}{4(k_0^2-1)}\mathrm{sin}(2j_0x-2j_0k_0t).
\end{align}
An inclusion of 
$\theta_2(x,t)$ shall be given later. 
The zeroth order of $K$ is given by 
\begin{align}
\label{eqn4.19}
K_0=&\frac{1}{2}\int dt \int_L dx \!\ [(\partial_t\delta\theta)^2-(\partial_x\delta\theta)^2]\nonumber \\
=& \frac{1}{2}\sum_{q} \int \frac{d\omega}{2\pi} \!\ \delta\theta_{q,\omega}^\dagger (\omega^2-q^2)\delta\theta_{q,\omega},
\end{align}
A substitution of Eq.~(\ref{eqn4.17}) into ${\cal O}(\alpha)$-terms in Eq.~(\ref{eqn4.14}) gives $K_1$ and $K_2$,  
\begin{align}
\label{eqn4.18}
& \frac{\alpha}{2} (\partial_x  \delta\theta)^2  \mathrm{cos} (2\theta_0+2\theta_1)  
+\alpha(\delta\theta)^2(\partial_x\theta_0+\partial_x\theta_1)^2
\nonumber \\
& \hspace{0.5cm}  \times \mathrm{cos}(2\theta_0+2\theta_1) 
+ \alpha(\delta\theta)^2 (\partial_x^2\theta_0  +\partial_x^2\theta_1)  \mathrm{sin} (2\theta_0+2\theta_1)
\nonumber\\
& =\frac{\alpha}{2}(\partial_x\delta\theta)^2\mathrm{cos}(2\theta_0)+\alpha(\delta\theta)^2(\partial_x\theta_0)^2\mathrm{cos}(2\theta_0)\nonumber\\
&\ \ -\alpha(\partial_x\delta\theta)^2\mathrm{sin}(2\theta_0)\theta_1-2\alpha(\delta\theta)^2(\partial_x\theta_0)(\partial_x\theta_1)\mathrm{cos}(2\theta_0)\nonumber\\
& \ \ -
2\alpha(\delta\theta)^2(\partial_x\theta_0)^2\mathrm{sin}(2\theta_0)\theta_1-\alpha(\delta\theta)^2(\partial_x^2\theta_1)\mathrm{sin}(2\theta_0)+\mathcal{O}(\alpha^3)\nonumber\\
& =\frac{\alpha}{2}(\partial_x\delta\theta)^2\mathrm{cos}(2j_0x-2k_0j_0t)+\alpha j_0^2(\delta\theta)^2\mathrm{cos}(2j_0x-2j_0k_0t)\nonumber\\
& \hspace{0.5cm} 
+\frac{\alpha^2(\partial_x\delta\theta)^2}{4(k_0^2-1)}\mathrm{sin}^2(2j_0x-2j_0k_0t) \nonumber \\
& \hspace{1.5cm} -\frac{\alpha^2j_0^2(\delta\theta)^2}{k_0^2-1}\mathrm{cos}^2(2j_0x-2j_0k_0t)\nonumber\\
&\hspace{0.5cm} 
+\frac{\alpha^2j_0^2(\delta\theta)^2}{2(k_0^2-1)}\mathrm{sin^2}(2j_0x-2j_0k_0t) \nonumber \\
& \hspace{1.5cm}  
+\frac{\alpha^2j_0^2(\delta\theta)^2}{k_0^2-1}\mathrm{sin}^2(2j_0x-2j_0k_0t)+\mathrm{O}(\alpha^3).
\end{align}
Equivalently, we have
\begin{widetext}
\begin{align}
\label{eqn4.20}
K_1=&\alpha\int dt \int_L dx \!\ [\frac{1}{2}(\partial_x\delta\theta)^2\mathrm{cos}(2j_0x-2j_0k_0t) 
+j_0^2(\delta\theta)^2\mathrm{cos}(2j_0x-2j_0k_0t)]\nonumber\\
=&\alpha\int dt \int_L dx \!\ [\frac{1}{4 L}\sum_{q,q'} \int_{\omega, \omega'} qq'\delta\theta_{q,\omega}^\dagger\delta\theta_{q',\omega'} +\frac{j_0^2}{2L}\sum_{q,q^{\prime}}
\int_{\omega,\omega'} \!\ \delta\theta_{q,\omega}^\dagger\delta\theta_{q',\omega'}] 
e^{-iqx+i\omega t}e^{iq'x-i\omega' t}e^{2ij_0x-2ij_0k_0t}+\mathrm{H.c.}\nonumber\\
=&\frac{\alpha}{4}\sum_{q} \int_{\omega} \!\ \delta\theta^\dagger_{q+2j_0,\omega+2j_0k_0}\delta\theta_{q,\omega}[q(q+2j_0)+2j_0^2]+\mathrm{H.c.} 
=\frac{\alpha}{4}\sum_{q} \int_{\omega} \!\ \delta\theta^\dagger_{q+j_0,\omega+j_0k_0}\delta\theta_{q-j_0,\omega-j_0k_0}  (q^2+j_0^2)e^{-2ij_0k_0t}+\mathrm{H.c.},
\end{align}
\begin{align}
\label{eqn4.21}
K_2=&\alpha^2 \int dt \int_L dx \!\  \Big\{
 \frac{(\partial_x\delta\theta)^2}{8(k_0^2-1)}[1-\mathrm{cos}(4j_0x-4j_0k_0t)]  
 -\frac{j_0^2(\delta\theta)^2}{k_0^2-1}\mathrm{cos}(4j_0x-4j_0k_0t) 
 +\frac{j_0^2(\delta\theta)^2}{4(k_0^2-1)}[1-\mathrm{cos}(4j_0x-4j_0k_0t)] \Big\}\nonumber\\
=&\frac{\alpha^2}{8(k_0^2-1)}\sum_{q} \int\frac{d\omega}{2\pi} \!\ 
\delta\theta^\dagger_{q,\omega}\delta\theta_{q,\omega}q^2 
+\frac{\alpha^2j_0^2}{4(k_0^2-1)}\sum_{q} \int \frac{d\omega}{2\pi} \!\ \delta\theta^\dagger_{q,\omega}\delta\theta_{q,\omega} \nonumber \\
& -\frac{\alpha^2}{16(k_0^2-1)}\sum_{q} \int \frac{d\omega}{2\pi} \!\  
\Big\{\delta 
\theta^\dagger_{q+2j_0,\omega+2j_0k_0}
\delta\theta_{q-2j_0,\omega-2j_0k_0} 
\big[(q^2-4j_0^2)+8j_0^2+2j_0^2 \big]e^{-4ij_0k_0t}
+\mathrm{H.c.} \Big\}\nonumber\\
=&\frac{\alpha^2}{8(k_0^2-1)}\sum_{q} 
\int \frac{d\omega}{2\pi} \!\ 
\delta\theta^\dagger_{q,\omega}\delta\theta_{q,\omega}(2j_0^2+q^2) 
-\frac{\alpha^2}{16(k_0^2-1)}\sum_{q} \int \frac{d\omega}{2\pi} \!\ 
\big[
\delta\theta^\dagger_{q+2j_0,\omega+2j_0k_0}\delta\theta_{q-2j_0,\omega-2j_0k_0}(q^2+6j_0^2)+\mathrm{H.c.} \big].
\end{align}
\end{widetext}
Taking Eqs.~(\ref{eqn4.19},\ref{eqn4.20},\ref{eqn4.21}) into Eq.~(\ref{eqn4.15}), we obtain the matrix elements,
\begin{align}
\label{eqn4.22}
& \frac{1}{2\pi} 
K_{q+p,\omega+\nu;q-p,\omega-\nu}=(\omega^2-q^2)\delta_{p,0} \delta(\nu) 
\nonumber \\
& +\frac{\alpha(q^2+j_0^2)}{2}(\delta_{p,j_0} 
\delta(\nu-j_0k_0)  
+\delta_{p,-j_0} \delta(\nu+j_0k_0)) \nonumber\\
&-\frac{\alpha^2}{8(k_0^2-1)}[-2(2j_0^2+q^2)\delta_{p,0}  \delta(\nu) 
\nonumber \\
& + (q^2+6j_0^2)(\delta_{p,2j_0} 
\delta(\nu-2j_0k_0) 
+\delta_{p,-2j_0}\delta(\nu+2j_0k_0))]\nonumber \\
& +\mathcal{O}(\alpha^3).
\end{align}

To find $\delta \theta$ that satisfies $\delta S_{\delta\theta}[\delta \theta]=0$, 
we only have to find an eigenmode of $K$ in Eq.~(\ref{B-eqK}) that belongs to 
zero eigenvalue (``eigenenergy"). At the zeroth order in $\alpha$, eigenmodes of $K$ 
are characterized by $q$ and $\omega$, and the ``zero-energy" eigenmodes 
are obtained by setting $\omega$ to be $q$ (on-shell condition). When 
$\alpha$ is included perturbatively, eigenmodes 
at $q$ and $\omega$ hybridize with eigenmodes at $q\pm 2j_0$ and $\omega \pm 2j_0 k_0$ as well as eigenmodes at $q\pm 4j_0$ and 
$\omega \pm 4j_0 k_0$ in terms of off-diagonal mixing terms. Due to the off-diagonal mixing terms, eigenmodes of $K$ are characterized by $q$ and $\omega$ 
modulo $2j_0$ and $2j_0k_0$ respectively, and 
$(q,\omega) \in [-j_0,j_0]\times [-k_0 j_0,k_0j_0]$ plays a role of a 
first Brillouin zone. In the Brillouin zone, eigenmodes at the same 
$(q,\omega)$ are distinguished by a band index $n$, 
\begin{align}
\label{eqn4.26}
K= \frac{1}{2} \sum_{n \in \mathbb{N}} \sum_{-j_0\leq q<j_0} 
\int^{k_0j_0}_{-k_0j_0} \frac{d\omega}{2\pi}  \!\  \delta\varphi^\dagger_{q,\omega,n}\Lambda_{q,\omega,n} 
\delta\varphi_{q,\omega,n}+\mathcal{O}(\alpha^3),  
\end{align}
with  
\begin{align}
\label{eqn4.27}
\delta\theta_{q+2j_0m_1,\omega+2k_0j_0m_2}=\sum_{n \in \mathbb{N}}c_{q,\omega,n;m_1,m_2} \delta\varphi_{q,\omega,n},
\end{align}
$m_1\in\mathbb{Z}$, $m_2\in\mathbb{Z}$. Here, $c_{q,\omega,n;m_1,m_2}$ is analogous to the periodic part of a Bloch wavefunction 
in the band theory. 
From Eq.~(\ref{eqn4.22}) $K_{q,\omega;q,\omega'}$ is real symmetric, so that $c_{q,\omega,n;m_1,m_2}$ are real. An eigenstate of the lowest energy, 
say $\delta \varphi_{q,\omega,n=0}$, must approach $\delta\theta_{q,\omega}$ 
in the limit of 
$\alpha\rightarrow 0$. Such lowest eigenmode ($n=0$) is calculated up to the first order in $\alpha$ as follows, 
\begin{align}
\label{eqn4.28}
&c_{q,\omega,0;m_1,m_2}=\delta_{m_1,0}\delta_{m_2,0} \nonumber \\
& -\frac{\alpha[\frac{(q+j_0)^2+j_0^2}{2}]}{(\omega+2j_0k_0)^2-(q+2j_0)^2-\omega^2+q^2}\delta_{m_1,1}\delta_{m_2,1}\nonumber\\
& -\frac{\alpha[\frac{(q-j_0)^2+j_0^2}{2}]}{(\omega-2j_0k_0)^2-(q-2j_0)^2-\omega^2+q^2}\delta_{m_1,-1}\delta_{m_2,-1}  +\mathcal{O}(\alpha^2)\nonumber\\
& \ \ =\delta_{m_1,0}\delta_{m_2,0}-\alpha\frac{1+\mathcal{O}(q,\omega)}{4(k_0^2-1)}\delta_{m_1,1}\delta_{m_2,1} \nonumber \\
& \ \ -\alpha\frac{1+\mathcal{O}(q,\omega)}{4(k_0^2-1)}\delta_{m_1,-1}\delta_{m_2,-1}+\mathcal{O}(\alpha^2).
\end{align}
A corresponding ``eigenenergy" is calculated up to the 2nd order, 
\begin{align}
\label{eqn4.23}
\Lambda_{q,\omega,0}=&\omega^2-q^2+\frac{\alpha^2(2j_0^2+q^2)}{4(k_0^2-1)}\nonumber\\
&-\frac{\alpha^2[\frac{(q+j_0)^2+j_0^2}{2}]^2}{(\omega+2j_0k_0)^2-(q+2j_0)^2-\omega^2+q^2} \nonumber \\
& -\frac{\alpha^2[\frac{(q-j_0)^2+j_0^2}{2}]^2}{(\omega-2j_0k_0)^2-(q-2j_0)^2-\omega^2+q^2}+\mathcal{O}(\alpha^3)\nonumber\\
=&\omega^2-q^2+\frac{\alpha^2(2j_0^2+q^2)}{4(k_0^2-1)} \nonumber \\
& -\frac{\alpha^2(2j_0^2+2qj_0+q^2)^2}{4[4j_0^2(k_0^2-1)+4qj_0-4\omega j_0k_0]}\nonumber\\
&-\frac{\alpha^2(2j_0^2-2qj_0+q^2)^2}{4[4j_0^2(k_0^2-1)-4qj_0+4\omega j_0k_0]}+\mathcal{O}(\alpha^3),
\end{align}
namely
\begin{align}
\label{eqn4.24}
\Lambda_{q,\omega,0}=&\omega^2+\frac{\alpha^2j_0^2}{2}\frac{1}{k_0^2-1}-q^2[1-\frac{\alpha^2}{4(k_0^2-1)}]\nonumber\\
&-\frac{\alpha^2}{16j_0^2(k_0^2-1)}(4j_0^4+8qj_0^3+8q^2j_0^2) \nonumber \\
&\hspace{-0.5cm} \times 
[1-\frac{q-\omega k_0}{j_0(k_0^2-1)}+\frac{(q-\omega k_0)^2}{j_0^2(k_0^2-1)^2}+\mathcal{O}((q+\omega k_0)^3)]\nonumber\\
&-\frac{\alpha^2}{16j_0^2(k_0^2-1)}(4j_0^4-8qj_0^3+8q^2j_0^2) \nonumber \\
& \hspace{-0.5cm} \times [1+\frac{q-\omega k_0}{j_0(k_0^2-1)}+\frac{(q-\omega k_0)^2}{j_0^2(k_0^2-1)^2}+\mathcal{O}((q+\omega k_0)^3)] \nonumber \\
& \ \ +\mathcal{O}(\alpha^3)\nonumber\\
=&\omega^2+\frac{\alpha^2 j_0^2}{2}\frac{1}{k_0^2-1}-q^2[1-\frac{\alpha^2}{4(k_0^2-1)}]
\nonumber \\
& -\frac{\alpha^2 j_0^2}{2(k_0^2-1)}+\mathcal{O}(\alpha^2\omega^2,\alpha^2q^2,\alpha^2\omega q,\alpha^3)\nonumber\\
=&\omega^2-q^2[1-\frac{\alpha^2}{4(k_0^2-1)}]+\mathcal{O}(\alpha^2\omega^2,\alpha^2q^2,\alpha^2\omega q,\alpha^3).
\end{align} 
The lowest energy band indicates that $\delta \theta$ evaluated on shell, $\Lambda_{q,\omega,0}=0$, behaves like a gapless classical wave. 
This is because the original theory, Eq.~(\ref{eqn5}), has a spacetime translational symmetry. Thus, for any $\theta$, one can choose $\delta \theta$ as a  translation of $\theta$, and such $\delta\theta$ does not change the Lagrangian. For a general $k_0$ (off the resonance point; $k_0 \ne 1$), $\alpha$ can be treated perturbatively, and the classical wave up to the 2nd order in 
$\alpha$ has a well-defined (i.e. real-valued) velocity $v$, 
\begin{align}
\label{eqn4.25}
1-\frac{\alpha^2}{4(k_0^2-1)}=1+\mathcal{O}(\alpha^2) \equiv v^2>0.
\end{align}
By evaluating the eigenmode on shell ($|\omega| = |q| + {\cal O}(\alpha^2) |q|$), we finally determine the first-order $\delta \theta_1$ for an arbitrary form of $\delta \theta_0$ given by Eq.~(\ref{B-start}),  
\begin{align}
\label{eqn4.30}
\delta \theta_1(x,t) &= \frac{1}{\sqrt{L}} \sum^{|q|<j_0}_q 
\int^{k_0j_0}_{-k_0 j_0} d\omega 
\Big[\delta(q-\omega)d_q +\delta(q+\omega)d'_q\Big] \nonumber \\ 
&\hspace{0.2cm} 
\times\Big[-\alpha\frac{1+\mathcal{O}(q)}{4(k_0^2-1)}e^{i(q+2j_0)x-i(\omega+2j_0k_0)t} \nonumber \\
& \hspace{0.2cm} 
-\alpha\frac{1+\mathcal{O}(q)}{4(k_0^2-1)}e^{i(q-2j_0)x-i(\omega-2j_0k_0)t}\Big].
\end{align}

Finally, let us include $\theta_2(x,t)$ into Eq.~(\ref{eqn4.16}), 
\begin{align}
\label{eqn4.31}
\theta(x,t)=\theta_0(x,t)+\theta_1(x,t)+\theta_2(x,t)+\mathcal{O}(\alpha^2),
\end{align}
\begin{align}
\label{eqn4.32}
\theta_2(x,t)=&2\alpha\mathrm{Re}(g)\mathrm{cos}[2k_0j_0(t-x)] \nonumber \\
& +2\alpha\mathrm{Re}(g')\mathrm{cos}[2k_0j_0(t+x)]\nonumber\\
&-2\alpha\mathrm{Im}(g)\mathrm{sin}[2k_0j_0(t-x)] \nonumber \\
& -2\alpha\mathrm{Im}(g')\mathrm{sin}[2k_0j_0(t+x)].
\end{align}
Eq.~(\ref{eqn4.21}) has an additional $\mathcal{O}(\alpha^2)$ contribution, 
\begin{align}
\label{eqn4.33}
\Delta K_2=&\int dtdx[-\alpha(\partial_x\delta\theta)^2\mathrm{sin}(2\theta_0)\theta_2 \nonumber \\
& \ \ +2\alpha(\delta\theta)^2(\partial_x\theta_0)(\partial_x\theta_2)\mathrm{cos}(2\theta_0)\nonumber\\
&\ \ -2\alpha(\delta\theta)^2(\partial_x\theta_0)^2\mathrm{sin}(2\theta_0)\theta_2 \nonumber \\
& \ \ \ \ +\alpha(\delta\theta)^2(\partial_x^2\theta_2)\mathrm{sin}(2\theta_0)].
\end{align}
For $k_0\neq 1$, $\Delta K_2$ contributes only to off-diagonal matrix elements of $K_{q,\omega;q,\omega}$, so that it changes neither Eq.~(\ref{eqn4.24}) 
nor Eq.~(\ref{eqn4.30}) at their respective sub-leading order.


\bibliography{Superfluid_SOC}

\providecommand{\noopsort}[1]{}\providecommand{\singleletter}[1]{#1}%
\begin{thebibliography}{48}%
\makeatletter
\providecommand \@ifxundefined [1]{%
 \@ifx{#1\undefined}
}%
\providecommand \@ifnum [1]{%
 \ifnum #1\expandafter \@firstoftwo
 \else \expandafter \@secondoftwo
 \fi
}%
\providecommand \@ifx [1]{%
 \ifx #1\expandafter \@firstoftwo
 \else \expandafter \@secondoftwo
 \fi
}%
\providecommand \natexlab [1]{#1}%
\providecommand \enquote  [1]{``#1''}%
\providecommand \bibnamefont  [1]{#1}%
\providecommand \bibfnamefont [1]{#1}%
\providecommand \citenamefont [1]{#1}%
\providecommand \href@noop [0]{\@secondoftwo}%
\providecommand \href [0]{\begingroup \@sanitize@url \@href}%
\providecommand \@href[1]{\@@startlink{#1}\@@href}%
\providecommand \@@href[1]{\endgroup#1\@@endlink}%
\providecommand \@sanitize@url [0]{\catcode `\\12\catcode `\$12\catcode
  `\&12\catcode `\#12\catcode `\^12\catcode `\_12\catcode `\%12\relax}%
\providecommand \@@startlink[1]{}%
\providecommand \@@endlink[0]{}%
\providecommand \url  [0]{\begingroup\@sanitize@url \@url }%
\providecommand \@url [1]{\endgroup\@href {#1}{\urlprefix }}%
\providecommand \urlprefix  [0]{URL }%
\providecommand \Eprint [0]{\href }%
\providecommand \doibase [0]{https://doi.org/}%
\providecommand \selectlanguage [0]{\@gobble}%
\providecommand \bibinfo  [0]{\@secondoftwo}%
\providecommand \bibfield  [0]{\@secondoftwo}%
\providecommand \translation [1]{[#1]}%
\providecommand \BibitemOpen [0]{}%
\providecommand \bibitemStop [0]{}%
\providecommand \bibitemNoStop [0]{.\EOS\space}%
\providecommand \EOS [0]{\spacefactor3000\relax}%
\providecommand \BibitemShut  [1]{\csname bibitem#1\endcsname}%
\let\auto@bib@innerbib\@empty
\bibitem [{\citenamefont {Kapitza}(1938)}]{Kapitza1938}%
  \BibitemOpen
  \bibfield  {author} {\bibinfo {author} {\bibfnamefont {P.}~\bibnamefont
  {Kapitza}},\ }\bibfield  {title} {\bibinfo {title} {Viscosity of {L}iquid
  {H}elium below the $\lambda$-{P}oint},\ }\href
  {https://doi.org/10.1038/141074a0} {\bibfield  {journal} {\bibinfo  {journal}
  {Nature}\ }\textbf {\bibinfo {volume} {141}},\ \bibinfo {pages} {74}
  (\bibinfo {year} {1938})}\BibitemShut {NoStop}%
\bibitem [{\citenamefont {Allen}\ and\ \citenamefont
  {Misener}(1938)}]{Allen1938}%
  \BibitemOpen
  \bibfield  {author} {\bibinfo {author} {\bibfnamefont {J.~F.}\ \bibnamefont
  {Allen}}\ and\ \bibinfo {author} {\bibfnamefont {A.~D.}\ \bibnamefont
  {Misener}},\ }\bibfield  {title} {\bibinfo {title} {Flow {P}henomena in
  {L}iquid {H}elium {II}},\ }\href {https://doi.org/10.1038/142643a0}
  {\bibfield  {journal} {\bibinfo  {journal} {Nature}\ }\textbf {\bibinfo
  {volume} {142}},\ \bibinfo {pages} {643} (\bibinfo {year}
  {1938})}\BibitemShut {NoStop}%
\bibitem [{\citenamefont {London}(1964)}]{London1964}%
  \BibitemOpen
  \bibfield  {author} {\bibinfo {author} {\bibfnamefont {F.}~\bibnamefont
  {London}},\ }\href@noop {} {\emph {\bibinfo {title} {Superfluids, Volume II:
  Macroscopic Theory of Superfluid Helium}}}\ (\bibinfo  {publisher} {Dover
  Publications},\ \bibinfo {address} {New York},\ \bibinfo {year}
  {1964})\BibitemShut {NoStop}%
\bibitem [{\citenamefont {Anderson}\ \emph {et~al.}(1995)\citenamefont
  {Anderson}, \citenamefont {Ensher}, \citenamefont {Matthews}, \citenamefont
  {Wieman},\ and\ \citenamefont {Cornell}}]{Anderson1995}%
  \BibitemOpen
  \bibfield  {author} {\bibinfo {author} {\bibfnamefont {M.~H.}\ \bibnamefont
  {Anderson}}, \bibinfo {author} {\bibfnamefont {J.~R.}\ \bibnamefont
  {Ensher}}, \bibinfo {author} {\bibfnamefont {M.~R.}\ \bibnamefont
  {Matthews}}, \bibinfo {author} {\bibfnamefont {C.~E.}\ \bibnamefont
  {Wieman}},\ and\ \bibinfo {author} {\bibfnamefont {E.~A.}\ \bibnamefont
  {Cornell}},\ }\bibfield  {title} {\bibinfo {title} {Observation of
  {B}ose-{E}instein {C}ondensation in a {D}ilute {A}tomic {V}apor},\ }\href
  {https://doi.org/10.1126/science.269.5221.198} {\bibfield  {journal}
  {\bibinfo  {journal} {Science}\ }\textbf {\bibinfo {volume} {269}},\ \bibinfo
  {pages} {198} (\bibinfo {year} {1995})}\BibitemShut {NoStop}%
\bibitem [{\citenamefont {Davis}\ \emph {et~al.}(1995)\citenamefont {Davis},
  \citenamefont {Mewes}, \citenamefont {Andrews}, \citenamefont {van Druten},
  \citenamefont {Durfee}, \citenamefont {Kurn},\ and\ \citenamefont
  {Ketterle}}]{Davis1995}%
  \BibitemOpen
  \bibfield  {author} {\bibinfo {author} {\bibfnamefont {K.~B.}\ \bibnamefont
  {Davis}}, \bibinfo {author} {\bibfnamefont {M.~O.}\ \bibnamefont {Mewes}},
  \bibinfo {author} {\bibfnamefont {M.~R.}\ \bibnamefont {Andrews}}, \bibinfo
  {author} {\bibfnamefont {N.~J.}\ \bibnamefont {van Druten}}, \bibinfo
  {author} {\bibfnamefont {D.~S.}\ \bibnamefont {Durfee}}, \bibinfo {author}
  {\bibfnamefont {D.~M.}\ \bibnamefont {Kurn}},\ and\ \bibinfo {author}
  {\bibfnamefont {W.}~\bibnamefont {Ketterle}},\ }\bibfield  {title} {\bibinfo
  {title} {Bose-{E}instein {C}ondensation in a {G}as of {S}odium {A}toms},\
  }\href {https://doi.org/10.1103/PhysRevLett.75.3969} {\bibfield  {journal}
  {\bibinfo  {journal} {Phys. Rev. Lett.}\ }\textbf {\bibinfo {volume} {75}},\
  \bibinfo {pages} {3969} (\bibinfo {year} {1995})}\BibitemShut {NoStop}%
\bibitem [{\citenamefont {Zwierlein}\ \emph {et~al.}(2005)\citenamefont
  {Zwierlein}, \citenamefont {Abo-Shaeer}, \citenamefont {Schirotzek},
  \citenamefont {Schunck},\ and\ \citenamefont {Ketterle}}]{Zwierlein2005}%
  \BibitemOpen
  \bibfield  {author} {\bibinfo {author} {\bibfnamefont {M.~W.}\ \bibnamefont
  {Zwierlein}}, \bibinfo {author} {\bibfnamefont {J.~R.}\ \bibnamefont
  {Abo-Shaeer}}, \bibinfo {author} {\bibfnamefont {A.}~\bibnamefont
  {Schirotzek}}, \bibinfo {author} {\bibfnamefont {C.~H.}\ \bibnamefont
  {Schunck}},\ and\ \bibinfo {author} {\bibfnamefont {W.}~\bibnamefont
  {Ketterle}},\ }\bibfield  {title} {\bibinfo {title} {Vortices and
  superfluidity in a strongly interacting fermi gas},\ }\href
  {https://doi.org/10.1038/nature03858} {\bibfield  {journal} {\bibinfo
  {journal} {Nature}\ }\textbf {\bibinfo {volume} {435}},\ \bibinfo {pages}
  {1047} (\bibinfo {year} {2005})}\BibitemShut {NoStop}%
\bibitem [{\citenamefont {Goldstone}(1961)}]{Goldstone1961}%
  \BibitemOpen
  \bibfield  {author} {\bibinfo {author} {\bibfnamefont {J.}~\bibnamefont
  {Goldstone}},\ }\bibfield  {title} {\bibinfo {title} {Field theories with
  {S}uperconductor solutions},\ }\href {https://doi.org/10.1007/bf02812722}
  {\bibfield  {journal} {\bibinfo  {journal} {Il Nuovo Cimento}\ }\textbf
  {\bibinfo {volume} {19}},\ \bibinfo {pages} {154} (\bibinfo {year}
  {1961})}\BibitemShut {NoStop}%
\bibitem [{\citenamefont {Nambu}\ and\ \citenamefont
  {Jona-Lasinio}(1961{\natexlab{a}})}]{Nambu1961_1}%
  \BibitemOpen
  \bibfield  {author} {\bibinfo {author} {\bibfnamefont {Y.}~\bibnamefont
  {Nambu}}\ and\ \bibinfo {author} {\bibfnamefont {G.}~\bibnamefont
  {Jona-Lasinio}},\ }\bibfield  {title} {\bibinfo {title} {Dynamical {M}odel of
  {E}lementary {P}articles {B}ased on an {A}nalogy with {S}uperconductivity.
  {I}},\ }\href {https://doi.org/10.1103/PhysRev.122.345} {\bibfield  {journal}
  {\bibinfo  {journal} {Phys. Rev.}\ }\textbf {\bibinfo {volume} {122}},\
  \bibinfo {pages} {345} (\bibinfo {year} {1961}{\natexlab{a}})}\BibitemShut
  {NoStop}%
\bibitem [{\citenamefont {Nambu}\ and\ \citenamefont
  {Jona-Lasinio}(1961{\natexlab{b}})}]{Nambu1961_2}%
  \BibitemOpen
  \bibfield  {author} {\bibinfo {author} {\bibfnamefont {Y.}~\bibnamefont
  {Nambu}}\ and\ \bibinfo {author} {\bibfnamefont {G.}~\bibnamefont
  {Jona-Lasinio}},\ }\bibfield  {title} {\bibinfo {title} {Dynamical {M}odel of
  {E}lementary {P}articles {B}ased on an {A}nalogy with {S}uperconductivity.
  {II}},\ }\href {https://doi.org/10.1103/PhysRev.124.246} {\bibfield
  {journal} {\bibinfo  {journal} {Phys. Rev.}\ }\textbf {\bibinfo {volume}
  {124}},\ \bibinfo {pages} {246} (\bibinfo {year}
  {1961}{\natexlab{b}})}\BibitemShut {NoStop}%
\bibitem [{\citenamefont {Vuorio}(1974)}]{Vuorio1974}%
  \BibitemOpen
  \bibfield  {author} {\bibinfo {author} {\bibfnamefont {M.}~\bibnamefont
  {Vuorio}},\ }\bibfield  {title} {\bibinfo {title} {Condensate spin currents
  in helium-3},\ }\href {https://doi.org/10.1088/0022-3719/7/1/002} {\bibfield
  {journal} {\bibinfo  {journal} {Journal of Physics C: Solid State Physics}\
  }\textbf {\bibinfo {volume} {7}},\ \bibinfo {pages} {L5} (\bibinfo {year}
  {1974})}\BibitemShut {NoStop}%
\bibitem [{\citenamefont {Sonin}(1978{\natexlab{a}})}]{Sonin1978_1}%
  \BibitemOpen
  \bibfield  {author} {\bibinfo {author} {\bibfnamefont {E.~B.}\ \bibnamefont
  {Sonin}},\ }\bibfield  {title} {\bibinfo {title} {Analogs of superfluid flows
  for spins and electron-hole pairs},\ }\href@noop {} {\bibfield  {journal}
  {\bibinfo  {journal} {Sov. Phys. JETP}\ }\textbf {\bibinfo {volume} {47}},\
  \bibinfo {pages} {1091} (\bibinfo {year} {1978}{\natexlab{a}})}\BibitemShut
  {NoStop}%
\bibitem [{\citenamefont {Grein}\ \emph {et~al.}(2009)\citenamefont {Grein},
  \citenamefont {Eschrig}, \citenamefont {Metalidis},\ and\ \citenamefont
  {Sch\"on}}]{Grein2009}%
  \BibitemOpen
  \bibfield  {author} {\bibinfo {author} {\bibfnamefont {R.}~\bibnamefont
  {Grein}}, \bibinfo {author} {\bibfnamefont {M.}~\bibnamefont {Eschrig}},
  \bibinfo {author} {\bibfnamefont {G.}~\bibnamefont {Metalidis}},\ and\
  \bibinfo {author} {\bibfnamefont {G.}~\bibnamefont {Sch\"on}},\ }\bibfield
  {title} {\bibinfo {title} {Spin-{D}ependent {C}ooper {P}air {P}hase and
  {P}ure {S}pin {S}upercurrents in {S}trongly {P}olarized {F}erromagnets},\
  }\href {https://doi.org/10.1103/PhysRevLett.102.227005} {\bibfield  {journal}
  {\bibinfo  {journal} {Phys. Rev. Lett.}\ }\textbf {\bibinfo {volume} {102}},\
  \bibinfo {pages} {227005} (\bibinfo {year} {2009})}\BibitemShut {NoStop}%
\bibitem [{\citenamefont {Sonin}(2010)}]{Sonin2010}%
  \BibitemOpen
  \bibfield  {author} {\bibinfo {author} {\bibfnamefont {E.~B.}\ \bibnamefont
  {Sonin}},\ }\bibfield  {title} {\bibinfo {title} {Spin currents and spin
  superfluidity},\ }\href {https://doi.org/10.1080/00018731003739943}
  {\bibfield  {journal} {\bibinfo  {journal} {Advances in Physics}\ }\textbf
  {\bibinfo {volume} {59}},\ \bibinfo {pages} {181} (\bibinfo {year}
  {2010})}\BibitemShut {NoStop}%
\bibitem [{\citenamefont {Takei}\ and\ \citenamefont
  {Tserkovnyak}(2014)}]{Takei2014_1}%
  \BibitemOpen
  \bibfield  {author} {\bibinfo {author} {\bibfnamefont {S.}~\bibnamefont
  {Takei}}\ and\ \bibinfo {author} {\bibfnamefont {Y.}~\bibnamefont
  {Tserkovnyak}},\ }\bibfield  {title} {\bibinfo {title} {Superfluid {S}pin
  {T}ransport {T}hrough {E}asy-{P}lane {F}erromagnetic {I}nsulators},\ }\href
  {https://doi.org/10.1103/PhysRevLett.112.227201} {\bibfield  {journal}
  {\bibinfo  {journal} {Phys. Rev. Lett.}\ }\textbf {\bibinfo {volume} {112}},\
  \bibinfo {pages} {227201} (\bibinfo {year} {2014})}\BibitemShut {NoStop}%
\bibitem [{\citenamefont {Takei}\ \emph {et~al.}(2014)\citenamefont {Takei},
  \citenamefont {Halperin}, \citenamefont {Yacoby},\ and\ \citenamefont
  {Tserkovnyak}}]{Takei2014_2}%
  \BibitemOpen
  \bibfield  {author} {\bibinfo {author} {\bibfnamefont {S.}~\bibnamefont
  {Takei}}, \bibinfo {author} {\bibfnamefont {B.~I.}\ \bibnamefont {Halperin}},
  \bibinfo {author} {\bibfnamefont {A.}~\bibnamefont {Yacoby}},\ and\ \bibinfo
  {author} {\bibfnamefont {Y.}~\bibnamefont {Tserkovnyak}},\ }\bibfield
  {title} {\bibinfo {title} {Superfluid spin transport through
  antiferromagnetic insulators},\ }\href
  {https://doi.org/10.1103/PhysRevB.90.094408} {\bibfield  {journal} {\bibinfo
  {journal} {Phys. Rev. B}\ }\textbf {\bibinfo {volume} {90}},\ \bibinfo
  {pages} {094408} (\bibinfo {year} {2014})}\BibitemShut {NoStop}%
\bibitem [{\citenamefont {Yuan}\ \emph {et~al.}(2018)\citenamefont {Yuan},
  \citenamefont {Zhu}, \citenamefont {Su}, \citenamefont {Yao}, \citenamefont
  {Xing}, \citenamefont {Chen}, \citenamefont {Ma}, \citenamefont {Lin},
  \citenamefont {Shi}, \citenamefont {Shindou}, \citenamefont {Xie},\ and\
  \citenamefont {Han}}]{Yuan2018}%
  \BibitemOpen
  \bibfield  {author} {\bibinfo {author} {\bibfnamefont {W.}~\bibnamefont
  {Yuan}}, \bibinfo {author} {\bibfnamefont {Q.}~\bibnamefont {Zhu}}, \bibinfo
  {author} {\bibfnamefont {T.}~\bibnamefont {Su}}, \bibinfo {author}
  {\bibfnamefont {Y.}~\bibnamefont {Yao}}, \bibinfo {author} {\bibfnamefont
  {W.}~\bibnamefont {Xing}}, \bibinfo {author} {\bibfnamefont {Y.}~\bibnamefont
  {Chen}}, \bibinfo {author} {\bibfnamefont {Y.}~\bibnamefont {Ma}}, \bibinfo
  {author} {\bibfnamefont {X.}~\bibnamefont {Lin}}, \bibinfo {author}
  {\bibfnamefont {J.}~\bibnamefont {Shi}}, \bibinfo {author} {\bibfnamefont
  {R.}~\bibnamefont {Shindou}}, \bibinfo {author} {\bibfnamefont {X.~C.}\
  \bibnamefont {Xie}},\ and\ \bibinfo {author} {\bibfnamefont {W.}~\bibnamefont
  {Han}},\ }\bibfield  {title} {\bibinfo {title} {Experimental signatures of
  spin superfluid ground state in canted antiferromagnet {C}r${}_2${O}${}_3$
  via nonlocal spin transport},\ }\href
  {https://doi.org/10.1126/sciadv.aat1098} {\bibfield  {journal} {\bibinfo
  {journal} {Science Advances}\ }\textbf {\bibinfo {volume} {4}},\ \bibinfo
  {pages} {eaat1098} (\bibinfo {year} {2018})}\BibitemShut {NoStop}%
\bibitem [{\citenamefont {Mao}\ and\ \citenamefont {Sun}(2022)}]{Mao2022}%
  \BibitemOpen
  \bibfield  {author} {\bibinfo {author} {\bibfnamefont {Y.}~\bibnamefont
  {Mao}}\ and\ \bibinfo {author} {\bibfnamefont {Q.-F.}\ \bibnamefont {Sun}},\
  }\bibfield  {title} {\bibinfo {title} {Spin phase regulated spin josephson
  supercurrent in topological superconductor},\ }\href
  {https://doi.org/10.1103/PhysRevB.105.184511} {\bibfield  {journal} {\bibinfo
   {journal} {Phys. Rev. B}\ }\textbf {\bibinfo {volume} {105}},\ \bibinfo
  {pages} {184511} (\bibinfo {year} {2022})}\BibitemShut {NoStop}%
\bibitem [{\citenamefont {Lozovik}\ and\ \citenamefont
  {Yudson}(1975)}]{Lozovik1975}%
  \BibitemOpen
  \bibfield  {author} {\bibinfo {author} {\bibfnamefont {Y.~E.}\ \bibnamefont
  {Lozovik}}\ and\ \bibinfo {author} {\bibfnamefont {V.~I.}\ \bibnamefont
  {Yudson}},\ }\bibfield  {title} {\bibinfo {title} {Feasibility of
  superfluidity of paired spatially separated electrons and holes; a new
  superconductivity mechanism},\ }\href@noop {} {\bibfield  {journal} {\bibinfo
   {journal} {JETP Lett.}\ }\textbf {\bibinfo {volume} {22}},\ \bibinfo {pages}
  {274} (\bibinfo {year} {1975})}\BibitemShut {NoStop}%
\bibitem [{\citenamefont {Sonin}(1977)}]{Sonin1977}%
  \BibitemOpen
  \bibfield  {author} {\bibinfo {author} {\bibfnamefont {E.~B.}\ \bibnamefont
  {Sonin}},\ }\bibfield  {title} {\bibinfo {title} {Superfluidity of bose
  condensate of electron-hole pairs},\ }\href@noop {} {\bibfield  {journal}
  {\bibinfo  {journal} {JETP Lett.}\ }\textbf {\bibinfo {volume} {25}},\
  \bibinfo {pages} {84} (\bibinfo {year} {1977})}\BibitemShut {NoStop}%
\bibitem [{\citenamefont {Sonin}(1978{\natexlab{b}})}]{Sonin1978_2}%
  \BibitemOpen
  \bibfield  {author} {\bibinfo {author} {\bibfnamefont {E.~B.}\ \bibnamefont
  {Sonin}},\ }\bibfield  {title} {\bibinfo {title} {Phase fixation, excitonic
  and spin superfluidity of electron-hole pairs and antiferromagnetic
  chromium},\ }\href
  {https://doi.org/https://doi.org/10.1016/0038-1098(78)90225-9} {\bibfield
  {journal} {\bibinfo  {journal} {Solid State Communication}\ }\textbf
  {\bibinfo {volume} {25}},\ \bibinfo {pages} {253} (\bibinfo {year}
  {1978}{\natexlab{b}})}\BibitemShut {NoStop}%
\bibitem [{\citenamefont {Eisenstein}\ and\ \citenamefont
  {MacDonald}(2004)}]{Eisenstein2004}%
  \BibitemOpen
  \bibfield  {author} {\bibinfo {author} {\bibfnamefont {J.~P.}\ \bibnamefont
  {Eisenstein}}\ and\ \bibinfo {author} {\bibfnamefont {A.~H.}\ \bibnamefont
  {MacDonald}},\ }\bibfield  {title} {\bibinfo {title}
  {Bose{\textendash}{E}instein condensation of excitons in bilayer electron
  systems},\ }\href {https://doi.org/10.1038/nature03081} {\bibfield  {journal}
  {\bibinfo  {journal} {Nature}\ }\textbf {\bibinfo {volume} {432}},\ \bibinfo
  {pages} {691} (\bibinfo {year} {2004})}\BibitemShut {NoStop}%
\bibitem [{\citenamefont {Li}\ \emph {et~al.}(2017)\citenamefont {Li},
  \citenamefont {Taniguchi}, \citenamefont {Watanabe}, \citenamefont {Hone},\
  and\ \citenamefont {Dean}}]{Li2017}%
  \BibitemOpen
  \bibfield  {author} {\bibinfo {author} {\bibfnamefont {J.~I.~A.}\
  \bibnamefont {Li}}, \bibinfo {author} {\bibfnamefont {T.}~\bibnamefont
  {Taniguchi}}, \bibinfo {author} {\bibfnamefont {K.}~\bibnamefont {Watanabe}},
  \bibinfo {author} {\bibfnamefont {J.}~\bibnamefont {Hone}},\ and\ \bibinfo
  {author} {\bibfnamefont {C.~R.}\ \bibnamefont {Dean}},\ }\bibfield  {title}
  {\bibinfo {title} {Excitonic superfluid phase in double bilayer graphene},\
  }\href {https://doi.org/10.1038/nphys4140} {\bibfield  {journal} {\bibinfo
  {journal} {Nature Physics}\ }\textbf {\bibinfo {volume} {13}},\ \bibinfo
  {pages} {751} (\bibinfo {year} {2017})}\BibitemShut {NoStop}%
\bibitem [{\citenamefont {Zhang}\ and\ \citenamefont
  {Shindou}(2022)}]{Zhang2022}%
  \BibitemOpen
  \bibfield  {author} {\bibinfo {author} {\bibfnamefont {Y.}~\bibnamefont
  {Zhang}}\ and\ \bibinfo {author} {\bibfnamefont {R.}~\bibnamefont
  {Shindou}},\ }\bibfield  {title} {\bibinfo {title} {Dissipationless
  {S}pin-{C}harge {C}onversion in {E}xcitonic {P}seudospin {S}uperfluid},\
  }\href {https://doi.org/10.1103/PhysRevLett.128.066601} {\bibfield  {journal}
  {\bibinfo  {journal} {Phys. Rev. Lett.}\ }\textbf {\bibinfo {volume} {128}},\
  \bibinfo {pages} {066601} (\bibinfo {year} {2022})}\BibitemShut {NoStop}%
\bibitem [{\citenamefont {Nambu}(2004)}]{Nambu2004}%
  \BibitemOpen
  \bibfield  {author} {\bibinfo {author} {\bibfnamefont {Y.}~\bibnamefont
  {Nambu}},\ }\bibfield  {title} {\bibinfo {title} {Spontaneous {B}reaking of
  {L}ie and {C}urrent {A}lgebras},\ }\href
  {https://doi.org/10.1023/b:joss.0000019827.74407.2d} {\bibfield  {journal}
  {\bibinfo  {journal} {Journal of Statistical Physics}\ }\textbf {\bibinfo
  {volume} {115}},\ \bibinfo {pages} {7} (\bibinfo {year} {2004})}\BibitemShut
  {NoStop}%
\bibitem [{\citenamefont {Watanabe}\ and\ \citenamefont
  {Brauner}(2011)}]{Watanabe2011}%
  \BibitemOpen
  \bibfield  {author} {\bibinfo {author} {\bibfnamefont {H.}~\bibnamefont
  {Watanabe}}\ and\ \bibinfo {author} {\bibfnamefont {T.}~\bibnamefont
  {Brauner}},\ }\bibfield  {title} {\bibinfo {title} {Number of
  {N}ambu-{G}oldstone bosons and its relation to charge densities},\ }\href
  {https://doi.org/10.1103/PhysRevD.84.125013} {\bibfield  {journal} {\bibinfo
  {journal} {Phys. Rev. D}\ }\textbf {\bibinfo {volume} {84}},\ \bibinfo
  {pages} {125013} (\bibinfo {year} {2011})}\BibitemShut {NoStop}%
\bibitem [{\citenamefont {Watanabe}\ and\ \citenamefont
  {Murayama}(2012)}]{Watanabe2012}%
  \BibitemOpen
  \bibfield  {author} {\bibinfo {author} {\bibfnamefont {H.}~\bibnamefont
  {Watanabe}}\ and\ \bibinfo {author} {\bibfnamefont {H.}~\bibnamefont
  {Murayama}},\ }\bibfield  {title} {\bibinfo {title} {Unified {D}escription of
  {N}ambu-{G}oldstone {B}osons without {L}orentz {I}nvariance},\ }\href
  {https://doi.org/10.1103/PhysRevLett.108.251602} {\bibfield  {journal}
  {\bibinfo  {journal} {Phys. Rev. Lett.}\ }\textbf {\bibinfo {volume} {108}},\
  \bibinfo {pages} {251602} (\bibinfo {year} {2012})}\BibitemShut {NoStop}%
\bibitem [{\citenamefont {Watanabe}\ and\ \citenamefont
  {Murayama}(2014)}]{Watanabe2014}%
  \BibitemOpen
  \bibfield  {author} {\bibinfo {author} {\bibfnamefont {H.}~\bibnamefont
  {Watanabe}}\ and\ \bibinfo {author} {\bibfnamefont {H.}~\bibnamefont
  {Murayama}},\ }\bibfield  {title} {\bibinfo {title} {Effective {L}agrangian
  for {N}onrelativistic {S}ystems},\ }\href
  {https://doi.org/10.1103/PhysRevX.4.031057} {\bibfield  {journal} {\bibinfo
  {journal} {Phys. Rev. X}\ }\textbf {\bibinfo {volume} {4}},\ \bibinfo {pages}
  {031057} (\bibinfo {year} {2014})}\BibitemShut {NoStop}%
\bibitem [{\citenamefont {Peskin}\ and\ \citenamefont
  {Schroeder}(2018)}]{Peskin2018}%
  \BibitemOpen
  \bibfield  {author} {\bibinfo {author} {\bibfnamefont {M.~E.}\ \bibnamefont
  {Peskin}}\ and\ \bibinfo {author} {\bibfnamefont {D.~V.}\ \bibnamefont
  {Schroeder}},\ }\href@noop {} {\emph {\bibinfo {title} {An Introduction to
  Quantum Field Theory}}}\ (\bibinfo  {publisher} {CRC Press},\ \bibinfo
  {address} {Boca Raton},\ \bibinfo {year} {2018})\BibitemShut {NoStop}%
\bibitem [{\citenamefont {Stanescu}\ \emph {et~al.}(2008)\citenamefont
  {Stanescu}, \citenamefont {Anderson},\ and\ \citenamefont
  {Galitski}}]{Stanescu2008}%
  \BibitemOpen
  \bibfield  {author} {\bibinfo {author} {\bibfnamefont {T.~D.}\ \bibnamefont
  {Stanescu}}, \bibinfo {author} {\bibfnamefont {B.}~\bibnamefont {Anderson}},\
  and\ \bibinfo {author} {\bibfnamefont {V.}~\bibnamefont {Galitski}},\
  }\bibfield  {title} {\bibinfo {title} {Spin-orbit coupled {B}ose-{E}instein
  condensates},\ }\href {https://doi.org/10.1103/PhysRevA.78.023616} {\bibfield
   {journal} {\bibinfo  {journal} {Phys. Rev. A}\ }\textbf {\bibinfo {volume}
  {78}},\ \bibinfo {pages} {023616} (\bibinfo {year} {2008})}\BibitemShut
  {NoStop}%
\bibitem [{\citenamefont {Wang}\ \emph {et~al.}(2010)\citenamefont {Wang},
  \citenamefont {Gao}, \citenamefont {Jian},\ and\ \citenamefont
  {Zhai}}]{Wang2010}%
  \BibitemOpen
  \bibfield  {author} {\bibinfo {author} {\bibfnamefont {C.}~\bibnamefont
  {Wang}}, \bibinfo {author} {\bibfnamefont {C.}~\bibnamefont {Gao}}, \bibinfo
  {author} {\bibfnamefont {C.-M.}\ \bibnamefont {Jian}},\ and\ \bibinfo
  {author} {\bibfnamefont {H.}~\bibnamefont {Zhai}},\ }\bibfield  {title}
  {\bibinfo {title} {Spin-{O}rbit {C}oupled {S}pinor {B}ose-{E}instein
  {C}ondensates},\ }\href {https://doi.org/10.1103/PhysRevLett.105.160403}
  {\bibfield  {journal} {\bibinfo  {journal} {Phys. Rev. Lett.}\ }\textbf
  {\bibinfo {volume} {105}},\ \bibinfo {pages} {160403} (\bibinfo {year}
  {2010})}\BibitemShut {NoStop}%
\bibitem [{\citenamefont {Jian}\ and\ \citenamefont {Zhai}(2011)}]{Jian2011}%
  \BibitemOpen
  \bibfield  {author} {\bibinfo {author} {\bibfnamefont {C.-M.}\ \bibnamefont
  {Jian}}\ and\ \bibinfo {author} {\bibfnamefont {H.}~\bibnamefont {Zhai}},\
  }\bibfield  {title} {\bibinfo {title} {Paired superfluidity and
  fractionalized vortices in systems of spin-orbit coupled bosons},\ }\href
  {https://doi.org/10.1103/PhysRevB.84.060508} {\bibfield  {journal} {\bibinfo
  {journal} {Phys. Rev. B}\ }\textbf {\bibinfo {volume} {84}},\ \bibinfo
  {pages} {060508} (\bibinfo {year} {2011})}\BibitemShut {NoStop}%
\bibitem [{\citenamefont {Ozawa}\ and\ \citenamefont {Baym}(2013)}]{Ozawa2013}%
  \BibitemOpen
  \bibfield  {author} {\bibinfo {author} {\bibfnamefont {T.}~\bibnamefont
  {Ozawa}}\ and\ \bibinfo {author} {\bibfnamefont {G.}~\bibnamefont {Baym}},\
  }\bibfield  {title} {\bibinfo {title} {Condensation {T}ransition of
  {U}ltracold {B}ose {G}ases with {R}ashba {S}pin-{O}rbit {C}oupling},\ }\href
  {https://doi.org/10.1103/PhysRevLett.110.085304} {\bibfield  {journal}
  {\bibinfo  {journal} {Phys. Rev. Lett.}\ }\textbf {\bibinfo {volume} {110}},\
  \bibinfo {pages} {085304} (\bibinfo {year} {2013})}\BibitemShut {NoStop}%
\bibitem [{\citenamefont {Zhou}\ and\ \citenamefont {Cui}(2013)}]{Zhou2013}%
  \BibitemOpen
  \bibfield  {author} {\bibinfo {author} {\bibfnamefont {Q.}~\bibnamefont
  {Zhou}}\ and\ \bibinfo {author} {\bibfnamefont {X.}~\bibnamefont {Cui}},\
  }\bibfield  {title} {\bibinfo {title} {Fate of a {B}ose-{E}instein
  {C}ondensate in the {P}resence of {S}pin-{O}rbit {C}oupling},\ }\href
  {https://doi.org/10.1103/PhysRevLett.110.140407} {\bibfield  {journal}
  {\bibinfo  {journal} {Phys. Rev. Lett.}\ }\textbf {\bibinfo {volume} {110}},\
  \bibinfo {pages} {140407} (\bibinfo {year} {2013})}\BibitemShut {NoStop}%
\bibitem [{\citenamefont {Zhai}(2015)}]{Zhai2015}%
  \BibitemOpen
  \bibfield  {author} {\bibinfo {author} {\bibfnamefont {H.}~\bibnamefont
  {Zhai}},\ }\bibfield  {title} {\bibinfo {title} {Degenerate quantum gases
  with spin-orbit coupling: a review},\ }\href
  {https://doi.org/10.1088/0034-4885/78/2/026001} {\bibfield  {journal}
  {\bibinfo  {journal} {Reports on Progress in Physics}\ }\textbf {\bibinfo
  {volume} {78}},\ \bibinfo {pages} {026001} (\bibinfo {year}
  {2015})}\BibitemShut {NoStop}%
\bibitem [{\citenamefont {Halperin}\ and\ \citenamefont
  {Rice}(1968)}]{Halperin1968}%
  \BibitemOpen
  \bibfield  {author} {\bibinfo {author} {\bibfnamefont {B.~I.}\ \bibnamefont
  {Halperin}}\ and\ \bibinfo {author} {\bibfnamefont {T.~M.}\ \bibnamefont
  {Rice}},\ }\bibfield  {title} {\bibinfo {title} {Possible {A}nomalies at a
  {S}emimetal-{S}emiconductor {T}ransistion},\ }\href
  {https://doi.org/10.1103/RevModPhys.40.755} {\bibfield  {journal} {\bibinfo
  {journal} {Rev. Mod. Phys.}\ }\textbf {\bibinfo {volume} {40}},\ \bibinfo
  {pages} {755} (\bibinfo {year} {1968})}\BibitemShut {NoStop}%
\bibitem [{\citenamefont {Hakio\u{g}lu}\ and\ \citenamefont
  {\c{S}ahin}(2007)}]{Hakioglu2007}%
  \BibitemOpen
  \bibfield  {author} {\bibinfo {author} {\bibfnamefont {T.}~\bibnamefont
  {Hakio\u{g}lu}}\ and\ \bibinfo {author} {\bibfnamefont {M.}~\bibnamefont
  {\c{S}ahin}},\ }\bibfield  {title} {\bibinfo {title} {Excitonic condensation
  under spin-orbit coupling and {BEC-BCS} crossover},\ }\href
  {https://doi.org/10.1103/PhysRevLett.98.166405} {\bibfield  {journal}
  {\bibinfo  {journal} {Phys. Rev. Lett.}\ }\textbf {\bibinfo {volume} {98}},\
  \bibinfo {pages} {166405} (\bibinfo {year} {2007})}\BibitemShut {NoStop}%
\bibitem [{\citenamefont {Pikulin}\ and\ \citenamefont
  {Hyart}(2014)}]{Pikulin2014}%
  \BibitemOpen
  \bibfield  {author} {\bibinfo {author} {\bibfnamefont {D.~I.}\ \bibnamefont
  {Pikulin}}\ and\ \bibinfo {author} {\bibfnamefont {T.}~\bibnamefont
  {Hyart}},\ }\bibfield  {title} {\bibinfo {title} {Interplay of {E}xciton
  {C}ondensation and the {Q}uantum {S}pin {H}all {E}ffect in
  $\mathrm{InAs}/\mathrm{GaSb}$ {B}ilayers},\ }\href
  {https://doi.org/10.1103/PhysRevLett.112.176403} {\bibfield  {journal}
  {\bibinfo  {journal} {Phys. Rev. Lett.}\ }\textbf {\bibinfo {volume} {112}},\
  \bibinfo {pages} {176403} (\bibinfo {year} {2014})}\BibitemShut {NoStop}%
\bibitem [{\citenamefont {Sonin}(1981)}]{Sonin1981}%
  \BibitemOpen
  \bibfield  {author} {\bibinfo {author} {\bibfnamefont {E.~B.}\ \bibnamefont
  {Sonin}},\ }\bibfield  {title} {\bibinfo {title} {Hydrodynamic theory of the
  sine-{G}ordon soliton lattice and soliton propagation in ${}^3${H}e-{A}},\
  }\href {https://doi.org/10.1016/0375-9601(81)90179-1} {\bibfield  {journal}
  {\bibinfo  {journal} {Physics Letters A}\ }\textbf {\bibinfo {volume} {86}},\
  \bibinfo {pages} {113} (\bibinfo {year} {1981})}\BibitemShut {NoStop}%
\bibitem [{\citenamefont {Milde}\ \emph {et~al.}(2013)\citenamefont {Milde},
  \citenamefont {Köhler}, \citenamefont {Seidel}, \citenamefont {Eng},
  \citenamefont {Bauer}, \citenamefont {Chacon}, \citenamefont {Kindervater},
  \citenamefont {Mühlbauer}, \citenamefont {Pfleiderer}, \citenamefont
  {Buhrandt}, \citenamefont {Schütte},\ and\ \citenamefont
  {Rosch}}]{Milde2013}%
  \BibitemOpen
  \bibfield  {author} {\bibinfo {author} {\bibfnamefont {P.}~\bibnamefont
  {Milde}}, \bibinfo {author} {\bibfnamefont {D.}~\bibnamefont {Köhler}},
  \bibinfo {author} {\bibfnamefont {J.}~\bibnamefont {Seidel}}, \bibinfo
  {author} {\bibfnamefont {L.~M.}\ \bibnamefont {Eng}}, \bibinfo {author}
  {\bibfnamefont {A.}~\bibnamefont {Bauer}}, \bibinfo {author} {\bibfnamefont
  {A.}~\bibnamefont {Chacon}}, \bibinfo {author} {\bibfnamefont
  {J.}~\bibnamefont {Kindervater}}, \bibinfo {author} {\bibfnamefont
  {S.}~\bibnamefont {Mühlbauer}}, \bibinfo {author} {\bibfnamefont
  {C.}~\bibnamefont {Pfleiderer}}, \bibinfo {author} {\bibfnamefont
  {S.}~\bibnamefont {Buhrandt}}, \bibinfo {author} {\bibfnamefont
  {C.}~\bibnamefont {Schütte}},\ and\ \bibinfo {author} {\bibfnamefont
  {A.}~\bibnamefont {Rosch}},\ }\bibfield  {title} {\bibinfo {title} {Unwinding
  of a {S}kyrmion {L}attice by {M}agnetic {M}onopoles},\ }\href
  {https://doi.org/10.1126/science.1234657} {\bibfield  {journal} {\bibinfo
  {journal} {Science}\ }\textbf {\bibinfo {volume} {340}},\ \bibinfo {pages}
  {1076} (\bibinfo {year} {2013})}\BibitemShut {NoStop}%
\bibitem [{\citenamefont {Seddon}\ \emph {et~al.}(2021)\citenamefont {Seddon},
  \citenamefont {Dogaru}, \citenamefont {Holt}, \citenamefont {Rusu},
  \citenamefont {Peters}, \citenamefont {Sanchez},\ and\ \citenamefont
  {Alexe}}]{Seddon2021}%
  \BibitemOpen
  \bibfield  {author} {\bibinfo {author} {\bibfnamefont {S.~D.}\ \bibnamefont
  {Seddon}}, \bibinfo {author} {\bibfnamefont {D.~E.}\ \bibnamefont {Dogaru}},
  \bibinfo {author} {\bibfnamefont {S.~J.~R.}\ \bibnamefont {Holt}}, \bibinfo
  {author} {\bibfnamefont {D.}~\bibnamefont {Rusu}}, \bibinfo {author}
  {\bibfnamefont {J.~J.~P.}\ \bibnamefont {Peters}}, \bibinfo {author}
  {\bibfnamefont {A.~M.}\ \bibnamefont {Sanchez}},\ and\ \bibinfo {author}
  {\bibfnamefont {M.}~\bibnamefont {Alexe}},\ }\bibfield  {title} {\bibinfo
  {title} {Real-space observation of ferroelectrically induced magnetic spin
  crystal in {SrRuO}${}_3$},\ }\href
  {https://doi.org/10.1038/s41467-021-22165-5} {\bibfield  {journal} {\bibinfo
  {journal} {Nature Communications}\ }\textbf {\bibinfo {volume} {12}},\
  \bibinfo {pages} {2007} (\bibinfo {year} {2021})}\BibitemShut {NoStop}%
\bibitem [{\citenamefont {Li}\ \emph {et~al.}(2016)\citenamefont {Li},
  \citenamefont {Shelford}, \citenamefont {Shafer}, \citenamefont {Tan},
  \citenamefont {Deng}, \citenamefont {Keatley}, \citenamefont {Hwang},
  \citenamefont {Arenholz}, \citenamefont {van~der Laan}, \citenamefont
  {Hicken},\ and\ \citenamefont {Qiu}}]{Li2016}%
  \BibitemOpen
  \bibfield  {author} {\bibinfo {author} {\bibfnamefont {J.}~\bibnamefont
  {Li}}, \bibinfo {author} {\bibfnamefont {L.~R.}\ \bibnamefont {Shelford}},
  \bibinfo {author} {\bibfnamefont {P.}~\bibnamefont {Shafer}}, \bibinfo
  {author} {\bibfnamefont {A.}~\bibnamefont {Tan}}, \bibinfo {author}
  {\bibfnamefont {J.~X.}\ \bibnamefont {Deng}}, \bibinfo {author}
  {\bibfnamefont {P.~S.}\ \bibnamefont {Keatley}}, \bibinfo {author}
  {\bibfnamefont {C.}~\bibnamefont {Hwang}}, \bibinfo {author} {\bibfnamefont
  {E.}~\bibnamefont {Arenholz}}, \bibinfo {author} {\bibfnamefont
  {G.}~\bibnamefont {van~der Laan}}, \bibinfo {author} {\bibfnamefont {R.~J.}\
  \bibnamefont {Hicken}},\ and\ \bibinfo {author} {\bibfnamefont {Z.~Q.}\
  \bibnamefont {Qiu}},\ }\bibfield  {title} {\bibinfo {title} {Direct
  {D}etection of {P}ure ac {S}pin {C}urrent by {X}-{R}ay {P}ump-{P}robe
  {M}easurements},\ }\href {https://doi.org/10.1103/PhysRevLett.117.076602}
  {\bibfield  {journal} {\bibinfo  {journal} {Phys. Rev. Lett.}\ }\textbf
  {\bibinfo {volume} {117}},\ \bibinfo {pages} {076602} (\bibinfo {year}
  {2016})}\BibitemShut {NoStop}%
\bibitem [{\citenamefont {Landau}(1941)}]{Landau1941}%
  \BibitemOpen
  \bibfield  {author} {\bibinfo {author} {\bibfnamefont {L.}~\bibnamefont
  {Landau}},\ }\bibfield  {title} {\bibinfo {title} {Theory of the
  {S}uperfluidity of {H}elium {II}},\ }\href
  {https://doi.org/10.1103/PhysRev.60.356} {\bibfield  {journal} {\bibinfo
  {journal} {Phys. Rev.}\ }\textbf {\bibinfo {volume} {60}},\ \bibinfo {pages}
  {356} (\bibinfo {year} {1941})}\BibitemShut {NoStop}%
\bibitem [{\citenamefont {Girvin}\ and\ \citenamefont
  {Yang}(2019)}]{Girvin2019}%
  \BibitemOpen
  \bibfield  {author} {\bibinfo {author} {\bibfnamefont {S.~M.}\ \bibnamefont
  {Girvin}}\ and\ \bibinfo {author} {\bibfnamefont {K.}~\bibnamefont {Yang}},\
  }\href@noop {} {\emph {\bibinfo {title} {Modern Condensed Matter Physics}}}\
  (\bibinfo  {publisher} {Cambridge University Press},\ \bibinfo {address}
  {Cambridge},\ \bibinfo {year} {2019})\BibitemShut {NoStop}%
\bibitem [{\citenamefont {Zhai}(2021)}]{Zhai2021}%
  \BibitemOpen
  \bibfield  {author} {\bibinfo {author} {\bibfnamefont {H.}~\bibnamefont
  {Zhai}},\ }\href@noop {} {\emph {\bibinfo {title} {Ultracold Atomic
  Physics}}}\ (\bibinfo  {publisher} {Cambridge University Press},\ \bibinfo
  {address} {Cambridge},\ \bibinfo {year} {2021})\BibitemShut {NoStop}%
\bibitem [{\citenamefont {Srednicki}(2007)}]{Srednicki2007}%
  \BibitemOpen
  \bibfield  {author} {\bibinfo {author} {\bibfnamefont {M.}~\bibnamefont
  {Srednicki}},\ }\href@noop {} {\emph {\bibinfo {title} {Quantum Firld
  Theory}}}\ (\bibinfo  {publisher} {Cambridge University Press},\ \bibinfo
  {address} {Cambridge},\ \bibinfo {year} {2007})\BibitemShut {NoStop}%
\bibitem [{\citenamefont {Taylor}(2004)}]{Taylor2004}%
  \BibitemOpen
  \bibfield  {author} {\bibinfo {author} {\bibfnamefont {J.~R.}\ \bibnamefont
  {Taylor}},\ }\href@noop {} {\emph {\bibinfo {title} {Classical Mechanics}}}\
  (\bibinfo  {publisher} {University Science Books},\ \bibinfo {address}
  {Melville},\ \bibinfo {year} {2004})\BibitemShut {NoStop}%
\bibitem [{\citenamefont {Blankschtein}\ \emph {et~al.}(1984)\citenamefont
  {Blankschtein}, \citenamefont {Ma}, \citenamefont {Berker}, \citenamefont
  {Grest},\ and\ \citenamefont {Soukoulis}}]{blankschtein84}%
  \BibitemOpen
  \bibfield  {author} {\bibinfo {author} {\bibfnamefont {D.}~\bibnamefont
  {Blankschtein}}, \bibinfo {author} {\bibfnamefont {M.}~\bibnamefont {Ma}},
  \bibinfo {author} {\bibfnamefont {A.~N.}\ \bibnamefont {Berker}}, \bibinfo
  {author} {\bibfnamefont {G.~S.}\ \bibnamefont {Grest}},\ and\ \bibinfo
  {author} {\bibfnamefont {C.~M.}\ \bibnamefont {Soukoulis}},\ }\bibfield
  {title} {\bibinfo {title} {Orderings of a stacked frustrated triangular
  system in three dimensions},\ }\href
  {https://doi.org/10.1103/PhysRevB.29.5250} {\bibfield  {journal} {\bibinfo
  {journal} {Phys. Rev. B}\ }\textbf {\bibinfo {volume} {29}},\ \bibinfo
  {pages} {5250} (\bibinfo {year} {1984})}\BibitemShut {NoStop}%
\bibitem [{\citenamefont {Oshikawa}(2000)}]{oshikawa00}%
  \BibitemOpen
  \bibfield  {author} {\bibinfo {author} {\bibfnamefont {M.}~\bibnamefont
  {Oshikawa}},\ }\bibfield  {title} {\bibinfo {title} {Ordered phase and
  scaling in ${Z}_{n}$ models and the three-state antiferromagnetic {P}otts
  model in three dimensions},\ }\href
  {https://doi.org/10.1103/PhysRevB.61.3430} {\bibfield  {journal} {\bibinfo
  {journal} {Phys. Rev. B}\ }\textbf {\bibinfo {volume} {61}},\ \bibinfo
  {pages} {3430} (\bibinfo {year} {2000})}\BibitemShut {NoStop}%
\end{thebibliography}%

\end{document}